\documentclass[longauth]{aaEC}

\usepackage{graphicx}
\usepackage{natbib}
\usepackage{scalerel}
\usepackage{comment}
\usepackage{array}
\usepackage[table]{xcolor}

\usepackage{txfonts}
\usepackage[pdfencoding=auto,psdextra]{hyperref}
\hypersetup{
    colorlinks=true,
    linkcolor=blue,
    filecolor=magenta,      
    urlcolor=blue,
    citecolor=blue
}
\makeatletter
\renewcommand*\aa@pageof{, page \thepage{} of \pageref*{LastPage}}
\makeatother

\usepackage[utf8]{inputenc}

\usepackage[switch, modulo]{lineno}

\usepackage{euclid}

\usepackage{multirow}
\usepackage{placeins}
\usepackage{xspace}
\newcommand{\DESKIDS}{DES$\&$KiDs\xspace}
\newcommand{\Kai}{H26\xspace}

\definecolor{darkraspberry}{rgb}{0.53, 0.15, 0.34}

\newcommand{\referee}{}

\def\der{\mathrm{d}}

\begin{document} 
\title{\Euclid preparation}
\subtitle{CVII. Testing analytic models of galaxy intrinsic alignments in the \Euclid Flagship simulation}

\newcommand{\orcid}[1]{} 
\author{Euclid Collaboration: R.~Paviot\orcid{0009-0002-8108-3460}\thanks{\email{romain.paviot@cea.fr}}\inst{\ref{aff1},\ref{aff2}}
\and B.~Joachimi\orcid{0000-0001-7494-1303}\inst{\ref{aff3}}
\and K.~Hoffmann\orcid{0000-0002-7885-5274}\inst{\ref{aff4}}
\and S.~Codis\inst{\ref{aff1}}
\and I.~Tutusaus\orcid{0000-0002-3199-0399}\inst{\ref{aff4},\ref{aff5},\ref{aff6}}
\and D.~Navarro-Giron\'{e}s\orcid{0000-0003-0507-372X}\inst{\ref{aff7}}
\and J.~Blazek\orcid{0000-0002-4687-4657}\inst{\ref{aff8}}
\and F.~Hervas-Peters\orcid{0009-0008-1839-2969}\inst{\ref{aff1},\ref{aff9}}
\and B.~Altieri\orcid{0000-0003-3936-0284}\inst{\ref{aff10}}
\and S.~Andreon\orcid{0000-0002-2041-8784}\inst{\ref{aff11}}
\and N.~Auricchio\orcid{0000-0003-4444-8651}\inst{\ref{aff12}}
\and C.~Baccigalupi\orcid{0000-0002-8211-1630}\inst{\ref{aff13},\ref{aff14},\ref{aff15},\ref{aff16}}
\and M.~Baldi\orcid{0000-0003-4145-1943}\inst{\ref{aff17},\ref{aff12},\ref{aff18}}
\and S.~Bardelli\orcid{0000-0002-8900-0298}\inst{\ref{aff12}}
\and A.~Biviano\orcid{0000-0002-0857-0732}\inst{\ref{aff14},\ref{aff13}}
\and E.~Branchini\orcid{0000-0002-0808-6908}\inst{\ref{aff19},\ref{aff20},\ref{aff11}}
\and M.~Brescia\orcid{0000-0001-9506-5680}\inst{\ref{aff21},\ref{aff22}}
\and S.~Camera\orcid{0000-0003-3399-3574}\inst{\ref{aff23},\ref{aff24},\ref{aff25}}
\and G.~Ca\~nas-Herrera\orcid{0000-0003-2796-2149}\inst{\ref{aff26},\ref{aff27},\ref{aff7}}
\and V.~Capobianco\orcid{0000-0002-3309-7692}\inst{\ref{aff25}}
\and C.~Carbone\orcid{0000-0003-0125-3563}\inst{\ref{aff28}}
\and V.~F.~Cardone\inst{\ref{aff29},\ref{aff30}}
\and J.~Carretero\orcid{0000-0002-3130-0204}\inst{\ref{aff31},\ref{aff32}}
\and S.~Casas\orcid{0000-0002-4751-5138}\inst{\ref{aff33}}
\and F.~J.~Castander\orcid{0000-0001-7316-4573}\inst{\ref{aff4},\ref{aff5}}
\and M.~Castellano\orcid{0000-0001-9875-8263}\inst{\ref{aff29}}
\and G.~Castignani\orcid{0000-0001-6831-0687}\inst{\ref{aff12}}
\and S.~Cavuoti\orcid{0000-0002-3787-4196}\inst{\ref{aff22},\ref{aff34}}
\and K.~C.~Chambers\orcid{0000-0001-6965-7789}\inst{\ref{aff35}}
\and A.~Cimatti\inst{\ref{aff36}}
\and C.~Colodro-Conde\inst{\ref{aff37}}
\and G.~Congedo\orcid{0000-0003-2508-0046}\inst{\ref{aff38}}
\and L.~Conversi\orcid{0000-0002-6710-8476}\inst{\ref{aff39},\ref{aff10}}
\and Y.~Copin\orcid{0000-0002-5317-7518}\inst{\ref{aff40}}
\and F.~Courbin\orcid{0000-0003-0758-6510}\inst{\ref{aff41},\ref{aff42}}
\and H.~M.~Courtois\orcid{0000-0003-0509-1776}\inst{\ref{aff43}}
\and A.~Da~Silva\orcid{0000-0002-6385-1609}\inst{\ref{aff44},\ref{aff45}}
\and H.~Degaudenzi\orcid{0000-0002-5887-6799}\inst{\ref{aff46}}
\and S.~de~la~Torre\inst{\ref{aff47}}
\and G.~De~Lucia\orcid{0000-0002-6220-9104}\inst{\ref{aff14}}
\and H.~Dole\orcid{0000-0002-9767-3839}\inst{\ref{aff48}}
\and F.~Dubath\orcid{0000-0002-6533-2810}\inst{\ref{aff46}}
\and C.~A.~J.~Duncan\orcid{0009-0003-3573-0791}\inst{\ref{aff38},\ref{aff49}}
\and X.~Dupac\inst{\ref{aff10}}
\and S.~Dusini\orcid{0000-0002-1128-0664}\inst{\ref{aff50}}
\and S.~Escoffier\orcid{0000-0002-2847-7498}\inst{\ref{aff51}}
\and M.~Farina\orcid{0000-0002-3089-7846}\inst{\ref{aff52}}
\and R.~Farinelli\inst{\ref{aff12}}
\and S.~Farrens\orcid{0000-0002-9594-9387}\inst{\ref{aff1}}
\and S.~Ferriol\inst{\ref{aff40}}
\and F.~Finelli\orcid{0000-0002-6694-3269}\inst{\ref{aff12},\ref{aff53}}
\and P.~Fosalba\orcid{0000-0002-1510-5214}\inst{\ref{aff5},\ref{aff4}}
\and M.~Frailis\orcid{0000-0002-7400-2135}\inst{\ref{aff14}}
\and E.~Franceschi\orcid{0000-0002-0585-6591}\inst{\ref{aff12}}
\and S.~Galeotta\orcid{0000-0002-3748-5115}\inst{\ref{aff14}}
\and K.~George\orcid{0000-0002-1734-8455}\inst{\ref{aff54}}
\and B.~Gillis\orcid{0000-0002-4478-1270}\inst{\ref{aff38}}
\and C.~Giocoli\orcid{0000-0002-9590-7961}\inst{\ref{aff12},\ref{aff18}}
\and J.~Gracia-Carpio\inst{\ref{aff55}}
\and A.~Grazian\orcid{0000-0002-5688-0663}\inst{\ref{aff56}}
\and F.~Grupp\inst{\ref{aff55},\ref{aff57}}
\and S.~V.~H.~Haugan\orcid{0000-0001-9648-7260}\inst{\ref{aff58}}
\and H.~Hoekstra\orcid{0000-0002-0641-3231}\inst{\ref{aff7}}
\and W.~Holmes\inst{\ref{aff59}}
\and F.~Hormuth\inst{\ref{aff60}}
\and A.~Hornstrup\orcid{0000-0002-3363-0936}\inst{\ref{aff61},\ref{aff62}}
\and K.~Jahnke\orcid{0000-0003-3804-2137}\inst{\ref{aff63}}
\and M.~Jhabvala\inst{\ref{aff64}}
\and E.~Keih\"anen\orcid{0000-0003-1804-7715}\inst{\ref{aff65}}
\and S.~Kermiche\orcid{0000-0002-0302-5735}\inst{\ref{aff51}}
\and A.~Kiessling\orcid{0000-0002-2590-1273}\inst{\ref{aff59}}
\and M.~Kilbinger\orcid{0000-0001-9513-7138}\inst{\ref{aff1}}
\and B.~Kubik\orcid{0009-0006-5823-4880}\inst{\ref{aff40}}
\and M.~K\"ummel\orcid{0000-0003-2791-2117}\inst{\ref{aff57}}
\and M.~Kunz\orcid{0000-0002-3052-7394}\inst{\ref{aff66}}
\and H.~Kurki-Suonio\orcid{0000-0002-4618-3063}\inst{\ref{aff67},\ref{aff68}}
\and A.~M.~C.~Le~Brun\orcid{0000-0002-0936-4594}\inst{\ref{aff69}}
\and S.~Ligori\orcid{0000-0003-4172-4606}\inst{\ref{aff25}}
\and P.~B.~Lilje\orcid{0000-0003-4324-7794}\inst{\ref{aff58}}
\and V.~Lindholm\orcid{0000-0003-2317-5471}\inst{\ref{aff67},\ref{aff68}}
\and I.~Lloro\orcid{0000-0001-5966-1434}\inst{\ref{aff70}}
\and G.~Mainetti\orcid{0000-0003-2384-2377}\inst{\ref{aff71}}
\and D.~Maino\inst{\ref{aff72},\ref{aff28},\ref{aff73}}
\and E.~Maiorano\orcid{0000-0003-2593-4355}\inst{\ref{aff12}}
\and O.~Mansutti\orcid{0000-0001-5758-4658}\inst{\ref{aff14}}
\and S.~Marcin\inst{\ref{aff74}}
\and O.~Marggraf\orcid{0000-0001-7242-3852}\inst{\ref{aff75}}
\and M.~Martinelli\orcid{0000-0002-6943-7732}\inst{\ref{aff29},\ref{aff30}}
\and N.~Martinet\orcid{0000-0003-2786-7790}\inst{\ref{aff47}}
\and F.~Marulli\orcid{0000-0002-8850-0303}\inst{\ref{aff76},\ref{aff12},\ref{aff18}}
\and R.~J.~Massey\orcid{0000-0002-6085-3780}\inst{\ref{aff77}}
\and S.~Maurogordato\inst{\ref{aff78}}
\and E.~Medinaceli\orcid{0000-0002-4040-7783}\inst{\ref{aff12}}
\and S.~Mei\orcid{0000-0002-2849-559X}\inst{\ref{aff79},\ref{aff80}}
\and Y.~Mellier\thanks{Deceased}\inst{\ref{aff81},\ref{aff82}}
\and M.~Meneghetti\orcid{0000-0003-1225-7084}\inst{\ref{aff12},\ref{aff18}}
\and E.~Merlin\orcid{0000-0001-6870-8900}\inst{\ref{aff29}}
\and G.~Meylan\inst{\ref{aff83}}
\and A.~Mora\orcid{0000-0002-1922-8529}\inst{\ref{aff84}}
\and M.~Moresco\orcid{0000-0002-7616-7136}\inst{\ref{aff76},\ref{aff12}}
\and L.~Moscardini\orcid{0000-0002-3473-6716}\inst{\ref{aff76},\ref{aff12},\ref{aff18}}
\and C.~Neissner\orcid{0000-0001-8524-4968}\inst{\ref{aff85},\ref{aff32}}
\and S.-M.~Niemi\orcid{0009-0005-0247-0086}\inst{\ref{aff26}}
\and C.~Padilla\orcid{0000-0001-7951-0166}\inst{\ref{aff85}}
\and S.~Paltani\orcid{0000-0002-8108-9179}\inst{\ref{aff46}}
\and F.~Pasian\orcid{0000-0002-4869-3227}\inst{\ref{aff14}}
\and K.~Pedersen\inst{\ref{aff86}}
\and V.~Pettorino\orcid{0000-0002-4203-9320}\inst{\ref{aff26}}
\and S.~Pires\orcid{0000-0002-0249-2104}\inst{\ref{aff1}}
\and G.~Polenta\orcid{0000-0003-4067-9196}\inst{\ref{aff87}}
\and M.~Poncet\inst{\ref{aff2}}
\and L.~A.~Popa\inst{\ref{aff88}}
\and L.~Pozzetti\orcid{0000-0001-7085-0412}\inst{\ref{aff12}}
\and F.~Raison\orcid{0000-0002-7819-6918}\inst{\ref{aff55}}
\and R.~Rebolo\orcid{0000-0003-3767-7085}\inst{\ref{aff37},\ref{aff89},\ref{aff90}}
\and A.~Renzi\orcid{0000-0001-9856-1970}\inst{\ref{aff91},\ref{aff50}}
\and J.~Rhodes\orcid{0000-0002-4485-8549}\inst{\ref{aff59}}
\and G.~Riccio\inst{\ref{aff22}}
\and E.~Romelli\orcid{0000-0003-3069-9222}\inst{\ref{aff14}}
\and M.~Roncarelli\orcid{0000-0001-9587-7822}\inst{\ref{aff12}}
\and R.~Saglia\orcid{0000-0003-0378-7032}\inst{\ref{aff57},\ref{aff55}}
\and Z.~Sakr\orcid{0000-0002-4823-3757}\inst{\ref{aff92},\ref{aff6},\ref{aff93}}
\and A.~G.~S\'anchez\orcid{0000-0003-1198-831X}\inst{\ref{aff55}}
\and D.~Sapone\orcid{0000-0001-7089-4503}\inst{\ref{aff94}}
\and B.~Sartoris\orcid{0000-0003-1337-5269}\inst{\ref{aff57},\ref{aff14}}
\and P.~Schneider\orcid{0000-0001-8561-2679}\inst{\ref{aff75}}
\and T.~Schrabback\orcid{0000-0002-6987-7834}\inst{\ref{aff95}}
\and A.~Secroun\orcid{0000-0003-0505-3710}\inst{\ref{aff51}}
\and E.~Sefusatti\orcid{0000-0003-0473-1567}\inst{\ref{aff14},\ref{aff13},\ref{aff15}}
\and G.~Seidel\orcid{0000-0003-2907-353X}\inst{\ref{aff63}}
\and S.~Serrano\orcid{0000-0002-0211-2861}\inst{\ref{aff5},\ref{aff96},\ref{aff4}}
\and P.~Simon\inst{\ref{aff75}}
\and C.~Sirignano\orcid{0000-0002-0995-7146}\inst{\ref{aff91},\ref{aff50}}
\and G.~Sirri\orcid{0000-0003-2626-2853}\inst{\ref{aff18}}
\and A.~Spurio~Mancini\orcid{0000-0001-5698-0990}\inst{\ref{aff97}}
\and L.~Stanco\orcid{0000-0002-9706-5104}\inst{\ref{aff50}}
\and J.~Steinwagner\orcid{0000-0001-7443-1047}\inst{\ref{aff55}}
\and P.~Tallada-Cresp\'{i}\orcid{0000-0002-1336-8328}\inst{\ref{aff31},\ref{aff32}}
\and A.~N.~Taylor\inst{\ref{aff38}}
\and I.~Tereno\orcid{0000-0002-4537-6218}\inst{\ref{aff44},\ref{aff98}}
\and N.~Tessore\orcid{0000-0002-9696-7931}\inst{\ref{aff3}}
\and S.~Toft\orcid{0000-0003-3631-7176}\inst{\ref{aff99},\ref{aff100}}
\and R.~Toledo-Moreo\orcid{0000-0002-2997-4859}\inst{\ref{aff101}}
\and F.~Torradeflot\orcid{0000-0003-1160-1517}\inst{\ref{aff32},\ref{aff31}}
\and L.~Valenziano\orcid{0000-0002-1170-0104}\inst{\ref{aff12},\ref{aff53}}
\and J.~Valiviita\orcid{0000-0001-6225-3693}\inst{\ref{aff67},\ref{aff68}}
\and T.~Vassallo\orcid{0000-0001-6512-6358}\inst{\ref{aff14}}
\and G.~Verdoes~Kleijn\orcid{0000-0001-5803-2580}\inst{\ref{aff102}}
\and A.~Veropalumbo\orcid{0000-0003-2387-1194}\inst{\ref{aff11},\ref{aff20},\ref{aff19}}
\and Y.~Wang\orcid{0000-0002-4749-2984}\inst{\ref{aff103}}
\and J.~Weller\orcid{0000-0002-8282-2010}\inst{\ref{aff57},\ref{aff55}}
\and A.~Zacchei\orcid{0000-0003-0396-1192}\inst{\ref{aff14},\ref{aff13}}
\and G.~Zamorani\orcid{0000-0002-2318-301X}\inst{\ref{aff12}}
\and F.~M.~Zerbi\inst{\ref{aff11}}
\and E.~Zucca\orcid{0000-0002-5845-8132}\inst{\ref{aff12}}
\and E.~Bozzo\orcid{0000-0002-8201-1525}\inst{\ref{aff46}}
\and C.~Burigana\orcid{0000-0002-3005-5796}\inst{\ref{aff104},\ref{aff53}}
\and R.~Cabanac\orcid{0000-0001-6679-2600}\inst{\ref{aff6}}
\and M.~Calabrese\orcid{0000-0002-2637-2422}\inst{\ref{aff105},\ref{aff28}}
\and J.~A.~Escartin~Vigo\inst{\ref{aff55}}
\and L.~Gabarra\orcid{0000-0002-8486-8856}\inst{\ref{aff106}}
\and W.~G.~Hartley\inst{\ref{aff46}}
\and S.~Matthew\orcid{0000-0001-8448-1697}\inst{\ref{aff38}}
\and M.~Maturi\orcid{0000-0002-3517-2422}\inst{\ref{aff92},\ref{aff107}}
\and N.~Mauri\orcid{0000-0001-8196-1548}\inst{\ref{aff36},\ref{aff18}}
\and R.~B.~Metcalf\orcid{0000-0003-3167-2574}\inst{\ref{aff76},\ref{aff12}}
\and A.~Pezzotta\orcid{0000-0003-0726-2268}\inst{\ref{aff11}}
\and M.~P\"ontinen\orcid{0000-0001-5442-2530}\inst{\ref{aff67}}
\and C.~Porciani\orcid{0000-0002-7797-2508}\inst{\ref{aff75}}
\and I.~Risso\orcid{0000-0003-2525-7761}\inst{\ref{aff11},\ref{aff20}}
\and V.~Scottez\orcid{0009-0008-3864-940X}\inst{\ref{aff81},\ref{aff108}}
\and M.~Sereno\orcid{0000-0003-0302-0325}\inst{\ref{aff12},\ref{aff18}}
\and M.~Tenti\orcid{0000-0002-4254-5901}\inst{\ref{aff18}}
\and M.~Viel\orcid{0000-0002-2642-5707}\inst{\ref{aff13},\ref{aff14},\ref{aff16},\ref{aff15},\ref{aff109}}
\and M.~Wiesmann\orcid{0009-0000-8199-5860}\inst{\ref{aff58}}
\and Y.~Akrami\orcid{0000-0002-2407-7956}\inst{\ref{aff110},\ref{aff111}}
\and I.~T.~Andika\orcid{0000-0001-6102-9526}\inst{\ref{aff112},\ref{aff113}}
\and S.~Anselmi\orcid{0000-0002-3579-9583}\inst{\ref{aff50},\ref{aff91},\ref{aff114}}
\and M.~Archidiacono\orcid{0000-0003-4952-9012}\inst{\ref{aff72},\ref{aff73}}
\and F.~Atrio-Barandela\orcid{0000-0002-2130-2513}\inst{\ref{aff115}}
\and D.~Bertacca\orcid{0000-0002-2490-7139}\inst{\ref{aff91},\ref{aff56},\ref{aff50}}
\and M.~Bethermin\orcid{0000-0002-3915-2015}\inst{\ref{aff116}}
\and A.~Blanchard\orcid{0000-0001-8555-9003}\inst{\ref{aff6}}
\and L.~Blot\orcid{0000-0002-9622-7167}\inst{\ref{aff117},\ref{aff69}}
\and H.~B\"ohringer\orcid{0000-0001-8241-4204}\inst{\ref{aff55},\ref{aff54},\ref{aff118}}
\and M.~Bonici\orcid{0000-0002-8430-126X}\inst{\ref{aff119},\ref{aff28}}
\and S.~Borgani\orcid{0000-0001-6151-6439}\inst{\ref{aff120},\ref{aff13},\ref{aff14},\ref{aff15},\ref{aff109}}
\and M.~L.~Brown\orcid{0000-0002-0370-8077}\inst{\ref{aff49}}
\and S.~Bruton\orcid{0000-0002-6503-5218}\inst{\ref{aff121}}
\and A.~Calabro\orcid{0000-0003-2536-1614}\inst{\ref{aff29}}
\and B.~Camacho~Quevedo\orcid{0000-0002-8789-4232}\inst{\ref{aff13},\ref{aff16},\ref{aff14}}
\and F.~Caro\inst{\ref{aff29}}
\and C.~S.~Carvalho\inst{\ref{aff98}}
\and T.~Castro\orcid{0000-0002-6292-3228}\inst{\ref{aff14},\ref{aff15},\ref{aff13},\ref{aff109}}
\and F.~Cogato\orcid{0000-0003-4632-6113}\inst{\ref{aff76},\ref{aff12}}
\and S.~Conseil\orcid{0000-0002-3657-4191}\inst{\ref{aff40}}
\and A.~R.~Cooray\orcid{0000-0002-3892-0190}\inst{\ref{aff122}}
\and O.~Cucciati\orcid{0000-0002-9336-7551}\inst{\ref{aff12}}
\and S.~Davini\orcid{0000-0003-3269-1718}\inst{\ref{aff20}}
\and F.~De~Paolis\orcid{0000-0001-6460-7563}\inst{\ref{aff123},\ref{aff124},\ref{aff125}}
\and G.~Desprez\orcid{0000-0001-8325-1742}\inst{\ref{aff102}}
\and A.~D\'iaz-S\'anchez\orcid{0000-0003-0748-4768}\inst{\ref{aff126}}
\and J.~J.~Diaz\orcid{0000-0003-2101-1078}\inst{\ref{aff37}}
\and S.~Di~Domizio\orcid{0000-0003-2863-5895}\inst{\ref{aff19},\ref{aff20}}
\and J.~M.~Diego\orcid{0000-0001-9065-3926}\inst{\ref{aff127}}
\and P.~Dimauro\orcid{0000-0001-7399-2854}\inst{\ref{aff128},\ref{aff29}}
\and M.~Y.~Elkhashab\orcid{0000-0001-9306-2603}\inst{\ref{aff14},\ref{aff15},\ref{aff120},\ref{aff13}}
\and A.~Enia\orcid{0000-0002-0200-2857}\inst{\ref{aff17},\ref{aff12}}
\and Y.~Fang\inst{\ref{aff57}}
\and A.~G.~Ferrari\orcid{0009-0005-5266-4110}\inst{\ref{aff18}}
\and A.~Finoguenov\orcid{0000-0002-4606-5403}\inst{\ref{aff67}}
\and A.~Fontana\orcid{0000-0003-3820-2823}\inst{\ref{aff29}}
\and A.~Franco\orcid{0000-0002-4761-366X}\inst{\ref{aff124},\ref{aff123},\ref{aff125}}
\and K.~Ganga\orcid{0000-0001-8159-8208}\inst{\ref{aff79}}
\and J.~Garc\'ia-Bellido\orcid{0000-0002-9370-8360}\inst{\ref{aff110}}
\and T.~Gasparetto\orcid{0000-0002-7913-4866}\inst{\ref{aff29}}
\and V.~Gautard\inst{\ref{aff129}}
\and R.~Gavazzi\orcid{0000-0002-5540-6935}\inst{\ref{aff47},\ref{aff82}}
\and E.~Gaztanaga\orcid{0000-0001-9632-0815}\inst{\ref{aff4},\ref{aff5},\ref{aff130}}
\and F.~Giacomini\orcid{0000-0002-3129-2814}\inst{\ref{aff18}}
\and F.~Gianotti\orcid{0000-0003-4666-119X}\inst{\ref{aff12}}
\and G.~Gozaliasl\orcid{0000-0002-0236-919X}\inst{\ref{aff131},\ref{aff67}}
\and M.~Guidi\orcid{0000-0001-9408-1101}\inst{\ref{aff17},\ref{aff12}}
\and C.~M.~Gutierrez\orcid{0000-0001-7854-783X}\inst{\ref{aff132}}
\and A.~Hall\orcid{0000-0002-3139-8651}\inst{\ref{aff38}}
\and S.~Hemmati\orcid{0000-0003-2226-5395}\inst{\ref{aff133}}
\and H.~Hildebrandt\orcid{0000-0002-9814-3338}\inst{\ref{aff134}}
\and J.~Hjorth\orcid{0000-0002-4571-2306}\inst{\ref{aff86}}
\and S.~Joudaki\orcid{0000-0001-8820-673X}\inst{\ref{aff31}}
\and J.~J.~E.~Kajava\orcid{0000-0002-3010-8333}\inst{\ref{aff135},\ref{aff136}}
\and Y.~Kang\orcid{0009-0000-8588-7250}\inst{\ref{aff46}}
\and V.~Kansal\orcid{0000-0002-4008-6078}\inst{\ref{aff137},\ref{aff138}}
\and D.~Karagiannis\orcid{0000-0002-4927-0816}\inst{\ref{aff139},\ref{aff140}}
\and K.~Kiiveri\inst{\ref{aff65}}
\and J.~Kim\orcid{0000-0003-2776-2761}\inst{\ref{aff106}}
\and C.~C.~Kirkpatrick\inst{\ref{aff65}}
\and S.~Kruk\orcid{0000-0001-8010-8879}\inst{\ref{aff10}}
\and J.~Le~Graet\orcid{0000-0001-6523-7971}\inst{\ref{aff51}}
\and L.~Legrand\orcid{0000-0003-0610-5252}\inst{\ref{aff141},\ref{aff142}}
\and M.~Lembo\orcid{0000-0002-5271-5070}\inst{\ref{aff82}}
\and F.~Lepori\orcid{0009-0000-5061-7138}\inst{\ref{aff9}}
\and G.~Leroy\orcid{0009-0004-2523-4425}\inst{\ref{aff143},\ref{aff77}}
\and G.~F.~Lesci\orcid{0000-0002-4607-2830}\inst{\ref{aff76},\ref{aff12}}
\and J.~Lesgourgues\orcid{0000-0001-7627-353X}\inst{\ref{aff33}}
\and T.~I.~Liaudat\orcid{0000-0002-9104-314X}\inst{\ref{aff144}}
\and A.~Loureiro\orcid{0000-0002-4371-0876}\inst{\ref{aff145},\ref{aff146}}
\and J.~Macias-Perez\orcid{0000-0002-5385-2763}\inst{\ref{aff147}}
\and G.~Maggio\orcid{0000-0003-4020-4836}\inst{\ref{aff14}}
\and M.~Magliocchetti\orcid{0000-0001-9158-4838}\inst{\ref{aff52}}
\and F.~Mannucci\orcid{0000-0002-4803-2381}\inst{\ref{aff148}}
\and R.~Maoli\orcid{0000-0002-6065-3025}\inst{\ref{aff149},\ref{aff29}}
\and C.~J.~A.~P.~Martins\orcid{0000-0002-4886-9261}\inst{\ref{aff150},\ref{aff151}}
\and L.~Maurin\orcid{0000-0002-8406-0857}\inst{\ref{aff48}}
\and M.~Miluzio\inst{\ref{aff10},\ref{aff152}}
\and P.~Monaco\orcid{0000-0003-2083-7564}\inst{\ref{aff120},\ref{aff14},\ref{aff15},\ref{aff13},\ref{aff109}}
\and C.~Moretti\orcid{0000-0003-3314-8936}\inst{\ref{aff14},\ref{aff13},\ref{aff15},\ref{aff16}}
\and G.~Morgante\inst{\ref{aff12}}
\and S.~Nadathur\orcid{0000-0001-9070-3102}\inst{\ref{aff130}}
\and K.~Naidoo\orcid{0000-0002-9182-1802}\inst{\ref{aff130},\ref{aff3}}
\and P.~Natoli\orcid{0000-0003-0126-9100}\inst{\ref{aff139},\ref{aff153}}
\and A.~Navarro-Alsina\orcid{0000-0002-3173-2592}\inst{\ref{aff75}}
\and S.~Nesseris\orcid{0000-0002-0567-0324}\inst{\ref{aff110}}
\and L.~Pagano\orcid{0000-0003-1820-5998}\inst{\ref{aff139},\ref{aff153}}
\and D.~Paoletti\orcid{0000-0003-4761-6147}\inst{\ref{aff12},\ref{aff53}}
\and F.~Passalacqua\orcid{0000-0002-8606-4093}\inst{\ref{aff91},\ref{aff50}}
\and K.~Paterson\orcid{0000-0001-8340-3486}\inst{\ref{aff63}}
\and L.~Patrizii\inst{\ref{aff18}}
\and A.~Pisani\orcid{0000-0002-6146-4437}\inst{\ref{aff51}}
\and D.~Potter\orcid{0000-0002-0757-5195}\inst{\ref{aff9}}
\and S.~Quai\orcid{0000-0002-0449-8163}\inst{\ref{aff76},\ref{aff12}}
\and M.~Radovich\orcid{0000-0002-3585-866X}\inst{\ref{aff56}}
\and S.~Sacquegna\orcid{0000-0002-8433-6630}\inst{\ref{aff154},\ref{aff123},\ref{aff124}}
\and M.~Sahl\'en\orcid{0000-0003-0973-4804}\inst{\ref{aff155}}
\and D.~B.~Sanders\orcid{0000-0002-1233-9998}\inst{\ref{aff35}}
\and E.~Sarpa\orcid{0000-0002-1256-655X}\inst{\ref{aff16},\ref{aff109},\ref{aff15}}
\and A.~Schneider\orcid{0000-0001-7055-8104}\inst{\ref{aff9}}
\and M.~Schultheis\inst{\ref{aff78}}
\and D.~Sciotti\orcid{0009-0008-4519-2620}\inst{\ref{aff29},\ref{aff30}}
\and E.~Sellentin\inst{\ref{aff156},\ref{aff7}}
\and L.~C.~Smith\orcid{0000-0002-3259-2771}\inst{\ref{aff157}}
\and J.~G.~Sorce\orcid{0000-0002-2307-2432}\inst{\ref{aff158},\ref{aff48}}
\and K.~Tanidis\orcid{0000-0001-9843-5130}\inst{\ref{aff106}}
\and C.~Tao\orcid{0000-0001-7961-8177}\inst{\ref{aff51}}
\and G.~Testera\inst{\ref{aff20}}
\and R.~Teyssier\orcid{0000-0001-7689-0933}\inst{\ref{aff159}}
\and S.~Tosi\orcid{0000-0002-7275-9193}\inst{\ref{aff19},\ref{aff20},\ref{aff11}}
\and A.~Troja\orcid{0000-0003-0239-4595}\inst{\ref{aff91},\ref{aff50}}
\and M.~Tucci\inst{\ref{aff46}}
\and C.~Valieri\inst{\ref{aff18}}
\and A.~Venhola\orcid{0000-0001-6071-4564}\inst{\ref{aff160}}
\and D.~Vergani\orcid{0000-0003-0898-2216}\inst{\ref{aff12}}
\and F.~Vernizzi\orcid{0000-0003-3426-2802}\inst{\ref{aff161}}
\and G.~Verza\orcid{0000-0002-1886-8348}\inst{\ref{aff162}}
\and P.~Vielzeuf\orcid{0000-0003-2035-9339}\inst{\ref{aff51}}
\and N.~A.~Walton\orcid{0000-0003-3983-8778}\inst{\ref{aff157}}}
										   
\institute{Universit\'e Paris-Saclay, Universit\'e Paris Cit\'e, CEA, CNRS, AIM, 91191, Gif-sur-Yvette, France\label{aff1}
\and
Centre National d'Etudes Spatiales -- Centre spatial de Toulouse, 18 avenue Edouard Belin, 31401 Toulouse Cedex 9, France\label{aff2}
\and
Department of Physics and Astronomy, University College London, Gower Street, London WC1E 6BT, UK\label{aff3}
\and
Institute of Space Sciences (ICE, CSIC), Campus UAB, Carrer de Can Magrans, s/n, 08193 Barcelona, Spain\label{aff4}
\and
Institut d'Estudis Espacials de Catalunya (IEEC),  Edifici RDIT, Campus UPC, 08860 Castelldefels, Barcelona, Spain\label{aff5}
\and
Institut de Recherche en Astrophysique et Plan\'etologie (IRAP), Universit\'e de Toulouse, CNRS, UPS, CNES, 14 Av. Edouard Belin, 31400 Toulouse, France\label{aff6}
\and
Leiden Observatory, Leiden University, Einsteinweg 55, 2333 CC Leiden, The Netherlands\label{aff7}
\and
Department of Physics, Northeastern University, Boston, MA, 02115, USA\label{aff8}
\and
Department of Astrophysics, University of Zurich, Winterthurerstrasse 190, 8057 Zurich, Switzerland\label{aff9}
\and
ESAC/ESA, Camino Bajo del Castillo, s/n., Urb. Villafranca del Castillo, 28692 Villanueva de la Ca\~nada, Madrid, Spain\label{aff10}
\and
INAF-Osservatorio Astronomico di Brera, Via Brera 28, 20122 Milano, Italy\label{aff11}
\and
INAF-Osservatorio di Astrofisica e Scienza dello Spazio di Bologna, Via Piero Gobetti 93/3, 40129 Bologna, Italy\label{aff12}
\and
IFPU, Institute for Fundamental Physics of the Universe, via Beirut 2, 34151 Trieste, Italy\label{aff13}
\and
INAF-Osservatorio Astronomico di Trieste, Via G. B. Tiepolo 11, 34143 Trieste, Italy\label{aff14}
\and
INFN, Sezione di Trieste, Via Valerio 2, 34127 Trieste TS, Italy\label{aff15}
\and
SISSA, International School for Advanced Studies, Via Bonomea 265, 34136 Trieste TS, Italy\label{aff16}
\and
Dipartimento di Fisica e Astronomia, Universit\`a di Bologna, Via Gobetti 93/2, 40129 Bologna, Italy\label{aff17}
\and
INFN-Sezione di Bologna, Viale Berti Pichat 6/2, 40127 Bologna, Italy\label{aff18}
\and
Dipartimento di Fisica, Universit\`a di Genova, Via Dodecaneso 33, 16146, Genova, Italy\label{aff19}
\and
INFN-Sezione di Genova, Via Dodecaneso 33, 16146, Genova, Italy\label{aff20}
\and
Department of Physics "E. Pancini", University Federico II, Via Cinthia 6, 80126, Napoli, Italy\label{aff21}
\and
INAF-Osservatorio Astronomico di Capodimonte, Via Moiariello 16, 80131 Napoli, Italy\label{aff22}
\and
Dipartimento di Fisica, Universit\`a degli Studi di Torino, Via P. Giuria 1, 10125 Torino, Italy\label{aff23}
\and
INFN-Sezione di Torino, Via P. Giuria 1, 10125 Torino, Italy\label{aff24}
\and
INAF-Osservatorio Astrofisico di Torino, Via Osservatorio 20, 10025 Pino Torinese (TO), Italy\label{aff25}
\and
European Space Agency/ESTEC, Keplerlaan 1, 2201 AZ Noordwijk, The Netherlands\label{aff26}
\and
Institute Lorentz, Leiden University, Niels Bohrweg 2, 2333 CA Leiden, The Netherlands\label{aff27}
\and
INAF-IASF Milano, Via Alfonso Corti 12, 20133 Milano, Italy\label{aff28}
\and
INAF-Osservatorio Astronomico di Roma, Via Frascati 33, 00078 Monteporzio Catone, Italy\label{aff29}
\and
INFN-Sezione di Roma, Piazzale Aldo Moro, 2 - c/o Dipartimento di Fisica, Edificio G. Marconi, 00185 Roma, Italy\label{aff30}
\and
Centro de Investigaciones Energ\'eticas, Medioambientales y Tecnol\'ogicas (CIEMAT), Avenida Complutense 40, 28040 Madrid, Spain\label{aff31}
\and
Port d'Informaci\'{o} Cient\'{i}fica, Campus UAB, C. Albareda s/n, 08193 Bellaterra (Barcelona), Spain\label{aff32}
\and
Institute for Theoretical Particle Physics and Cosmology (TTK), RWTH Aachen University, 52056 Aachen, Germany\label{aff33}
\and
INFN section of Naples, Via Cinthia 6, 80126, Napoli, Italy\label{aff34}
\and
Institute for Astronomy, University of Hawaii, 2680 Woodlawn Drive, Honolulu, HI 96822, USA\label{aff35}
\and
Dipartimento di Fisica e Astronomia "Augusto Righi" - Alma Mater Studiorum Universit\`a di Bologna, Viale Berti Pichat 6/2, 40127 Bologna, Italy\label{aff36}
\and
Instituto de Astrof\'{\i}sica de Canarias, E-38205 La Laguna, Tenerife, Spain\label{aff37}
\and
Institute for Astronomy, University of Edinburgh, Royal Observatory, Blackford Hill, Edinburgh EH9 3HJ, UK\label{aff38}
\and
European Space Agency/ESRIN, Largo Galileo Galilei 1, 00044 Frascati, Roma, Italy\label{aff39}
\and
Universit\'e Claude Bernard Lyon 1, CNRS/IN2P3, IP2I Lyon, UMR 5822, Villeurbanne, F-69100, France\label{aff40}
\and
Institut de Ci\`{e}ncies del Cosmos (ICCUB), Universitat de Barcelona (IEEC-UB), Mart\'{i} i Franqu\`{e}s 1, 08028 Barcelona, Spain\label{aff41}
\and
Instituci\'o Catalana de Recerca i Estudis Avan\c{c}ats (ICREA), Passeig de Llu\'{\i}s Companys 23, 08010 Barcelona, Spain\label{aff42}
\and
UCB Lyon 1, CNRS/IN2P3, IUF, IP2I Lyon, 4 rue Enrico Fermi, 69622 Villeurbanne, France\label{aff43}
\and
Departamento de F\'isica, Faculdade de Ci\^encias, Universidade de Lisboa, Edif\'icio C8, Campo Grande, PT1749-016 Lisboa, Portugal\label{aff44}
\and
Instituto de Astrof\'isica e Ci\^encias do Espa\c{c}o, Faculdade de Ci\^encias, Universidade de Lisboa, Campo Grande, 1749-016 Lisboa, Portugal\label{aff45}
\and
Department of Astronomy, University of Geneva, ch. d'Ecogia 16, 1290 Versoix, Switzerland\label{aff46}
\and
Aix-Marseille Universit\'e, CNRS, CNES, LAM, Marseille, France\label{aff47}
\and
Universit\'e Paris-Saclay, CNRS, Institut d'astrophysique spatiale, 91405, Orsay, France\label{aff48}
\and
Jodrell Bank Centre for Astrophysics, Department of Physics and Astronomy, University of Manchester, Oxford Road, Manchester M13 9PL, UK\label{aff49}
\and
INFN-Padova, Via Marzolo 8, 35131 Padova, Italy\label{aff50}
\and
Aix-Marseille Universit\'e, CNRS/IN2P3, CPPM, Marseille, France\label{aff51}
\and
INAF-Istituto di Astrofisica e Planetologia Spaziali, via del Fosso del Cavaliere, 100, 00100 Roma, Italy\label{aff52}
\and
INFN-Bologna, Via Irnerio 46, 40126 Bologna, Italy\label{aff53}
\and
University Observatory, LMU Faculty of Physics, Scheinerstr.~1, 81679 Munich, Germany\label{aff54}
\and
Max Planck Institute for Extraterrestrial Physics, Giessenbachstr. 1, 85748 Garching, Germany\label{aff55}
\and
INAF-Osservatorio Astronomico di Padova, Via dell'Osservatorio 5, 35122 Padova, Italy\label{aff56}
\and
Universit\"ats-Sternwarte M\"unchen, Fakult\"at f\"ur Physik, Ludwig-Maximilians-Universit\"at M\"unchen, Scheinerstr.~1, 81679 M\"unchen, Germany\label{aff57}
\and
Institute of Theoretical Astrophysics, University of Oslo, P.O. Box 1029 Blindern, 0315 Oslo, Norway\label{aff58}
\and
Jet Propulsion Laboratory, California Institute of Technology, 4800 Oak Grove Drive, Pasadena, CA, 91109, USA\label{aff59}
\and
Felix Hormuth Engineering, Goethestr. 17, 69181 Leimen, Germany\label{aff60}
\and
Technical University of Denmark, Elektrovej 327, 2800 Kgs. Lyngby, Denmark\label{aff61}
\and
Cosmic Dawn Center (DAWN), Denmark\label{aff62}
\and
Max-Planck-Institut f\"ur Astronomie, K\"onigstuhl 17, 69117 Heidelberg, Germany\label{aff63}
\and
NASA Goddard Space Flight Center, Greenbelt, MD 20771, USA\label{aff64}
\and
Department of Physics and Helsinki Institute of Physics, Gustaf H\"allstr\"omin katu 2, University of Helsinki, 00014 Helsinki, Finland\label{aff65}
\and
Universit\'e de Gen\`eve, D\'epartement de Physique Th\'eorique and Centre for Astroparticle Physics, 24 quai Ernest-Ansermet, CH-1211 Gen\`eve 4, Switzerland\label{aff66}
\and
Department of Physics, P.O. Box 64, University of Helsinki, 00014 Helsinki, Finland\label{aff67}
\and
Helsinki Institute of Physics, Gustaf H{\"a}llstr{\"o}min katu 2, University of Helsinki, 00014 Helsinki, Finland\label{aff68}
\and
Laboratoire d'etude de l'Univers et des phenomenes eXtremes, Observatoire de Paris, Universit\'e PSL, Sorbonne Universit\'e, CNRS, 92190 Meudon, France\label{aff69}
\and
SKAO, Jodrell Bank, Lower Withington, Macclesfield SK11 9FT, UK\label{aff70}
\and
Centre de Calcul de l'IN2P3/CNRS, 21 avenue Pierre de Coubertin 69627 Villeurbanne Cedex, France\label{aff71}
\and
Dipartimento di Fisica "Aldo Pontremoli", Universit\`a degli Studi di Milano, Via Celoria 16, 20133 Milano, Italy\label{aff72}
\and
INFN-Sezione di Milano, Via Celoria 16, 20133 Milano, Italy\label{aff73}
\and
University of Applied Sciences and Arts of Northwestern Switzerland, School of Computer Science, 5210 Windisch, Switzerland\label{aff74}
\and
Universit\"at Bonn, Argelander-Institut f\"ur Astronomie, Auf dem H\"ugel 71, 53121 Bonn, Germany\label{aff75}
\and
Dipartimento di Fisica e Astronomia "Augusto Righi" - Alma Mater Studiorum Universit\`a di Bologna, via Piero Gobetti 93/2, 40129 Bologna, Italy\label{aff76}
\and
Department of Physics, Institute for Computational Cosmology, Durham University, South Road, Durham, DH1 3LE, UK\label{aff77}
\and
Universit\'e C\^{o}te d'Azur, Observatoire de la C\^{o}te d'Azur, CNRS, Laboratoire Lagrange, Bd de l'Observatoire, CS 34229, 06304 Nice cedex 4, France\label{aff78}
\and
Universit\'e Paris Cit\'e, CNRS, Astroparticule et Cosmologie, 75013 Paris, France\label{aff79}
\and
CNRS-UCB International Research Laboratory, Centre Pierre Bin\'etruy, IRL2007, CPB-IN2P3, Berkeley, USA\label{aff80}
\and
Institut d'Astrophysique de Paris, 98bis Boulevard Arago, 75014, Paris, France\label{aff81}
\and
Institut d'Astrophysique de Paris, UMR 7095, CNRS, and Sorbonne Universit\'e, 98 bis boulevard Arago, 75014 Paris, France\label{aff82}
\and
Institute of Physics, Laboratory of Astrophysics, Ecole Polytechnique F\'ed\'erale de Lausanne (EPFL), Observatoire de Sauverny, 1290 Versoix, Switzerland\label{aff83}
\and
Telespazio UK S.L. for European Space Agency (ESA), Camino bajo del Castillo, s/n, Urbanizacion Villafranca del Castillo, Villanueva de la Ca\~nada, 28692 Madrid, Spain\label{aff84}
\and
Institut de F\'{i}sica d'Altes Energies (IFAE), The Barcelona Institute of Science and Technology, Campus UAB, 08193 Bellaterra (Barcelona), Spain\label{aff85}
\and
DARK, Niels Bohr Institute, University of Copenhagen, Jagtvej 155, 2200 Copenhagen, Denmark\label{aff86}
\and
Space Science Data Center, Italian Space Agency, via del Politecnico snc, 00133 Roma, Italy\label{aff87}
\and
Institute of Space Science, Str. Atomistilor, nr. 409 M\u{a}gurele, Ilfov, 077125, Romania\label{aff88}
\and
Consejo Superior de Investigaciones Cientificas, Calle Serrano 117, 28006 Madrid, Spain\label{aff89}
\and
Universidad de La Laguna, Dpto. Astrof\'\i sica, E-38206 La Laguna, Tenerife, Spain\label{aff90}
\and
Dipartimento di Fisica e Astronomia "G. Galilei", Universit\`a di Padova, Via Marzolo 8, 35131 Padova, Italy\label{aff91}
\and
Institut f\"ur Theoretische Physik, University of Heidelberg, Philosophenweg 16, 69120 Heidelberg, Germany\label{aff92}
\and
Universit\'e St Joseph; Faculty of Sciences, Beirut, Lebanon\label{aff93}
\and
Departamento de F\'isica, FCFM, Universidad de Chile, Blanco Encalada 2008, Santiago, Chile\label{aff94}
\and
Universit\"at Innsbruck, Institut f\"ur Astro- und Teilchenphysik, Technikerstr. 25/8, 6020 Innsbruck, Austria\label{aff95}
\and
Satlantis, University Science Park, Sede Bld 48940, Leioa-Bilbao, Spain\label{aff96}
\and
Department of Physics, Royal Holloway, University of London, Surrey TW20 0EX, UK\label{aff97}
\and
Instituto de Astrof\'isica e Ci\^encias do Espa\c{c}o, Faculdade de Ci\^encias, Universidade de Lisboa, Tapada da Ajuda, 1349-018 Lisboa, Portugal\label{aff98}
\and
Cosmic Dawn Center (DAWN)\label{aff99}
\and
Niels Bohr Institute, University of Copenhagen, Jagtvej 128, 2200 Copenhagen, Denmark\label{aff100}
\and
Universidad Polit\'ecnica de Cartagena, Departamento de Electr\'onica y Tecnolog\'ia de Computadoras,  Plaza del Hospital 1, 30202 Cartagena, Spain\label{aff101}
\and
Kapteyn Astronomical Institute, University of Groningen, PO Box 800, 9700 AV Groningen, The Netherlands\label{aff102}
\and
Infrared Processing and Analysis Center, California Institute of Technology, Pasadena, CA 91125, USA\label{aff103}
\and
INAF, Istituto di Radioastronomia, Via Piero Gobetti 101, 40129 Bologna, Italy\label{aff104}
\and
Astronomical Observatory of the Autonomous Region of the Aosta Valley (OAVdA), Loc. Lignan 39, I-11020, Nus (Aosta Valley), Italy\label{aff105}
\and
Department of Physics, Oxford University, Keble Road, Oxford OX1 3RH, UK\label{aff106}
\and
Zentrum f\"ur Astronomie, Universit\"at Heidelberg, Philosophenweg 12, 69120 Heidelberg, Germany\label{aff107}
\and
ICL, Junia, Universit\'e Catholique de Lille, LITL, 59000 Lille, France\label{aff108}
\and
ICSC - Centro Nazionale di Ricerca in High Performance Computing, Big Data e Quantum Computing, Via Magnanelli 2, Bologna, Italy\label{aff109}
\and
Instituto de F\'isica Te\'orica UAM-CSIC, Campus de Cantoblanco, 28049 Madrid, Spain\label{aff110}
\and
CERCA/ISO, Department of Physics, Case Western Reserve University, 10900 Euclid Avenue, Cleveland, OH 44106, USA\label{aff111}
\and
Technical University of Munich, TUM School of Natural Sciences, Physics Department, James-Franck-Str.~1, 85748 Garching, Germany\label{aff112}
\and
Max-Planck-Institut f\"ur Astrophysik, Karl-Schwarzschild-Str.~1, 85748 Garching, Germany\label{aff113}
\and
Laboratoire Univers et Th\'eorie, Observatoire de Paris, Universit\'e PSL, Universit\'e Paris Cit\'e, CNRS, 92190 Meudon, France\label{aff114}
\and
Departamento de F{\'\i}sica Fundamental. Universidad de Salamanca. Plaza de la Merced s/n. 37008 Salamanca, Spain\label{aff115}
\and
Universit\'e de Strasbourg, CNRS, Observatoire astronomique de Strasbourg, UMR 7550, 67000 Strasbourg, France\label{aff116}
\and
Center for Data-Driven Discovery, Kavli IPMU (WPI), UTIAS, The University of Tokyo, Kashiwa, Chiba 277-8583, Japan\label{aff117}
\and
Max-Planck-Institut f\"ur Physik, Boltzmannstr. 8, 85748 Garching, Germany\label{aff118}
\and
Waterloo Centre for Astrophysics, University of Waterloo, Waterloo, Ontario N2L 3G1, Canada\label{aff119}
\and
Dipartimento di Fisica - Sezione di Astronomia, Universit\`a di Trieste, Via Tiepolo 11, 34131 Trieste, Italy\label{aff120}
\and
California Institute of Technology, 1200 E California Blvd, Pasadena, CA 91125, USA\label{aff121}
\and
Department of Physics \& Astronomy, University of California Irvine, Irvine CA 92697, USA\label{aff122}
\and
Department of Mathematics and Physics E. De Giorgi, University of Salento, Via per Arnesano, CP-I93, 73100, Lecce, Italy\label{aff123}
\and
INFN, Sezione di Lecce, Via per Arnesano, CP-193, 73100, Lecce, Italy\label{aff124}
\and
INAF-Sezione di Lecce, c/o Dipartimento Matematica e Fisica, Via per Arnesano, 73100, Lecce, Italy\label{aff125}
\and
Departamento F\'isica Aplicada, Universidad Polit\'ecnica de Cartagena, Campus Muralla del Mar, 30202 Cartagena, Murcia, Spain\label{aff126}
\and
Instituto de F\'isica de Cantabria, Edificio Juan Jord\'a, Avenida de los Castros, 39005 Santander, Spain\label{aff127}
\and
Observatorio Nacional, Rua General Jose Cristino, 77-Bairro Imperial de Sao Cristovao, Rio de Janeiro, 20921-400, Brazil\label{aff128}
\and
CEA Saclay, DFR/IRFU, Service d'Astrophysique, Bat. 709, 91191 Gif-sur-Yvette, France\label{aff129}
\and
Institute of Cosmology and Gravitation, University of Portsmouth, Portsmouth PO1 3FX, UK\label{aff130}
\and
Department of Computer Science, Aalto University, PO Box 15400, Espoo, FI-00 076, Finland\label{aff131}
\and
 Instituto de Astrof\'{\i}sica de Canarias, E-38205 La Laguna; Universidad de La Laguna, Dpto. Astrof\'\i sica, E-38206 La Laguna, Tenerife, Spain\label{aff132}
\and
Caltech/IPAC, 1200 E. California Blvd., Pasadena, CA 91125, USA\label{aff133}
\and
Ruhr University Bochum, Faculty of Physics and Astronomy, Astronomical Institute (AIRUB), German Centre for Cosmological Lensing (GCCL), 44780 Bochum, Germany\label{aff134}
\and
Department of Physics and Astronomy, Vesilinnantie 5, University of Turku, 20014 Turku, Finland\label{aff135}
\and
Serco for European Space Agency (ESA), Camino bajo del Castillo, s/n, Urbanizacion Villafranca del Castillo, Villanueva de la Ca\~nada, 28692 Madrid, Spain\label{aff136}
\and
ARC Centre of Excellence for Dark Matter Particle Physics, Melbourne, Australia\label{aff137}
\and
Centre for Astrophysics \& Supercomputing, Swinburne University of Technology,  Hawthorn, Victoria 3122, Australia\label{aff138}
\and
Dipartimento di Fisica e Scienze della Terra, Universit\`a degli Studi di Ferrara, Via Giuseppe Saragat 1, 44122 Ferrara, Italy\label{aff139}
\and
Department of Physics and Astronomy, University of the Western Cape, Bellville, Cape Town, 7535, South Africa\label{aff140}
\and
DAMTP, Centre for Mathematical Sciences, Wilberforce Road, Cambridge CB3 0WA, UK\label{aff141}
\and
Kavli Institute for Cosmology Cambridge, Madingley Road, Cambridge, CB3 0HA, UK\label{aff142}
\and
Department of Physics, Centre for Extragalactic Astronomy, Durham University, South Road, Durham, DH1 3LE, UK\label{aff143}
\and
IRFU, CEA, Universit\'e Paris-Saclay 91191 Gif-sur-Yvette Cedex, France\label{aff144}
\and
Oskar Klein Centre for Cosmoparticle Physics, Department of Physics, Stockholm University, Stockholm, SE-106 91, Sweden\label{aff145}
\and
Astrophysics Group, Blackett Laboratory, Imperial College London, London SW7 2AZ, UK\label{aff146}
\and
Univ. Grenoble Alpes, CNRS, Grenoble INP, LPSC-IN2P3, 53, Avenue des Martyrs, 38000, Grenoble, France\label{aff147}
\and
INAF-Osservatorio Astrofisico di Arcetri, Largo E. Fermi 5, 50125, Firenze, Italy\label{aff148}
\and
Dipartimento di Fisica, Sapienza Universit\`a di Roma, Piazzale Aldo Moro 2, 00185 Roma, Italy\label{aff149}
\and
Centro de Astrof\'{\i}sica da Universidade do Porto, Rua das Estrelas, 4150-762 Porto, Portugal\label{aff150}
\and
Instituto de Astrof\'isica e Ci\^encias do Espa\c{c}o, Universidade do Porto, CAUP, Rua das Estrelas, PT4150-762 Porto, Portugal\label{aff151}
\and
HE Space for European Space Agency (ESA), Camino bajo del Castillo, s/n, Urbanizacion Villafranca del Castillo, Villanueva de la Ca\~nada, 28692 Madrid, Spain\label{aff152}
\and
Istituto Nazionale di Fisica Nucleare, Sezione di Ferrara, Via Giuseppe Saragat 1, 44122 Ferrara, Italy\label{aff153}
\and
INAF - Osservatorio Astronomico d'Abruzzo, Via Maggini, 64100, Teramo, Italy\label{aff154}
\and
Theoretical astrophysics, Department of Physics and Astronomy, Uppsala University, Box 516, 751 37 Uppsala, Sweden\label{aff155}
\and
Mathematical Institute, University of Leiden, Einsteinweg 55, 2333 CA Leiden, The Netherlands\label{aff156}
\and
Institute of Astronomy, University of Cambridge, Madingley Road, Cambridge CB3 0HA, UK\label{aff157}
\and
Univ. Lille, CNRS, Centrale Lille, UMR 9189 CRIStAL, 59000 Lille, France\label{aff158}
\and
Department of Astrophysical Sciences, Peyton Hall, Princeton University, Princeton, NJ 08544, USA\label{aff159}
\and
Space physics and astronomy research unit, University of Oulu, Pentti Kaiteran katu 1, FI-90014 Oulu, Finland\label{aff160}
\and
Institut de Physique Th\'eorique, CEA, CNRS, Universit\'e Paris-Saclay 91191 Gif-sur-Yvette Cedex, France\label{aff161}
\and
Center for Computational Astrophysics, Flatiron Institute, 162 5th Avenue, 10010, New York, NY, USA\label{aff162}}      

\date{Received January 01, 2026; accepted September 23, 2026}

\abstract
{We model intrinsic alignment (IA) in \Euclid's Flagship simulation to investigate its impact on \Euclid's weak lensing signal. Our IA implementation in the Flagship simulation takes the photometric properties of galaxies into account, along with their dark matter host halos. The simulation parameters are calibrated using combined constraints from observations and cosmological hydrodynamical simulations. We compare simulations against theory predictions, determining the parameters of two of the most widely used IA models: the nonlinear alignment (NLA) and the tidal alignment and tidal torquing (TATT) models.
We measured the amplitude of the simulated IA signal as a function of galaxy magnitude and colour in the redshift range $0.1<z<2.1$, close that of \Euclid's main galaxy sample. We find that both NLA and TATT are able to accurately describe the IA signal in the simulation down to scales of $6$--$7 \,h^{-1}\,$Mpc. We measured the alignment amplitudes for red galaxies that are comparable to those of the observations, using samples that had not been used in the initial calibration procedure. For blue galaxies, our constraints are consistent with zero alignments in our first redshift bin $0.1 < z < 0.3$, but we detected a non-negligible signal at higher redshift; however, this signal is consistent with the upper limits set by observational constraints. Additionally, several hydrodynamical simulations have predicted alignments for spiral galaxies, in agreement with our findings. Finally, the evolution of alignment with redshift is realistic and comparable to what has been determined based on observations. However, we find that the commonly adopted redshift power law for IA fails to reproduce the simulation alignments above $z=1.1$. A significantly improved agreement can be obtained when the luminosity dependence is included, capturing the intrinsic luminosity evolution with redshift in magnitude-limited surveys.
We conclude that the Flagship IA simulation is a useful tool for translating current IA constraints into predictions for IA contamination of \Euclid-like samples.}

\keywords{cosmology -- weak lensing -- simulations -- deep surveys}

\titlerunning{Testing analytic alignment models of \Euclid in Flagship}
\authorrunning{Euclid Collaboration: R. Paviot et al.}

\maketitle
\nolinenumbers

\section{Introduction}\label{sec:intro}
As predicted by general relativity, light from distant galaxies is deflected by matter inhomogeneities along the line of sight. The sum of these small distortions leads to coherent alignment of galaxy images, usually referred to as weak gravitational lensing. This term includes two main distinct measurements. First, neighboring galaxies' light must pass through similar cross-sections of the Universe, leading to correlations among galaxy shapes. This correlation is referred to as cosmic shear and it is now widely known that 2-point shear-shear correlations can be used to provide direct constraints on the evolution of the matter distribution in the Universe and on the nature of dark energy itself \citep{Hildebrandt17,Troxel18,Hamana2020,Asgari2021,Amon21,Secco21,2023DESKIDS}. 

Furthermore, background galaxy shape distortions could be correlated with foreground galaxies that act as lenses, a measurement referred to as galaxy-galaxy lensing (GGL). Since galaxies trace the underlying matter distribution in a biased way, GGL measures effects similar to those probed by shear-shear correlations; namely, the distribution of matter and how structures grow in the Universe. Therefore, a correctly derived redshift distribution of the lenses combined with a proper model of galaxy bias can provide precise cosmological constraints given some GGL lensing measurements \citep{Sheldon2004,Baldauf2010,Mandelbaum2013,Prat2018,Blake2020,Pandey2022,Prat2022}. Many of these cosmological analyses also combine GGL lensing with galaxy clustering measurements to break the degeneracy between galaxy bias and the amplitude of matter fluctuations $\sigma_8$, a combination known as 2$\times$2pt \citep{Porredon2022,Dvornik2023}. Finally, it has become common to combine cosmic shear with GGL lensing and galaxy clustering in a so-called 3$\times$2pt analysis to determine cosmological constraints from weak lensing photometric surveys \citep{Vanuitert2018,DES21,Sugiyama2023}. 

However, constraining cosmological parameters from weak lensing surveys is challenging. Observationally, the main sources of bias come from uncertainties in the estimation of the source redshift distribution and the shape measurement algorithms, with great improvements achieved in the last decade due to the refinement of image simulations and shape measurement methods \citep{Mandelbaum2018,Kannadawi2019}. From the modelling side, the correlation between galaxy orientations does not arise solely from the lensing effect. Galaxies that form within overdense regions are affected by the tides generated by the quadrupole of the local gravitational field, which will shape the spatial distribution of its stars. This process starts at the initial stages of galaxy formation \citep{Catelan01} and persists over their entire lifetime, as galaxies have continuous interactions with their surroundings, leading to the effect of galaxy intrinsic alignments \citep[IA, see][for reviews]{Joachimi15,Kiessling15,Kirk15,Troxel15,2024OJAp....7E..14L}. These coherent orientations will lead to additional shape correlations that we must model precisely to extract unbiased cosmological measurements from a weak lensing survey. This is particularly important for \Euclid \citep{EuclidSkyOverview}, for which the alignment of red galaxies has already been observed \citep{Q1-SP028}, and which aims to measure cosmological parameters with sub-percent precision.

Consequently, it is now mandatory to mitigate the impact of IA on weak lensing analyses. This led the scientific community over the past two decades to directly constrain IA from observations. On this basis, a clear dichotomy has been detected. Red elliptical (pressure-supported) galaxies tend to stretch their shapes towards the direction of matter overdensities \citep{Catelan01}, which result in a non-negligible IA signal that has been constrained by a number of studies \citep{Mandelbaum06,Hirata07,Okumura09b,Joachimi11,Singh15,Johnston19,fortuna21b,Samuroff22,FABIBI2024,David2025}. Blue spiral (rotationally supported) galaxies on the other hand, are thought to preferentially align their spins via tidal torquing \citep{2009IJMPD..18..173S}. This has been observed in simulations \citep[e.g.][]{Codis18}, with predictions for the resulting intrinsic alignment contamination that vary based on the implementations of hydrodynamics and subgrid galaxy physics (including baryonic feedback) within the simulation \citep{codis_intrinsic_2015,Chisari15,Tenneti15,Codis18,Kraljic2020}; however, they remain low compared to red galaxies in all cases (although some simulations might indicate a higher signal at high redshift). In observations, an alignment of blue galaxies has not been detected at low redshift \citep{mandelbaum11,Johnston19} or at intermediate and high redshift \citep{Tonegawa2018,Samuroff19,Samuroff22}, although the error bars in such cases are large.

To maximise the signal-to-noise ratio (S/N), the majority of IA studies have therefore focussed their analyses on red galaxies at low and intermediate redshift. Results from observations show a luminosity dependence of the IA signal, best described with a broken power law, with the bright galaxy tail well described by an index $\approx$ 1.2 \citep{Hirata07,Joachimi11,Singh15} and a fainter tail showing nearly constant evolution with luminosity \citep{Johnston19,Samuroff19,fortuna21a,fortuna21b}. This luminosity dependence can be attributed to the evolution of halo IA with halo mass following a single power-law model \citep{Piras18,Fortuna2024}, whereas observations indicate a break in the stellar-to-halo mass relation \citep{Fortuna2024}. Recently, \cite{Georgiou2025} showed that this luminosity evolution also depends upon on the morphological properties of the considered red samples.

The redshift evolution of IA has been studied in \cite{Singh15}, 
 \cite{fortuna21b}, and in \cite{Samuroff22}, with no clear trend derived from these analyses. However, these relations were derived from samples with varying satellite fractions, which complicated the interpretation of these results. Indeed, constraints on IA up to today were mostly derived from the linear alignment (LA) model \citep[LA]{Hirata04}, which relates galaxy orientations to the strength of its local tidal field at the moment of its formation. While this model accurately described central galaxy alignments, which are predominant at high luminosity, observations \citep{Johnston19,Georgiou2019b} -- together with simulations \citep{2017arXiv171207818W,Chisari17,2020MNRAS.491.5330S} -- suggest a radial alignment of satellite galaxies at small scales, with a vanishing contribution at larger scales thus leading to random satellite orientation at linear scales. In this scenario, satellites would therefore boost the IA signal at small scales and suppress it at larger semi-linear and linear scales. 

Since most IA analyses are derived from the observations of red galaxies, a gap exists between the samples used to constrain IA and weak lensing surveys. These typically span a much broader redshift range and observe all galaxies above a magnitude threshold, thereby making no colour or luminosity selection. These surveys employ tomographic bins to quantify structure growth and more accurately constrain cosmological parameters. Being flux-limited, the highest tomographic bins will include the most luminous objects. Since satellites are fainter, this results in a varying overall satellite fraction within each tomographic bin. In addition, the fraction of red galaxies evolves with redshift. Therefore, it is not straightforward to extrapolate the results of IA analyses towards informative priors for a full weak lensing analysis. To investigate the impact of IA in future \Euclid cosmological analysis, it is therefore necessary to simulate a realistic galaxy population with intrinsic alignments, and with survey properties similar to future \Euclid observations. 

In this context, we developed a series of papers to accurately forecast the effect of IA in future \Euclid analyses. \citet[][hereafter \Kai]{EP-Hoffmann} presented a method developed to implement realistic IA within the \Euclid Flagship simulation \citep{EuclidSkyFlagship}. The current paper offers two main objectives. First, it validates the empirical implementation of IA in Flagship,  testing whether analytical IA models can accurately fit the IA correlation functions in the simulation and yield parameter constraints that are consistent with observational samples with precise IA measurements. Second, we investigate the evolution of galaxy intrinsic alignments with redshift, luminosity, and colour in the \Euclid Flagship simulation.
The results presented here serve as informative priors for the Flagship 3$\times$2pt analysis \citep{EP-NavarroGirones}, which forecast the contamination to \Euclid's weak-lensing measurements.
This paper is organised as follows. In Sect. \ref{sec:data}, we present the specifics of the simulation and IA statistics we measured. In Sects. \ref{sec:model} and \ref{sec:methodology}, we present the analytical IA models and the methodology that we developed to constrain IA evolution. Finally, we present our results in Sects. \ref{sec:mockvalid} and \ref{sec:result} and we our conclusions in Sect. \ref{sec:Conclusions}.

\section{\Euclid IA mock data}\label{sec:data}
\subsection{The Flagship simulation} \label{sec:flagship}

\begin{figure}
  \includegraphics[width=0.98\columnwidth]{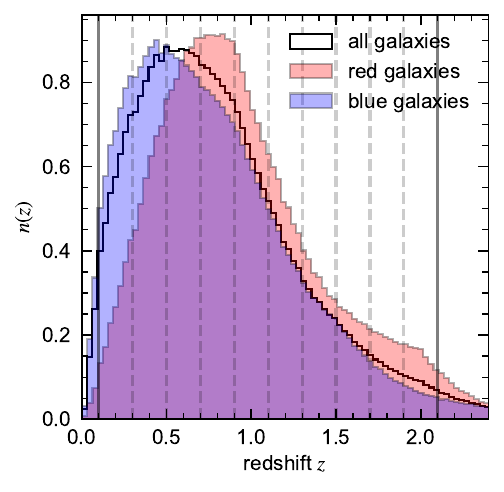}
  \caption{Flagship normalised redshift distribution. This distribution is computed for galaxies with visible magnitude $\IE < 24.5$. The two continuous vertical lines represent the range of our analysis, $0.1 < z < 2.1$, while each dashed line delimits the redshift bins we considered.}
 \label{fig:fig_fs2}
\end{figure}
The \Euclid Flagship simulation is one of the largest $N$-body simulations carried out thus far, with a box size  $L = 3.6 \, h^{-1} \,$Gpc and a particle mass resolution of $m_{\mathrm{p}} = 10^9 \,h^{-1} \, M_{\odot}$. The simulation was run with the \texttt{PKDGRAV3} code \citep{PKDGRAV3-potter2016}. The fiducial cosmology corresponds to a flat-$\Lambda$CDM cosmology, with cosmological parameters defined in \cite{EuclidSkyFlagship}. 
A light-cone up to $z$ = 3 was produced on the fly during the simulation, covering one octant of the sky, around 5157 deg$^2$, octant approximately centred at the North Galactic Pole ($145\, \text{deg} \,  < \text{RA} <235 \, \text{deg} \,, 0 \,\text{deg} < \text{Dec} <90\,\text{deg}$). Dark matter halos were identified with the \texttt{rockstar} algorithm \citep{behroozi2012}. Halos were populated given an improved halo occupation distribution (HOD) model \citep{Carretero15}. The HOD was combined with halo abundance matching techniques to reproduce the observed local galaxy luminosity function and the colour-magnitude distribution observed by the Sloan Digital Sky Survey \citep[SDSS;][]{York2000}. Satellite galaxies were placed within dark matter halos following a triaxial NFW profile \citep{Navarro1997}. Eventually, this HOD combined with halo abundance matching yielded a galaxy catalogue with physical properties such as stellar mass, and magnitude in several observational bands. In addition, lensing parameters are derived from convergence maps following the method introduced in \cite{lensing-fosalba2016}. 
 \subsection{Galaxy intrinsic shape}
The HOD model described above only yields luminosity-based physical properties. In addition, we needed to model consistent galaxy shapes with realistic alignment amplitudes. We quickly review the work presented in \Kai here. First, galaxies were modelled as 3D ellipsoids whose axis ratios were drawn from a Gaussian distribution. The parameters of this distribution vary with galaxy redshift, colour, and magnitude, calibrated in such a way that the distribution of projected 2D shapes matched the one observed by the COSMOS survey \citep{Cosmos2016}. Once a set of 3D shapes had been modelled, we needed to assign the orientations for galaxies.

\begin{table}
\centering
\caption{Parameters used to model galaxy misalignment amplitudes in the Flagship simulation, as a function of redshift, colour, and magnitude (see Eq. \ref{eq:kaimodel}) }
\begin{tabular}{c|cc}
\hline
& centrals & satellites \\
\hline
\hline
\noalign{\vskip 1pt}
$p_{0}$   & 1.02 & 4.74 \\
$p_{1}$  & $-$0.41 & 0.43 \\
$p_{2}$ & 3.46 & 5.72 \\
$p_{3}$ & 2.71 & 2.26 \\
$p_{4}$ & 4.28 & 4.27 \\
$p_{5}$ & $-$2.43 & $-$3.07 \\
\hline
\end{tabular}
\label{tab:params}
\end{table}
 Over the past few years, a variety of methods have been developed to implement intrinsic alignment within $N$-body simulations. Some methods \citep{Jagvaral2022b,Jagvaral2022} use deep generative models trained on hydrodynamical simulations to determine galaxy intrinsic shape given the estimated tidal field of the simulation. Other methods \citep{Hoffmann2022,VanAlfven2023} are based on empirical models, with parameters determined in a Bayesian approach and comparing the alignment with observations and/or hydrodynamical simulations. In \Kai, we used an empirical model, similar to the approach taken in the MICE simulation \citep{Hoffmann2022}, to determine intrinsic orientations based on dark matter halo and galaxy properties. Central galaxies are supposed to have their principal axis aligned with those of their host dark matter halos, while satellite galaxies are oriented by pointing their major axis towards their host halo’s centre. Then, the strength of the alignment is determined by misaligning the initial orientations by \referee{3D angle} $\theta$ given a von Mises--Fisher distribution,
\begin{equation}
P(\cos \theta)=\frac{1}{2 \sigma_{\mathrm{MF}}^2 \sinh \left(\sigma_{\mathrm{MF}}^{-2}\right)} \exp \left(\frac{\cos\theta}{\sigma_{\mathrm{MF}}^2}\right) \ ,
\end{equation}
which specifies the misalignment probability. \referee{We refer the reader to \Kai for technical details on the implementation of galaxy misalignments in the simulation}. The width of the distribution $\sigma_{\mathrm{MF}}$ is a free parameter that determines the strength of the alignment. Increasing  $\sigma_{\mathrm{MF}}$ leads to a higher randomisation of the initial orientations and therefore to a lower IA signal. $\sigma_{\mathrm{MF}}$ is parametrised as 
\begin{equation} \label{eq:kaimodel}
\sigma_{\mathrm{MF}}=p_0 \underbrace{\left(\frac{z}{z_0}+1\right)^{p_1}}_{\begin{matrix}{\sigma_z\left(p_1\right)}\end{matrix}} \underbrace{\left(\frac{M_r}{M_0}+p_2\right)^{p_3}}_{\begin{matrix}{\sigma_{\text {mag }}\left(p_2, p_3\right)}\end{matrix}} \underbrace{\left(\frac{u-r}{(u-r)_0}+p_4\right)^{p_5}}_{\begin{matrix}{\sigma_{\rm{c o l}}\left(p_4, p_5\right)}\end{matrix}} \ ,
\end{equation}
where three power laws have been introduced to model the dependence of galaxy alignment with galaxy colour, magnitude, and redshift. The fiducial values $z_0=1$, $(u-r)_0 = 1$, and $M_0 = -22$ were chosen to provide realistic amplitudes of misalignment. The alignment strength is determined separately for central and satellite galaxies, such that the model has a total of 12 parameters. These parameters have been calibrated against observed alignment strengths measured with the SDSS Main \citep{Blanton2005} and Baryon Oscillations Spectroscopic Survey (BOSS) LOWZ samples \citep{Dawson13,Reid16} at low redshifts ($z<0.36$), and with the HORIZON-AGN hydrodynamical simulation \citep{Dubois14,codis_intrinsic_2015, Chisari15, Chisari17} at redshift $z=1$. It is worth mentioning that the model used in Flagship is different from the one used in MICE \citep{Hoffmann2022}. Indeed, in \cite{Hoffmann2022}, a distinct calibration was performed on spiral and elliptical galaxies, with spiral galaxies having their spin aligned with the angular momentum of their host dark matter halo. However, here we specifically introduced a dependence on galaxy colour and thus galaxy type. \referee{It was shown in \cite{Hoffmann2022} that a simple redshift-independent cut 
at $u-r \equiv M_{u}-M_{r}=1.2$ provides a distinction between red and blue galaxies in the COSMOS
survey, where $M_{u}$ and $M_{r}$ are the absolute rest-frame magnitudes in the CFHT $u$-band and the Subaru $r$-band in COSMOS, respectively. \Kai showed, that a cut at $u-r=1.32$
in Flagship yields a fraction of blue galaxies that deviates by $\lesssim 10$ percent
from COSMOS observations at redshifts below $2.5$.}

\begin{figure}
  \includegraphics[width=0.98\columnwidth]{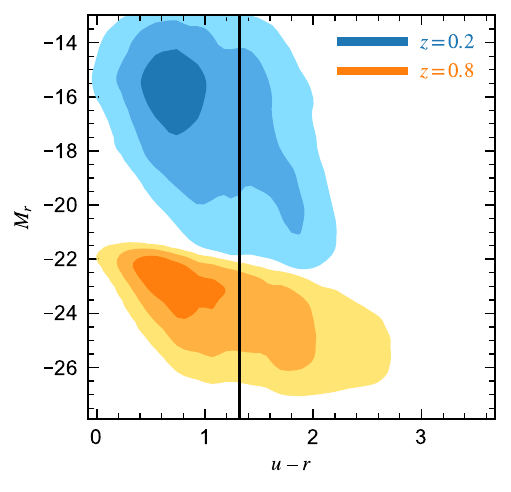}
  \caption{CMD of Flagship galaxies at redshift $z=0.2$ and $z=0.8$. We present the 68, 95, and 99$\%$ percentiles of the distributions. The \referee{solid} vertical line represents the cut that we used to define red and blue galaxy samples, i.e. $u-r = 1.32$. We shifted the rest-frame magnitude by $-4$ at redshift $z=0.8$ for clarity.}
  \label{fig:color_distribution}
\end{figure}

We present in Fig. \ref{fig:fig_fs2} the redshift distribution of red and blue galaxies, defined by this colour cut while we show in Fig. \ref{fig:color_distribution} the colour-magnitude diagram (CMD) of Flagship galaxies for two different redshift slices.
We present in Table \ref{tab:params} the parameters determined by the calibration procedure.
Given the galaxy intrinsic alignment implemented in the simulation, we were finally in a position to measure the 2D statistics and model those with conventional approaches.

\subsection{2-point function estimators}\label{sec:measure}

\begin{figure}
  \includegraphics[width=0.98\columnwidth]{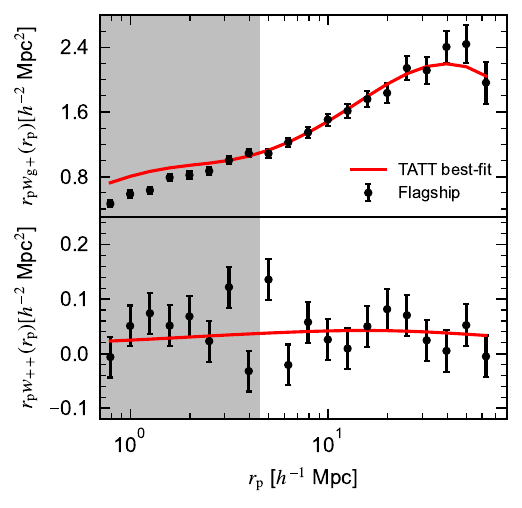}
  \caption{Example of IA measurements. The black points with error bars correspond to the measurements of $w_{\mathrm{g}+}$ (top) and $w_{++}$ (bottom), for the brightest red sample ($M_r = -22.8)$ in the redshift slice \referee{$ z \in [0.9, 1.1]$} in the Flagship simulation. The solid red lines correspond to the best fit using the TATT model given in Eqs. \eqref{eq:TATT1} and \eqref{eq:TATT2}. The shaded area corresponds to the scales that we do not use in the fitting procedure.}
    \label{fig:examplemeasurements}
\end{figure}

To determine constraints on IA observationally, it has been common to use projected correlation functions \citep{Joachimi11,Singh15,Johnston21,Samuroff22,FABIBI2024}. There are two main advantages associated with these statistics. \referee{First, integrating the 3D correlation function along the line of sight reduces sensitivity to radial separations, and therefore to redshift uncertainties. This is especially advantageous for photometric surveys, where the redshift scatter can be modelled within the projected signal \citep[e.g.][]{Joachimi11}.} Second, projected statistics are weakly sensitive to the detailed effect of redshift-space distortions (RSD) which we can thus model in a simple manner. Similar to 3$\times$2pt analysis, and for a reason that will become clear in Sect. \ref{sec:model_pgi_pii}, galaxies are divided into a clustering sample and a shape sample (that could be the same). Given these two populations, the cross-clustering 2-point correlation is estimated with the standard Landy--Szalay estimator \citep{Landy93} as
\begin{equation}
\xi_{\rm gg}(r_{\rm p},\Pi) = 1+ \frac{\rm{DS}(r_{\mathrm{p}},\Pi) - \rm{DR}_{\mathrm{S}}(r_{\mathrm{p}},\Pi) - \rm{SR}_{\mathrm{D}}(r_{\mathrm{p}},\Pi)}{\rm{R_\mathrm{D}R_\mathrm{S}}(r_{\mathrm{p}},\Pi)}\, ,
\end{equation}
with $r_{\mathrm{p}}$ and $\Pi$ the transverse and radial distance between the objects. D corresponds to the clustering sample while S corresponds the shape sample. $\rm{DS},\rm{DR}_{\mathrm{S}} \ (\rm{SR}_{\mathrm{D}})$, and $\rm{R_{\mathrm{D}}R_{\mathrm{S}}}$ represent the galaxy-galaxy, galaxy-random, and random-random pairs. Alternatively, clustering statistics can be described only in terms of the autocorrelation of the clustering sample. Similarly, we can estimate the shape-position and shape-shape correlations \citep{mandelbaum11} expressed as
\begin{equation}
\label{eq:measurements_clustering}
\xi_{\rm{g}+}= {\rm \frac{S_{+} D-S_{+} R_\mathrm{D}}{R_S R_D} \,; \quad \xi_{++}=\frac{S_{+} S_{+}}{R_\mathrm{S} R_\mathrm{S}} \,; \quad \xi_{\times\times}=\frac{S_{\times} S_{\times}}{R_\mathrm{S}R_\mathrm{S}}} \, .
\end{equation}
Here, we define
\begin{equation}
\label{eq:measurements_wgp}
{\rm S_{+} X}=\sum_{i \in S, j \in X} \gamma_{+} \left(j|i \right) \, ,
\end{equation}
and 
\begin{equation}
\label{eq:measurements_wpp}
{\rm S_{+} S_{+}}=\sum_{i \in S, j \in S} \gamma_{+} \left(j|i \right) \gamma_{+} \left(i|j \right), \quad {\rm S_{\times} S_{\times}}=\sum_{i \in S, j \in S} \gamma_{\times} \left(j|i \right) \gamma_{\times} \left(i|j \right) \ ,
\end{equation}
where \referee{$\mathrm{X}$ is either the clustering or random clustering sample}. $\gamma_{+, \times}\left(j|i \right)$\footnote{For consistency with previous works, we express the correlation function estimators in terms of the galaxy's observed shear. In this work, however, we only consider intrinsic galaxy shapes though their intrinsic components $e_{+,\times}$.} measures the component of the shear of galaxy $j$ along the line joining the pair of galaxy $i,j$ and at 45 degrees of that line, respectively. Throughout this paper, we will assume that positive $\gamma_+$ implies radial alignments, while negative $\gamma_+$ implies tangential alignments. We then integrate these 2-dimensional measurements along the line of sight (l.o.s.), expressed as
\begin{equation}
w_{AB}(\rm{r_p}) = \int_{-\Pi_{max}}^{+\Pi_{max}} \xi_{AB}(\rm{r_p},\Pi) \, \der \Pi \ ;
\end{equation}
where $A,B \in (\rm{g},+,\times)$. Finally, we define as in \cite{Blazek15},
\begin{equation}
W_{+}(r_{\mathrm{p}})  \equiv w_{++}(r_{\mathrm{p}}) + w_{\times\times}(r_{\mathrm{p}}) \ .
\end{equation}
We present in Fig. \ref{fig:examplemeasurements} an example of projected 2-pt statistics measured in Flagship, for a red galaxy population at an effective redshift $\langle z \rangle = 1.0$.
Given the measurements of these 2-point statistics, we can go on to describe their modelling.

\section{IA modelling}\label{sec:model}
\subsection{Galaxy power spectrum}\label{sec:model_pgg}
On large enough scales, we can relate the galaxy and matter density fields as $\delta_{\rm g} = b_1 \, \delta_{\mathrm{m}}$, with $b_1$ the so-called linear bias parameter \citep{1984ApJ...284L...9K}. The galaxy power spectrum can then be expressed as 
\begin{equation}
P_{{\rm gg},\textrm{lin}}(k)= b_1^2 \, P_{\delta\delta,\rm{lin}}(k) \, ,
\end{equation}
with $P_{\delta\delta,\rm{lin}}(k)$ the linear matter power spectrum.
At quasi-linear scales (around 10 $h^{-1}$Mpc), the galaxy overdensity field can be written as an expansion of the density field $\delta$ and the tidal field $s_{ij}$ \citep{McDonald2009,Baldauf12,Saito14}
\begin{equation}
\label{eq:nl_bias_pgg}
\delta_{\rm g}=b_1 \delta+\frac{1}{2} b_2\left(\delta^2-\left\langle\delta^2\right\rangle\right)+\frac{1}{2} b_{s^2}\left(s^2-\left\langle s^2\right\rangle\right)+b_{3 \mathrm{n} 1} \psi+\cdots \, ,
\end{equation}
where $s^2=s_{ij}s_{ji}$ and the tidal field, $s_{ij}(\mathbf{k})=\left(\hat{k}_i \hat{k}_j- 1/3 \ \delta^{\rm K}_{ij}\right) \delta(\mathbf{k})$,\footnote{the hat subscript indicates that $\hat{k}$ is normalised.} is a symmetric $3 \times 3$ tensor. The $\psi$ term corresponds to the sum of third-order terms scaled with the same bias factor $b_{3 \mathrm{nl}}$ \citep{Saito14}. The corresponding expression of the galaxy power spectrum is given in Appendix \ref{sec:appendix_theory}. 

\subsection{Intrinsic alignments}\label{sec:model_pgi_pii}
In this section, we briefly review how to model the IAs of galaxies, with a primary focus on perturbative models. In particular, we are interested in the so-called nonlinear Alignment (NLA) model \citep{Hirata04,Bridle07} and the tidal alignment and tidal torquing (TATT) model \citep{Blazek19}. The NLA model is a simple extension of the LA model \citep{Catelan01} that replaces the linear matter power spectrum in the prediction by its nonlinear counterpart and has been widely used to determine constraints on IA amplitude over the last decade \citep{Joachimi11,Singh15,Johnston19,fortuna21b}. Recently, the TATT model, based on an exact perturbative expansion, was used to model the IA signal of SDSS BOSS and extended (e)BOSS \citep{Dawson16} samples in the work of \citet{Samuroff22} and \citet{FABIBI2024}. Within $3\times2$pt cosmological analysis where IA is modelled as a nuisance term, the NLA model has been used in the Kilo-Degree Survey (KiDS) analysis \citep{Heymans21} and the TATT model in the Dark Energy Survey (DES) analysis \citep{DES21}. In the following, we dive into more details about these models.

The observed shear of a galaxy can be decomposed as the sum of the local gravitational shear and its intrinsic ellipticity $\gamma = \gamma^{\rm G} + \gamma^{\rm I}$. Hence, the observed 2-point shear correlation takes the following form, 
\begin{equation}
\langle\gamma \gamma\rangle=\left\langle\gamma^G \gamma^{\rm G}+\gamma^{\rm I} \gamma^{\rm G}+\gamma^{\rm G} \gamma^{\rm I}+\gamma^{\rm I} \gamma^{\rm I}\right\rangle=\xi_{\rm G G}+\xi_{\rm GI}+ \xi_{\rm IG} + \xi_{\rm I I} \, ,
\end{equation}
where $\xi_{\rm G G}$ corresponds to the lensing signal and $\xi_{\rm I I}$ corresponds to the correlation between intrinsic galaxy shapes affected by
the same tidal field. 
\referee{$\xi_{\rm I G}$ ($\xi_{\rm GI}$) refers to the correlation between the intrinsic shape of a foreground (background) galaxy and the gravitational shear experienced by a background (foreground) galaxy, where the first (second) index refers to the lower (higher) redshift bin.} The $\rm I G$ term arises because the local gravitational potential that causes alignment of foreground galaxies also contributes to the shear of background sources, see \cite{Hirata04}\footnote{Note that the cross-term $\xi_{\rm GI}$ can only be non-zero if the two samples of galaxies do overlap in redshift, i.e the shear of a galaxy is not correlated with the intrinsic shape of a background galaxy for sufficient radial separation.}.

\subsubsection{Linear model}
The linear (or tidal) alignment model relates the intrinsic shape of a galaxy to the primordial potential at the time of galaxy formation, $\Psi_{\rm p}$,  \citep{Catelan01} expressed as
\begin{equation}
\gamma^{\rm I}=\left(\gamma_{E}^{\rm I}, \gamma_{B}^{\rm I} \right)=-\frac{\bar{C}}{4 \pi G} \left(\nabla_x^2-\nabla_y^2, 2 \nabla_x \nabla_y\right) \Psi_{\rm p} \, ,
\end{equation}
 where the derivatives with respect to $x$ and 
$y$ are evaluated in the transverse plane. Here the $E$ and $B$ components are defined with respect to the direction of the local tidal field. $\bar{C}$ is a normalisation constant typically fixed to a value of $5 \times 10^{-14} h^{-2} \, M_{\odot}^{-1} \,  \mathrm{Mpc}^3$ \referee{\citep{Bridle07}}. The intrinsic alignment power spectra are then given by \cite{Hirata04},
\begin{equation}
P_{\delta\mathrm{I}}(k, z)=C_1(z) P_{\delta\delta,\rm{lin}}(k,z) \, , \quad P_{\mathrm{II}}(k, z)=C_1^2(z) P_{\delta\delta,\rm{lin}}(k) \, .
\label{eq:linear}
\end{equation}
In these equations, $C_1(z)$ is given by 
\begin{equation}
\label{eq:linear_A1}
C_1(z)= -A_1 \, \bar{C} \, \frac{\rho_{\mathrm{crit}} \Omega_{\mathrm{m}}}{D(z)} \, , 
\end{equation}
where $\rho_{\mathrm{crit}}$ is the critical density and $D$ is the linear growth function, normalised at $z=0$. The parameter $A_1$ controls the response of a galaxy's intrinsic shape to the tidal field.
The cross-power spectrum galaxy-IA can then be expressed as $P_{\mathrm{gI}}$ = $b_1 P_{\delta \mathrm{I}}$. This model only predicts $E$-modes (curl-free) contributions. $B$-modes (divergence-free) contributions can arise if we include density weighting contributions, the latter of which are typically neglected in the implementation \citep{Hirata04,Blazek15}. 

The so-called nonlinear alignment model (NLA) substitutes the linear power spectrum in Eq. \eqref{eq:linear} by the nonlinear power spectrum, which improves the performance on intermediate and small scales (around 5 to 20 $h^{-1}$Mpc) even if this modelling is not consistent by itself \citep{Blazek15,Singh15}. The redshift evolution of $A_1$ is usually modelled as a power law,
\begin{equation}
\label{eq:znla}
A_1(z) = A_{1z}\left(\frac{1+z}{1+z_0}\right)^{\eta_1} \, ,
\end{equation}
with $z_0$ an arbitrary pivot redshift usually fixed to $z_0 = 0.62$. This redshift extension of NLA is called the $z$-NLA model. Additionally, we can add a luminosity dependence on $A_1$ such that 
\begin{equation} \label{eq:enla}
A_1(z,L) = A_{1z}\left(\frac{1+z}{1+z_0}\right)^{\eta_1} \left(\frac{L}{L_0}\right)^{\beta_1} \, , 
\end{equation}
with $L_0$ as a pivot luminosity corresponding to a $r$-band magnitude of $M_r = -22$ in the so-called e-NLA model. The tidal alignment model has shown reasonable agreement with observations of elliptical galaxy alignments on linear and quasi-linear scales \citep[e.g.][]{Fortuna2024}. 
 \subsubsection{Tidal alignment and tidal torquing}
 \label{sec:TATT}
 In addition, theory suggests that the orientation of spiral galaxies can be correlated by means of a tidal torquing mechanism that tends to align their spins, namely, the so-called quadratic TATT alignment model \citep{Blazek19}.
In reality, intrinsic galaxy shapes are modulated by the galaxy overdensity field \citep{Blazek15}, as $\gamma^{\rm I}_{\rm g} \equiv \gamma^{\rm I} (1+\delta_{\rm g})$. The contribution of the density weighting, as well as higher order contributions such as tidal torquing, are included in a nonlinear framework proposed by \cite{Blazek19}. Similarly to the bias expansion, the intrinsic galaxy shape field can be written as an expansion of the density field, $\delta$, and tidal field, $s_{ij}$,
\begin{equation}\label{eq:IA_field}
\gamma_{i j}^{\rm I}= \underbrace{C_1 s_{i j}}_{\text{Tidal Alignment}}+ \underbrace{C_{1 \delta} \delta s_{i j}}_{\text{Density Weighting}} + \underbrace{C_2  \left(\sum_k s_{i k} s_{k j} - \frac{1}{3}\delta^{\rm k}_{ij}s^2\right)}_{\text{Tidal Torquing}} +\cdots \, 
\end{equation}
This expansion can be propagated at the 2-point level to yield expressions for the intrinsic power spectra, see Appendix \ref{sec:appendix_theory}. We note that those terms are modulated by the redshift-dependent amplitudes $A_1$, $A_2$, and $A_{1\delta}$, with the latter two being defined as 
\begin{equation}
\label{eq:nl_IA}
C_2(z)=5 A_2 \bar{C}_1 \frac{\rho_{\text {crit }} \Omega_{\mathrm{m}}}{D^2(z)} \, ,  \referee{\quad C_{1 \delta}(z)= - A_{1\delta} \bar{C}_1 \frac{\rho_{\text {crit }} \Omega_{\mathrm{m}}}{D(z)} \,.}
\end{equation}
The factor five in Eq. \eqref{eq:nl_IA} is included to account for the approximate difference in variance produced by the tidal alignment (TA) and tidal torquing (TT) power spectra. With this factor, the TA and TT contribution to $P_{\rm II}$ at $z=0$ should be roughly equal if $A_1 = A_2$. The redshift dependence of $A_2$ can be parametrised in a similar way as for $A_1$, 
\begin{equation}
\label{eq:A2_z}
A_2(z) =  A_{2z}\left(\frac{1+z}{1+z_0}\right)^{\eta_2} \,  . 
\end{equation}
As pointed out in \cite{Blazek19}, if $A_{1\delta}$ describes purely the density-weighting contribution, the linear galaxy bias approximation would imply $A_{1 \delta}(z)=b_{1} A_1(z)$. However, $A_{1 \delta}$ might capture any alignment physics that depends on $\delta s_{ij}$ and for this reason, it is common to add another free parameter $b_{\mathrm{TA}}$. The expression for the galaxy-IA power spectrum is then given by (\citealt{Blazek19}, Carter et al, in prep)
\begin{equation} \label{eq:TATT2}
P_{\mathrm{gI}} = b_1 P_{\delta \rm{I}} + \, \text{cross terms} \,  ,
\end{equation}
where the cross terms correlate galaxy bias terms with IA terms at perturbative level.
It is important to note that in addition to the NLA and TATT models, effective field theory (EFT) and Lagrangian perturbative models \citep{Vlah2019,Vlah2020,Bakx2023,Maion2023,Chen2023} that have been recently developed might provide better agreement to the observed IA signal in simulations at higher wave vector amplitude. However, these are more complex, with more free parameters than we might be able to constrain realistically, given the empirical IA implementation of Flagship. We leave to future works the investigation of the accuracy of these new models in a weak lensing cosmological analysis. 
 \begin{figure}
  \includegraphics[width=0.98\columnwidth]{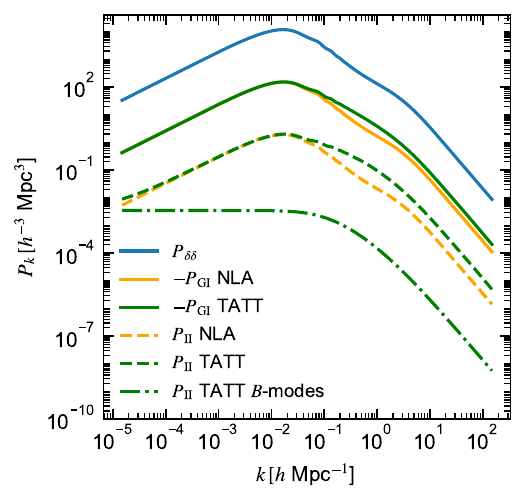}
  \caption{Example of IA power spectra. The power spectra are computed with the Flagship cosmology at an arbitrary redshift $z = 0.7$. The values of the IA parameters are $A_1 = 2$, $A_2 = 1$, and $b_{\mathrm{TA}}=1.$ \referee{We observe a constant $B$-modes amplitude on large scales arising from the $P_{0 B \mid 0 B}$ and $P_{B 2 \mid B 2}$ terms which can be subtracted and reduced into a effective shot noise contribution \citep{Blazek19}}.}
 \label{fig:fig_pks}
\end{figure}

We present the matter-intrinsic alignment power spectra of NLA and TATT in Fig. \ref{fig:fig_pks}. As we can see, the $\mathrm{GI}$ term has a higher amplitude than the $\mathrm{II}$ term, as the normalisation factor in Eq. \eqref{eq:linear_A1} has an amplitude below unity. The NLA power spectra are simply proportional to the matter power spectrum while the TATT spectra present different features due to the contribution of nonlinear alignments.
The $\mathrm{II}$ term, which correlates intrinsic shapes, can be measured in the observation given the auto-correlation of the shape sample, described by the projected 2-point functions $w_{++}$ and $w_{\times \times}$. The $\mathrm{\delta I}$ term (more precisely its biased version, $\mathrm{gI}$),  is instead effectively measured with $w_{\mathrm{g}+}$ by cross-correlating a density sample with a shape sample. Thus, we needed to model these statistics given the theoretical power spectra presented in this section, for which the expressions can be found in Appendix \ref{sec:appendix_theory}.

\begin{figure*}
    \centering
    \includegraphics[width=1.5\columnwidth]{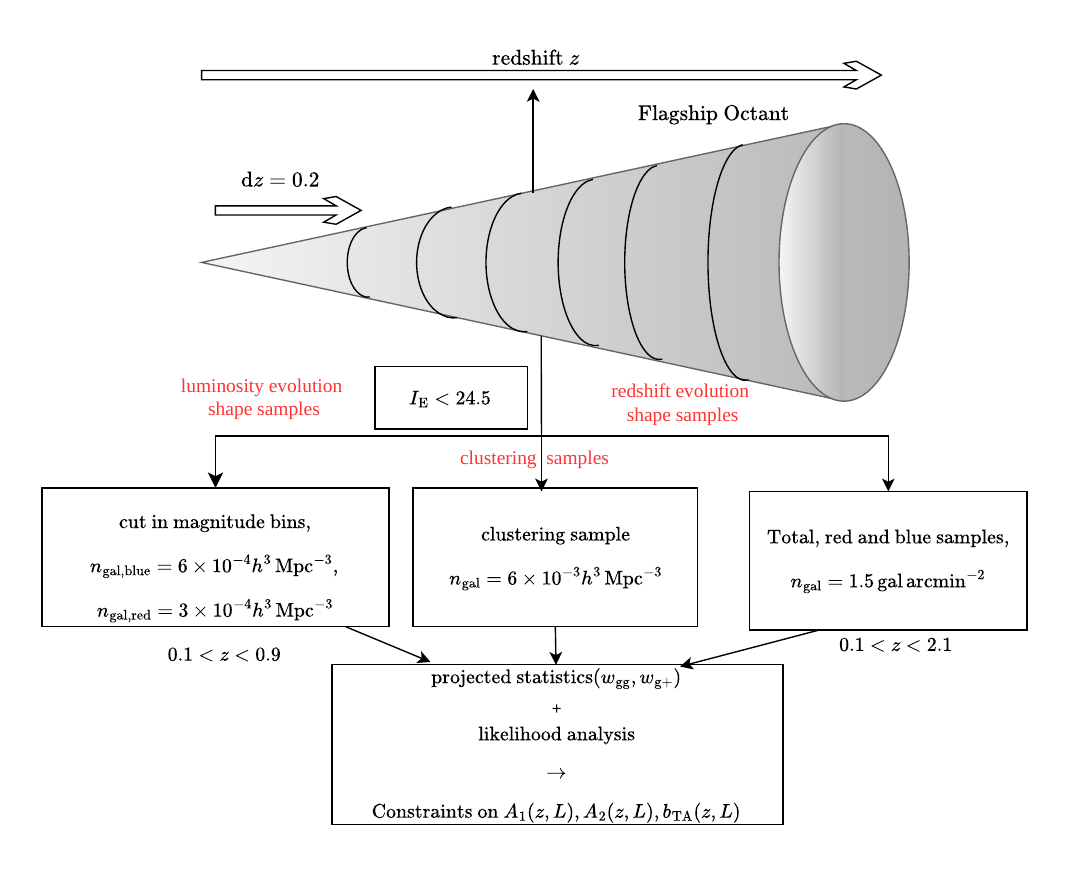}
    \caption{Illustration of the sample selection implemented in Flagship. The lightcone is divided into redshift slices of a size $\Delta z = 0.2$. From each slice, we constructed a density sample from which we computed $w_{\mathrm{gg}}$ and a variety of shape samples from which we computed projected IA statistics.}
    \label{fig:diagram}
\end{figure*}

\section{Parameter inference} \label{sec:methodology}
The parameter inference is performed by means of a likelihood analysis of the
data. The likelihood, $\mathcal{L}$, is assumed to be Gaussian,
\begin{equation}
-2 \ln \mathcal{L}(\theta)=\sum_{i, j}^{\rm{N_d}} \Delta_i(\theta) \hat{\Psi}_{i j} \Delta_j(\theta)  + \text{const}\,,
\end{equation}
where $\vec{\theta}$ is the vector of parameters, $\vec{\Delta}$ is the data-model difference vector, $N_{\rm d}$ is the total number of data points and $\hat{\Psi}_{i j}$ is the precision matrix, the inverse of the covariance matrix.

\subsection{Sample selection}
\referee{We sketch in Fig. \ref{fig:diagram} the methodology used in this work to investigate the evolution of IA amplitudes with colour, magnitude, and redshift. The Flagship light-cone is divided into redshift slices of a size $\Delta z = 0.2$
in the range $0 \leq z \leq 2.1$, and only galaxies with apparent magnitude $\IE<24.5$ are selected to mimic
\Euclid's DR1 observations.}

\referee{The clustering sample is constructed within each slice by selecting a sub-sample of the total galaxy catalogue with number density $n = 6 \times 10^{-3} \,h^{3}$ Mpc$^{-3}$ to avoid very time-consuming 2-point function estimations. We have checked that our clustering measurements $w_{\rm {gg}}$ remain unchanged for a higher density threshold.}

\referee{To investigate the redshift evolution of IA over $0.1 < z < 2.1$ three shape samples are constructed in each slice: a red and blue sample defined via the $u-r$ colour cut, and a total (red and blue combined) sample. Each shape sample is downsampled to a galaxy number density $n = 1.5 \, \text{gal} \,\text{arcmin}^{-2}$ where possible, ensuring high S/N measurements of $w_{\rm g+}$. This density threshold only affects the low-$z$ redshift bins.}

\referee{To probe the luminosity evolution of IA in the range $0.1 < z < 0.9 $, the red and blue galaxy populations are split into equi-populated bins of rest-frame absolute magnitude, with number densities chosen to probe the high-luminosity regime while maintaining sufficient S/N for the 2-point measurements of each magnitude bin sample. We only analyse galaxies brighter than $M_{r}= -19.5$; incorporating fainter galaxies increases the number of samples significantly while having only a minor impact on constraining the 
$A_1$-luminosity relationship.}

\referee{We note that the downsampling of the parent catalogue leads to larger error bars than those expected for the full Euclid DR2 like sample.\footnote{given the area of Flagship.} Higher number densities would increase the S/N of the 2-point measurements, potentially revealing model failures at smaller scales and thus modifying the scale cuts derived in Sect. \ref{sec:scalecuts}. However, the scale cuts presented in this work are not intended to be directly applied in weak-lensing cosmological analyses, which should derive their own scale cuts based on the full simulated sample. The IA parameter constraints derived here are valid within the scale ranges where our analytical models provide a consistent description of the Flagship simulation, and can therefore be used as informative, broads priors, for futures analyses.}

\subsection{Data vectors}

When computing the projected statistics described in Sect. \ref{sec:measure} for each sample, we used 21 logarithmic bins in $r_{\mathrm{p}}$, over the range $0.7$--$70 \, h^{-1}$Mpc with $\Pi_{\rm max} = 100 \, h^{-1}$Mpc, and $\der \Pi = 10 \, h^{-1}$Mpc. We have checked that our projected measurements are insensitive to the choice of $\der \Pi$. The random catalogues $R_{\rm D}$ and $R_{\rm S}$ -- for clustering and shapes respectively -- 
have the same radial and angular distribution as the galaxy catalogues and are oversampled by a factor of ten compared to the clustering and shape galaxy catalogue. The sampling value of $r_{\mathrm{p}}$ is taken at the average separation of the data-random clustering pairs within each bin, to consider survey geometry. This yields the same values of $r_{\mathrm{p}}$ across all redshift bins at the $1 \%$ level, independently of the random catalogue used. In the above equations, all position-position correlations are computed with the package \texttt{pycorr}\footnote{\url{https://github.com/cosmodesi/pycorr}} wrapped around \texttt{Corrfunc} \citep{corrfunc}, while position-shape and shape-shape correlations are estimated with \texttt{TreeCorr} \citep{TreeCorr}.\footnote{\url{https://github.com/rmjarvis/TreeCorr}}
When investigating the evolution of galaxy IA with luminosity, we considered for our data vector $\vec{y}$, the concatenation of $w_{\rm gg}$ and $w_{\rm{g}+}$, $ \vec{y} = (w_{\rm gg}, w_{\rm{g}+})$. We found that $w_{++}$ only provides non-negligible information for highly dense shape catalogues. This statistic was therefore only used when investigating the redshift evolution of galaxy IA, in which case $\vec{y} = (w_{\rm gg}, w_{\rm{g}+},w_{++})$.
\subsection{Covariance matrices}
In this section, we describe how we derived the covariance matrix. Given a data vector and a covariance matrix, we can perform a likelihood analysis to determine constraints on IA amplitude, as we show in the next section.

The covariance matrix is first determined through the Jackknife method, which has been widely used in past analyses \citep{Hirata07,mandelbaum11,Joachimi11,fortuna21b}. We split the octant in $n_{\mathrm{j}} = 100$ subvolumes following a K-means algorithm implemented in \texttt{Treecorr}. The jackknife realisations are built by removing one of the $n_{\rm j}$ sub-samples each time, and computing the correlation functions for the remaining sample. We therefore have $n_{\rm j}$ jackknife realisations, each of them with a volume fraction of $(n_{\mathrm{j}} - 1)/n_{\mathrm{j}}$. The Jackknife estimate of the covariance matrix can thus be written as 
\begin{equation}
\hat{C}_{ij} =\frac{n_{\mathrm{j}}-1}{n_{\mathrm{j}}} \sum_{k=1}^{n_{\mathrm{j}}}\left(y_{i}^{k}-\bar{y}_i\right)\left(y_{j}^{k}-\bar{y}_{j}\right) \, , 
\end{equation}
where $y_{i}^{k}$ is the value of the data vector in bin $i$ for the $k$-th jackknife realisation and $\bar{y_{i}}$ is the average of the jackknife realisations 
\begin{equation}
\bar{y}_i=\frac{1}{n_{\mathrm{j}}} \sum_{k=1}^{n_{\mathrm{j}}} y_i^k \,  . 
\end{equation}
\begin{table}
    \centering
    \caption{Priors used in this analysis.}
\begin{tabular}{c|cc}
\hline \text { Parameter }  & \text { TATT Prior } & \text { NLA Prior } \\
\hline \hline 
\noalign{\vskip 1pt}
$A_1$  & \textrm{U}[$-15$,15] & \textrm{U}[$-15$,15] \\
$A_2$  & \textrm{U}[$-8$,8] & $\delta^{\rm K}[0]$\ \\
$b_{\text {TA }}$ & \textrm{U}[$-6$,6] & $\delta^{\rm K}[0]$\\\
$b_1$ & \textrm{U}[0.1,4] & \textrm{U}[0.1,4] \\
$b_2$  & \textrm{U}[$-6$,6] & \textrm{U}[$-6$,6] / $\delta^{\rm K}[0]$\\
\hline
\end{tabular}
 \tablefoot{For the NLA model, we tried two configurations, see Sect. \ref{sec:mockvalid}. $\rm{U}$[$a$,$b$] denotes the uniform priors defined within the quoted ranges, while fixed parameters are denoted with the Kronecker symbol $\delta^{\rm K}$.}
\label{tab:tab_prior}
\end{table}
An unbiased estimate of the precision matrix is then given by \citep{Hartlap07}
\begin{equation}
\hat{\Psi}=\frac{n_{\rm j}-n_{\rm d}-2}{n_{\rm j}-1} \, \hat{C}^{-1} \, ,
\end{equation}
with $n_{\rm d}$ the number of elements in the data vector.
We then updated our covariance matrix estimator following analytical prescriptions. We assumed Gaussian covariance, which is the sum of two components: the cosmic variance and the shape and shot noise. The expression for the cosmic variance contribution to the covariance is given by \citep{Krause2017,Samuroff22,FABIBI2024} 
\begin{equation} 
\label{eq:gcovariance}
\begin{aligned} 
& \operatorname{Cov}\left[w_{\alpha \beta}\left(r_{\mathrm{p}, i}\right) w_{\gamma \epsilon}\left(r_{\mathrm{p}, j}\right)\right] =  \\ 
& \frac{1}{\mathcal{A}\left(z_c\right)} \int_{0}^{\infty} \frac{k \mathrm{~d} k}{2 \pi} \bar{\Theta}_{\alpha \beta}\left(k r_{\mathrm{p}, i}\right) \bar{\Theta}_{\gamma \epsilon} \left(k r_{\mathrm{p}, j}\right) \times \\ 
  \Big[\Big.&(P_{\alpha \gamma}(k) + \deltaK{\alpha \gamma} N^{\alpha}) \ (P_{\beta \epsilon}(k) + \deltaK{\beta \epsilon} N^{\beta})\\ 
 + & (P_{\alpha \epsilon}(k) + \deltaK{\alpha \epsilon} N^{\alpha}) \ (P_{\beta \gamma}(k) + \deltaK{\beta \gamma} N^{\beta}) \Big. \Big] \, .
\end{aligned} 
\end{equation}

In the above equation, $\deltaK{}$ corresponds to the Kronecker symbol, which means that there is no noise term in cross-correlations. $\mathcal{A}\left(z_c\right)$ is the comoving area at the effective redshift of the sample; in our case, given our thin redshift bin, this simply corresponds to $z_c = (z_{\mathrm{min}} + z_{\mathrm{max}})/2$. The $\bar \Theta$ functions correspond to the bin-averaged Bessel functions given by 
\begin{equation}
\bar{\Theta}_{\alpha \beta}\left(k r_{\mathrm{p}, i}\right) = \frac{2 \pi}{A_i}\int_{r_{\mathrm{p}, i}^{\mathrm{min}}}^{r_{\mathrm{p}, i}^{\mathrm{max}}}\  r\, \Theta_{\alpha \beta}\left(k r \right) \der r \, ,
\end{equation}
where $A_i$ is the surface element of bin $i$.
The Bessel functions $\Theta_{\alpha \beta}$ are given by  $J_0$, $J_2$, $(J_0 + J_4)/2$, and $(J_0 - J_4)/2$ for $\alpha \beta \in \{\mathrm{gg},\mathrm{g+},++,\times\times\}$, respectively. The shot noise terms are given by 
\begin{align}
  \deltaK{\alpha \beta}N^{\alpha} = \left\{
\begin{array}{ll}
1/n_{\mathrm{dens}} \text{ for $ij \in \{\textrm{gg}\}$} \, , \\
\sigma_e^2/n_{\mathrm{shape}} \text{ for $ij \in \{++,\times\times\}$} \, , \\
\end{array}
\right.
\end{align}
where $\sigma_e$ is the shape noise, computed 
given the definition of \cite{Heymans2012}
\begin{equation}
\sigma_e^2=\frac{1}{2}\left[\frac{\sum\left(w_i e_{i, 1}\right)^2}{\left(\sum w_i\right)^2}+\frac{\sum\left(w_i e_{i, 2}\right)^2}{\left(\sum w_i\right)^2}\right]\left[\frac{\left(\sum w_i\right)^2}{\sum w_i^2}\right] \, ,
\end{equation}
where $e_{i,1},e_{i,2}$ are the intrinsic shape component of galaxy $i$.
We assume weights equal to unity for our galaxies as we perfectly know the shapes in the simulation. The number densities (in $h^{3} \, $Mpc$^{-3}$) are computed in a thin redshift slice $\Delta z = 0.01$ around the effective redshift of the sample. We note that this implementation of Gaussian covariance is different from the one of \cite{Samuroff22}. We decided to use the averaged Bessel functions as it has been commonly used in galaxy-galaxy lensing and clustering \citep{Marian2015,Grieb2016} covariance estimation. It also provide us with more stable numerical results. \referee{As the analytical covariance is Gaussian, the power spectra appearing in Eq. \eqref{eq:gcovariance} are evaluated at the linear order. In practice, for a given data vector, we first obtain a preliminary estimate of $A_1$ and $b_1$
using the jackknife covariance, which are then used to evaluate the Gaussian covariance. This analytical covariance is subsequently used as the fiducial covariance matrix throughout the rest of the analysis. We have verified that the jackknife and analytical covariance estimates yield consistent results.}

\begin{figure*}
  \includegraphics[width=1.85\columnwidth]{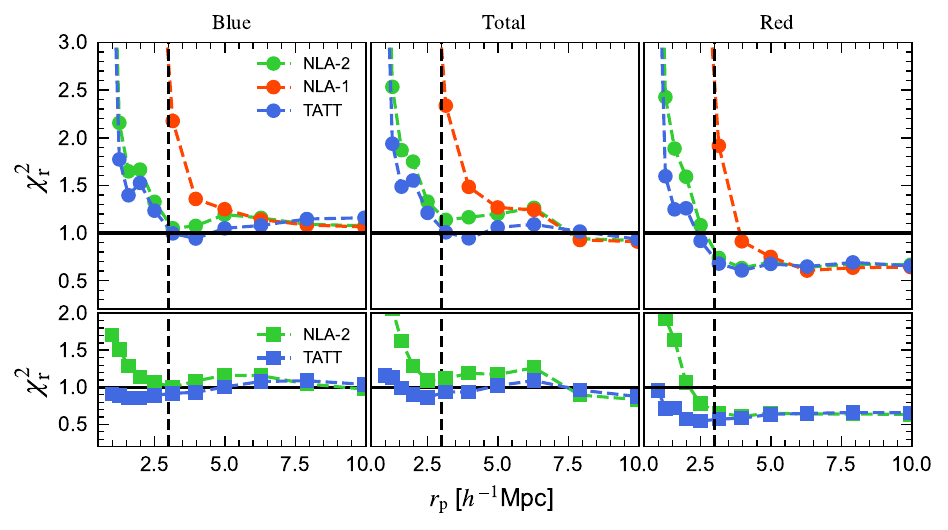}
  \caption{$\chi^2$ test for the NLA and TATT models for the blue population (\emph{left}), the total sample (\emph{middle}) and the red population (\emph{right}) at redshift $z=1$. We represent here the reduced $\chi^2$, $\chi_{\rm r}^2$ = $\chi^2$/$N_{\textrm{dof}}$ as a function of the minimum transverse separation $r_{\mathrm{p}}$ used in the fit. The top panel shows the evolution of $\chi_{\rm r}^2$ changing the scale cuts for the combined ($w_{\mathrm{gg}}$,$w_{\mathrm{g}+},w_{\mathrm{++}})$ data vector while in the bottom panel we only modify the minimum range of $w_{\mathrm{g}+}$, fixing the minimum scale of $w_{\mathrm{gg}}$ to 2 $h^{-1}$Mpc. The blue (red) population corresponds to a sample of galaxies with $u -r$ below (above) 1.32. In this test, each galaxy sample has the number density, $n = 1.5  \, \textrm{gal} \, \textrm{arcmin}^{-2}$. The total population has the same galaxy number density and is composed of both red and blue galaxies.}
  \label{fig:chi_testv1}
\end{figure*}

\subsection{Analysis choices}
Throughout this paper, the input cosmology used to measure and model the projected 2-point statistics described in previous sections is fixed to the one of the Flagship simulation, which is given in Sect. \ref{sec:flagship}. 
For the NLA model, we tried two different configurations. The first one follows the methodology described in \cite{Singh15}. Since the NLA model is purely linear (except with the substitution of the linear power spectrum by its nonlinear counterpart), we do not attempt to fit any nonlinear parameters, with $b_2$, $b_{s 2}$, and $b_{3 \mathrm{nl}}$ also fixed to zero. Our second configuration allows these bias parameters to be free, in the same formalism as described in \cite{Samuroff22}, in which $w_{\rm gg}$ is modelled including $b_2$ as a free parameter while $b_{s 2}$ and $b_{3 \mathrm{nl}}$ are fixed according to Eq. \eqref{eq:lagrangian_bias}. We have either two $\vec{\theta} = (b_{1},A_{1})$ or three $\vec{\theta} = (b_{1},b_{2},A_{1})$ parameters that we used for the so-called NLA-1 and NLA-2 models respectively. For the TATT model we have 5 parameters $\vec{\theta} = (b_{1},b_{2},A_{1},A_{2},b_{\text {TA}})$. To get constraints on IA parameters, we explore the likelihood of the data in a Bayesian approach with the nested sampler \texttt{nautilus}\footnote{\url{https://nautilus-sampler.readthedocs.io/en/latest/}} \citep{nautilus}, except when stated otherwise. Each chain is run until reaching 200 000 effective points with the exploration phase being removed. The priors used in this analysis are presented in Table \ref{tab:tab_prior}. Additionally, since our galaxy samples are defined on thin redshift slices, we do not integrate our model over the window functions in Eqs. \eqref{eq:w_gp_estimation} and \eqref{eq:w_gp_estimation2}, but instead estimate them at the effective redshift of our sample.\footnote{Except in Sect. \ref{sec:LOWZ} when deriving constraints for the LOWZ sample.} This yields negligible changes to parameters constraints and considerably improves the evaluation speed of the model. Given the measurements of 2-pt statistics, the estimation of theoretical models, covariance matrices, and an inference pipeline, we have now every ingredient to forecast the evolution of the IA amplitude with Flagship. But before doing so, we need to properly define scales at which we can model our data vectors and test the validity of our IA calibration within Flagship.

\section{Mock validation} \label{sec:mockvalid}
 The validation of the calibration of galaxy IA in Flagship performed in \Kai comes in several steps. First, analytical models should correctly describe the IA signal in the simulation. This step is essential because the calibration procedure in \Kai tuned the misalignment between galaxies and their host halos to match observed IA amplitudes, but did not directly fit NLA or TATT models to projected correlation functions. Therefore, we must verify that these widely used analytical frameworks can accurately reproduce the IA statistics measured in Flagship. Second, constraints on LOWZ Flagship samples should be in agreement with observations, as we used these samples to perform the IA calibration. This confirms that the calibration translates into consistent parameter values when the same model-fitting approach is applied to both simulated and observed data. Lastly, the evolution of galaxy IA with magnitude, colour, and redshift should be consistent with observations and hydrodynamical simulations. This final check is crucial: while LOWZ samples were used in the calibration, we must verify that Flagship also reproduces IA measurements from observational samples that were not included in the calibration procedure. The last point is discussed in Sect. 6 while we address in this section the first two.
\subsection{Scale cuts} \label{sec:scalecuts}

\begin{figure*}
  \includegraphics[width=1.8\columnwidth]{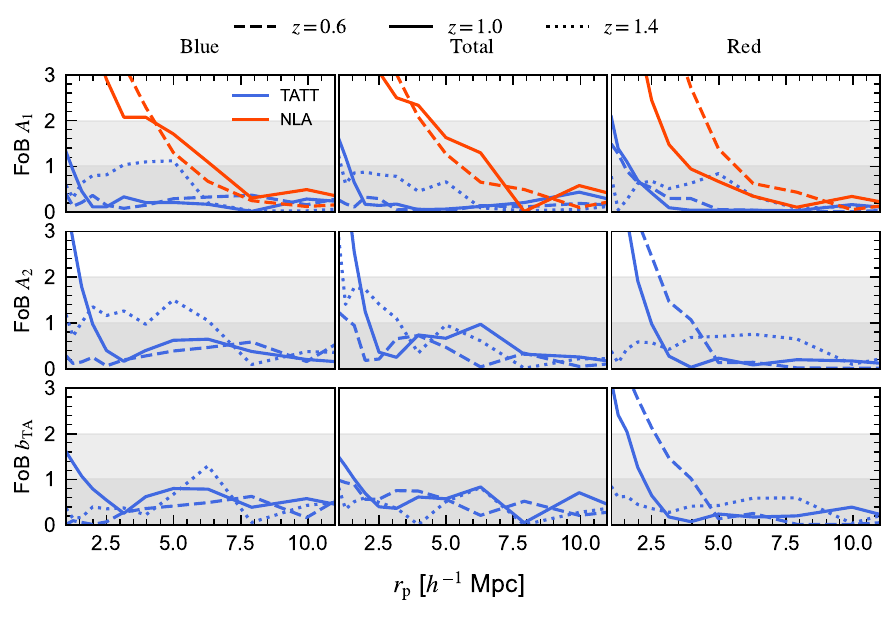}
  \caption{Evolution of the Figure of bias (FoB) at three different redshifts $z = 0.6$, $1.0$, and $z=1.4$ for $A_1$ (\emph{top}), $A_2$ (\emph{middle}), and $b_{\rm TA}$  (\emph{bottom}). For clarity, we do not represent NLA $A_1$ FoB at redshift $z=1.4$. We can observe a 1$\sigma$ shift on the FoB of IA parameters on scales below $7(6) \, h^{-1} \,\mathrm{Mpc}$ for the NLA(TATT) model.}
  \label{fig:evolution_constraints}
\end{figure*}

Based on observational data
\citep{Joachimi11,Singh15,Johnston19,fortuna21b,Samuroff22,FABIBI2024}, the NLA and TATT models have been used to model the observed $w_{\mathrm{g}+}$ measurements on scales above 6 and 2 $h^{-1}$Mpc, respectively. The majority of past IA studies were modelling projected statistics with the linear bias approximation. In the weakly nonlinear regime, the linear bias approximation should in principle break down. However \cite{Samuroff21} showed that linear bias could be extended up to $1 \, h^{-1}$Mpc with the hydrodynamical simulations they studied. These findings are nevertheless sample-dependent as a more strongly biased sample will exhibit nonlinear features on larger scales. It is the reason that an NLA scale cut of $6 \, h^{-1}$Mpc was first introduced by \cite{Joachimi11} as a conservative cut. The scale cuts for the TATT model were defined in \cite{Samuroff22} following the results of \cite{Blazek15}, which showed that a second-order perturbative framework could push the range of validity of the model down to $2 \, h^{-1}$Mpc. 

Before deriving constraints on intrinsic alignments, we first checked that these analytical models provided us with a correct description of the measured IA signals in the simulation on similar scales. This was done in two steps: first, we looked at the evolution of the reduced chi-square, $\chi_{\rm r}^2$, as a function of the minimum transverse scale for different samples of the simulation. Second, we investigated how IA constraints vary as a function of scale to estimate the level of uncertainties in their measurements. The best-fit parameters are determined in a frequentist approach, by minimising the $\chi^2$ with the minimum finder algorithm \texttt{iminuit}  \citep{James1975}\footnote{\url{https://iminuit.readthedocs.io/en/stable/}}. Parameter errors are found by computing the intervals where the $\chi^2$ increases by unity. 

The results are illustrated in Fig. \ref{fig:chi_testv1}. In the top panel, we represent the reduced $\chi^2$ as a function of scale for the combined \referee{($w_{\mathrm{gg}},w_{\mathrm{g}+}, w_{\mathrm++}$)} data vector at redshift $z=1$, for the blue, red, and total galaxy populations. We chose these populations at this particular redshift, as it provided us with a high S/N 2-pt measurement. The most widely used model in the literature, NLA-1, can properly model our measurements on scales above  $5$--$6 \, h^{-1}$Mpc. The modelling of nonlinear galaxy bias pushes this limit down to lower scales, with the NLA-2 and TATT models performing similarly well on our measurements on scales above $3 \, h^{-1}$Mpc. In addition, it is interesting to look at how NLA and TATT perform only when considering IA statistics for our data vector. This is illustrated in the bottom panel of Fig. \ref{fig:chi_testv1}. Here, we fix the minimum scale of $w_{\mathrm{gg}}$ at $r_{\mathrm{min}} = 6 \, h^{-1}$Mpc, and we only modify the minimum scale of ($w_{\mathrm{g}+},w_{++}$). We perform this test for NLA-2 only, which produces the exact same $w_{\mathrm{g}+}$ output as NLA-1. For the blue population, which has an IA amplitude $A_1$ close to 1, both models perform similarly well down to very small scales, $r_{\mathrm{min}}$ $\sim$ $1 \,h^{-1}$Mpc. As we start considering more strongly aligned galaxy samples, the  TATT model starts to outperform the NLA model. %

Additionally, we investigated how IA amplitudes evolve with scales. This was achieved by looking at the evolution of the figure of bias (FoB) defined as
\begin{equation}
    \textrm{FoB}(X,r_{\mathrm{p}}) = \frac{|X(r_{\mathrm{p}}) - X_{\mathrm{LS}}|}{\sqrt{\sigma(X,r_{\mathrm{p}})^2+\sigma(X,r_{\mathrm{LS}})^2}} \, ,
\end{equation}
where $X$ corresponds to the IA parameters $A_1$, $A_2$, and $b_{\rm TA}$. The large-scale values denoted by the subscript $\mathrm{LS}$ correspond to the mean and standard deviation of $X$ measured on large scales.\footnote{More specifically, the large-scale constraints are determined in the range $10$--$20 h^{-1}$ Mpc.} Our results are presented in Fig. \ref{fig:evolution_constraints} at redshift $z=0.6, \, z= 1.0$, and $z=1.4$. We can observe for the NLA models significant shifts in $A_1$ on scales below $7 \, h^{-1}$Mpc for the blue and total populations. For TATT, the FoB of $A_2$ for the blue population is biased by one sigma on scale below $5$--$6 \, h^{-1}$ Mpc at redshift $z=1.4$. For the red population, both $b_{\rm TA}$ and $A_2$ are biased on scales below 5 $h^{-1}$ Mpc. Based on these results, we impose the following conservative scale cuts for the data vector \referee{($w_{\mathrm{gg}},w_{\mathrm{g}+}$, $w_{++}$)}:
\begin{itemize}
    \item NLA-1 : $r_{\mathrm{min}} = (7,7,7)\, h^{-1}$Mpc ,
    \item NLA-2: $r_{\mathrm{min}} = (7,7,7)\, h^{-1}$Mpc ,
    \item TATT : $r_{\mathrm{min}} = (6,6,6)\, h^{-1}$Mpc .
\end{itemize}
These cuts yield degrees of freedom $N_{\textrm{dof}} = 28(18)$ and $28(17)$ for the NLA and TATT models, both including and not including $w_{++}$. 
Interestingly, \Kai showed that the Flagship satellite alignment contributions start to be a non-negligible source of alignment on scales below 5$\, h^{-1}$Mpc. Our perturbative models are therefore only capable of determining consistent IA constraints on scales at which central alignment dominates.
\referee{We emphasise that the FoB for $A_2$ and $b_{\mathrm{TA}}$ should be interpreted carefully. Since these parameters are weakly constrained on large scales and often consistent with zero, a shift at smaller scales may simply reflect scale-dependent parameter sensitivity rather than a actual failure of the model. The scale cuts we adopt are intentionally conservative: our goal is to provide informative priors for future Euclid 3$\times$2pt analyses, and pushing to smaller scales risks biasing the recovered parameters due to nonlinear contributions from satellite alignments and higher order tidal effects that become significant below below 5$\, h^{-1}$Mpc. We also note that weak lensing cosmological analyses typically apply their own scale cuts to deal with baryonic feedback and nonlinear modelling, so the small scales excluded here are unlikely to be used in practice anyway. The constraints derived within our scale cuts can therefore safely serve as priors for future analyses.}

The fact that NLA and TATT can model the observed signal down to such a small scale is surprising. Indeed, \cite{Bakx2023} showed that both models cannot reproduce the 3-dimensional power spectra of halo alignment above $k\approx 0.15 \, h$ Mpc$^{-1}$. However, the volume probed in their analysis is smaller than the volume we analysed in Fig. \ref{fig:chi_testv1} and in Fig. \ref{fig:evolution_constraints}, while they probed halo alignments. Even if the model of galaxy IA in Flagship has been calibrated against observations, it remains an empirical model that does not perfectly capture all the physical processes that shape galaxy orientation. This means that in future work, we would need to check the consistency of our results independently of the implementation of IA within Flagship. 
\subsection{LOWZ constraints}
\label{sec:LOWZ}
As a sanity check, given the pre-determined scale cuts, we also derived constraints on the IA amplitude of Flagship LOWZ-like samples, which are defined in the same way as in \Kai. Indeed, these samples have been used to provide calibration on the model used in \Kai. The results are presented in Table \ref{tab:tabLOWZ}. Since the area of  Flagship is smaller than the LOWZ footprint \citep{Reid16}, our errors on $A_1$ are slightly larger. 
\referee{We found that our results are in agreement at the $1\sigma$ level with the constraints of \cite{Singh15}, however with a small underestimation of the IA amplitude in the mock for the brightest samples.} 
Subsequently, one question remains regarding which NLA model we should use in the rest of the analysis. The TATT framework provides us with all the necessary ingredients (nonlinear bias, nonlinear IA amplitudes, and corresponding cross-terms) to probe nonlinear features. While we showed that NLA-2 can also bring similar performance compared to TATT on similar scales, we decided to adopt the NLA-1 configuration to investigate the luminosity and redshift dependence on linear alignment. The reason is simple: the majority of observational constraints investigating the luminosity$-A_1$ relationship were performed with the NLA-1 model. Therefore, using the same pipeline will ease the comparison with previous analyses.

\begin{table}
    \centering
    \caption{NLA best-fit for the Flagship LOWZ samples. We compare the results with observational constraints from \cite{Singh15}.}
\begin{tabular}{c|cccc}
\hline \text { Sample }   & $\logten{\frac{L}{L_0}}$& \text { $A_1$ NLA-1} &  \referee{\text{ $A_1$ NLA-2}} & \text {$A_1$ obs} \\
\hline \hline
\noalign{\vskip 1pt}
$L_1$& \phantom{0}0.44   & $6.8 \pm 2.0$ & $7.2 \pm 2.2$ & $8.5 \pm 0.9$ \\
$L_2$  &\phantom{0}0.03& $3.8 \pm 2.1$ & $4.5 \pm 2.6$ & $5.0 \pm 1.0$ \\
$L_3$  &$-0.14$& $3.7 \pm 2.2$ & $3.9 \pm 2.5$ & $4.7 \pm 1.0$ \\
$L_4$  &$-0.43$& $3.2 \pm 1.3$ & $3.2 \pm 1.4$ & $2.2 \pm 0.9$ \\
All &$-0.05$ & $4.3 \pm 0.7$ & $4.5 \pm 0.6$ & $4.6 \pm 0.5$ \\
\hline
\end{tabular}
 \tablefoot{The pivot luminosity $L_0$ corresponds to an absolute magnitude $M_{r}= -22$.}
 \label{tab:tabLOWZ}
\end{table}

\section{IA forecast}\label{sec:result}
For each topic, we first offer a brief overview of what we know from observations and hydrodynamical simulations before describing our results. Section \ref{sec:luminosity} presents constraints on the luminosity evolution of the IA parameters $A_1$, $A_2$, and $b_{\rm TA}$ while the redshift evolution is addressed in Sect. \ref{sec:redshiftsection}. All redshift- and luminosity-dependent power laws are determined using the local minima estimated with \texttt{iminuit}, with uncertainties defined by the $\Delta\chi^2=1$ criterion, unless stated otherwise.
 \subsection{Dependence on luminosity} \label{sec:luminosity}
 \subsubsection{Red galaxies}
The luminosity dependence of the red galaxy population has been widely studied in the literature. Results from observations \citep{Mandelbaum06,Hirata07,Joachimi11,Singh15,fortuna21b,FABIBI2024,Siegel2025} and hydrodynamical simulations \citep{Chisari15,Hilbert17,Bate2020} show that IA is correlated with luminosity, with brighter galaxies having a stronger IA amplitude. In its simplest representation, this relationship can be modelled as in Eq. \eqref{eq:enla} as a single power law, without considering the redshift dependence, the so-called L-NLA model,
\begin{equation}
\label{eq:l_nla}
A(L)=A_\beta\left(\frac{L}{L_{0}}\right)^\beta \, ,
\end{equation}
\cite{Joachimi11} and \cite{Singh15} found similar values for the amplitude and the power-law index for bright central galaxies, $A_{\mathrm{MegaZ}}=5.76_{-0.62}^{+0.60} \, , \beta_{\mathrm{MegaZ}}=1.13_{-0.27}^{+0.25} \, , A_{\mathrm{LOWZ}}=4.5_{-0.6}^{+0.6} \,, \beta_{\mathrm{LOWZ}}=1.27_{-0.27}^{+0.27}$ for the MegaZ and LOWZ samples. However, \cite{Johnston19} found $A_{\mathrm{G}+\mathrm{S}}=3.17_{-0.54}^{+0.55} \, ,\beta=0.09_{-0.33}^{+0.32}$ for the GAMA \citep{Gama2011} and SDSS sample, suggesting that the IA amplitude is independent of luminosity. The latter sample contains a larger fraction of satellites compared to MegaZ and LOWZ samples, which might explain this difference. However, the way it impacts the luminosity dependence is non-trivial, as we show in Appendix. \ref{sec:fsat}. In addition, results from hydrodynamical simulations \citep{Samuroff21} and from observations \citep{fortuna21b} show that a single power law badly describes the whole galaxy population when we consider galaxies with low luminosity $L < L_0$. It has been advocated that a better description of the IA-luminosity relation might be achieved with a broken power law given by 
 \begin{equation}
 A(L)=A_\beta\left(\frac{L}{L_{\text {break }}}\right)^\beta \text { with } \begin{cases}\beta=\beta_1 & \text { for } L \le L_{\text {break }}, \\ \beta=\beta_2 & \text { for } L>L_{\text {break }}.\end{cases}
 \end{equation}
\cite{fortuna21b} found that at low luminosity, the IA amplitude is nearly constant with a power-law index $\beta_1= 0.26^{+0.42}_{-0.77}$. In the high-luminosity regime, the authors found $\beta_2 = 1.17^{+0.21}_{-0.17}$ in agreement with the results of \cite{Joachimi11} and \cite{Singh15}. The transition between the two regimes occurs at $L_{\text{break}} = 0.3 - 0.6$ $L_0$ in agreement with the results of \cite{Samuroff21}. The origin of this break was investigated in \cite{Fortuna2024}, where the evolution of intrinsic alignments of KiDS LRGs as a function of halo mass was studied. They found that a single power law accurately describes the halo mass-intrinsic alignment relation, suggesting that the observed break in the intrinsic alignment-luminosity relation directly results from the break of the stellar-to-halo mass relation.
 \begin{figure*}
  \includegraphics[width=1.85\columnwidth]{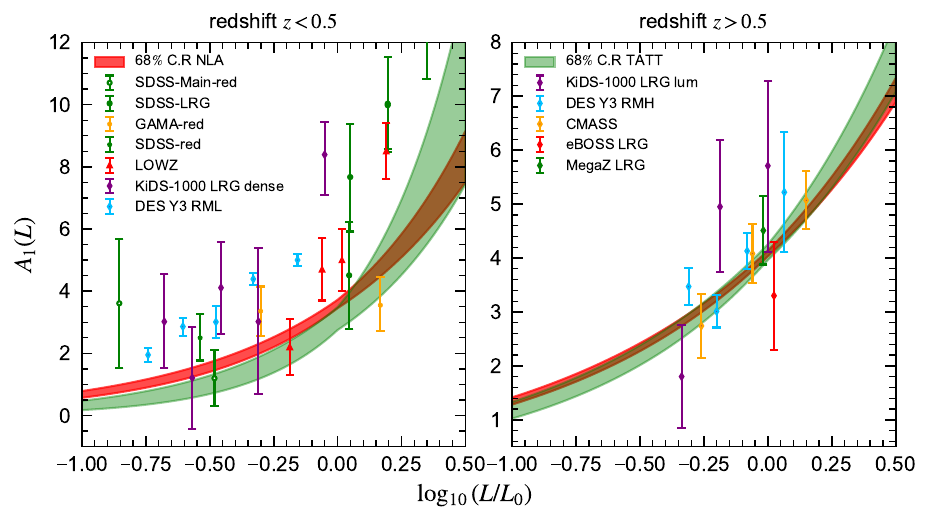}
  \caption{IA linear amplitude versus luminosity for the NLA model. The low-$z$ sample corresponds to the redshift slices below $z = 0.5$, while the high-$z$ sample corresponds to the redshift slices between $0.5 < z < 0.9$. The luminosity is computed as $\logten( L / L_0) = -(M_\text{r} - M_0)/2.5$. The shaded areas correspond to the 68$\%$ confidence regions, determined by propagating the errors on the power-law amplitudes and indices. The observational measurements are given in: SDSS-Main-red, SDSS-LRG and MegaZ LRG \citep{Joachimi11}, Gama-red and SDSS-red \citep{Johnston19}, LOWZ \citep{Singh15}, KiDS-1000 LRG dense and lum \citep{fortuna21b}, DES Y3 RML and RMH \citep{Samuroff22}, CMASS and eBOSS LRG \citep{FABIBI2024}.}
    \label{fig:LR_vs_A1_obs}
\end{figure*}
In Flagship, the IA misalignment model is modelled as a function of absolute magnitude, colour, and redshift. Table \ref{tab:params} presents the best-fit model of Eq. \eqref{eq:kaimodel}, used to implement the alignment in the simulation.  Negative values for $p_1$ and $p_5$ for centrals indicate that red, high-redshift galaxies will align more strongly with their corresponding dark matter halos. 
The positive value of $p_3$ for centrals means that brighter galaxies are misaligned more strongly with respect to their host halos compared to the dimmer ones in the simulation. However, it was shown in \Kai (see Fig. 13) that the misalignment amplitude as a function of magnitude is nearly constant. Since massive dark matter halos, which host bright red galaxies in the simulation, align more strongly with the large-scale structure \citep{Kurita2021}, our calibrated model will still produce a consistent luminosity evolution in the simulation.

To test the validity of Flagship, we determined constraints on parameters of Eq. \eqref{eq:l_nla}.
Since Eq. \eqref{eq:kaimodel} also predicts an evolution with redshift, we analyse low redshift ($0.1 < z < 0.5$) and high redshift ($0.5 < z  < 0.9)$ independently. 
The power laws are determined by minimising the $\chi^2$ defined as
\begin{equation}
\label{eq:chisquare}
\chi^2 = \sum_i \frac{(A_i - A_{\textrm{model}})^2}{\sigma_i^2} \, , 
\end{equation}
where the $\sigma_i$ correspond to the standard deviation on each individual measurements.
We present in Table \ref{tab:tab_LNLA} the corresponding constraints, and we compare the power laws we obtained with observations, as illustrated in Fig. \ref{fig:LR_vs_A1_obs}. For completeness, we present in Appendix \ref{sec:Appendixb} all samples used to derive these luminosity dependencies.

As expected, a single power law provides us with an accurate match to our IA measurements with a $\chi_{\rm r}^2$ of $0.46$ and $1.03$ at low and high redshift, respectively.  At low redshift (left panel), where most of the observations of IA have been carried out, our forecast on IA amplitude is below the majority of the observations but does agree with IA measurements on KiDS, GAMA, and SDSS. Indeed our best-fit value of $A_\beta$ is $20 \%$ and $38 \%$ below the amplitude determined in \cite{Singh15}, and \cite{Joachimi11}, respectively, while being consistent with the amplitude observed in \cite{Johnston19}. The power-law index is instead larger by $32\%$ and $48\%$ compared to \cite{Singh15}, and \cite{Joachimi11}, respectively. 
At higher redshift (right panel) our results are in better agreement with observations, and the power-law index we derived matches the ones measured in \cite{Joachimi11}, \cite{Singh15}, and \cite{fortuna21b}. Additionally, we can observe a similar trend between the observations and the simulation: \referee{the IA amplitude at pivot luminosity} increases with redshift while the power-law index decreases. Indeed the constraints given in \cite{Joachimi11} include the MegaZ sample, which has an effective redshift $\langle z \rangle = 0.54$ well above LOWZ, with $ \langle z \rangle = 0.28$. 
\begin{table}
\centering
\caption{Constraints on the evolution of $A_1$ with luminosity for the NLA model.}
\label{tab:tab_LNLA}
\begin{tabular}{p{2cm}|ccc}  
\hline
\hline
\text{Sample} & $A_{1\beta}$ & \phantom{0}$\beta_1$ & $\chi_{\rm r}^2$ \\
\hline
Red $z < 0.5$  & $3.59 \pm 0.15$ & $1.67 \pm 0.14$ & 0.46 \\
Red $z > 0.5$  & $4.08 \pm 0.06$ & $1.11 \pm 0.04$ & 1.03 \\
Blue $z < 0.5$ & $1.70 \pm 0.38$ & $1.29 \pm 0.41$ & 0.93 \\
Blue $z > 0.5$ & $1.70 \pm 0.11$ & $0.58 \pm 0.11$ & 1.79 \\
\hline
\end{tabular}
\end{table}

Even if we underestimate the signal at low $z$, the observed scatter in the observations suggests that the IA amplitude depends on factors other than galaxy luminosity, such as colour, with a trend recently measured in \cite{Siegel2025}. Satellite fraction can also provide significant changes in observed amplitude, as we explain in Appendix \ref{sec:fsat}. Therefore, the observed differences might arise from poorly controlled and understood selection effects. Observationally,  distinct shape measurement algorithms can provide different strengths of alignment \citep{Singh16,Georgiou19,Macmahon2024}, which can also explain the observed scatter in Fig. \ref{fig:LR_vs_A1_obs}. This is for example illustrated in \cite{FABIBI2024} where the authors found statistically significant shifts in IA amplitude for the BOSS and eBOSS samples when using either DES or UNIONS shape estimates for these galaxies. At redshift 0.5 < $z$ < 0.9 our results are in better agreement with observations. We, therefore, conclude that the intrinsic alignments implemented in Flagship provide an approximate description of the observed IA-luminosity relationship of the red population. However, further analyses are needed to understand how selection and observational effects induced the observed scatter in the observations. 

Additionally, we performed a fit to our IA measurements with broken power laws. In that case, we have four parameters: one amplitude, two power-law indices, and one pivot luminosity. We found that this more complex model does not provide us with a better fit than a single power law, with the same resulting $\chi^2$, but with two additional parameters. This is consistent with how the calibration was performed in Flagship: galaxy misalignment as a function of $r$-band magnitude is modelled as a single power law.

\subsubsection{Blue galaxies}

\begin{figure}
  \includegraphics[width=0.98\columnwidth]{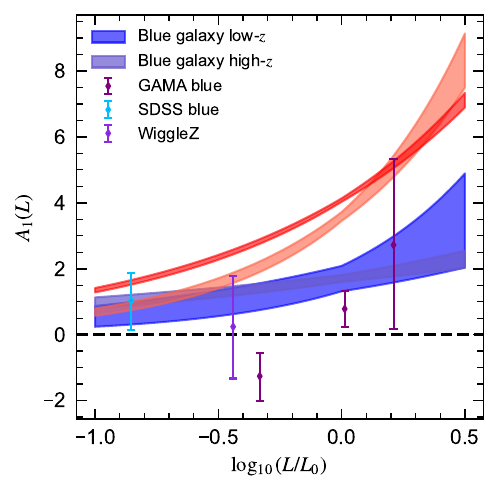}
\caption{Same as Fig. \ref{fig:LR_vs_A1_obs}, but for the blue galaxies of Flagship. The constraints on $A_1$ are determined with the NLA-1 model. We observe a non-zero IA amplitude over the whole considered luminosity range. This is compared with $A_1$ measurements of SDSS Blue, GAMA blue \citep{Johnston19}, and WiggleZ \citep{mandelbaum11}. The IA amplitude $A_1$ is consistent with zero for the five considered blue samples. We also show in red the luminosity evolution of red galaxy alignments presented in Fig. \ref{fig:LR_vs_A1_obs}.}
  \label{fig:LR_vs_A1_blue}
\end{figure}

While observations have shown that pressure-supported elliptical galaxies align their shape with the large-scale structure, there is no observational evidence for any linear alignment of spiral galaxies up to very high redshift, $z \approx 1.4$ \citep{mandelbaum11,Johnston19,Tonegawa2018,Samuroff19,Samuroff22} with $A_1$ measurements consistent with zero, albeit with large errors bars. This led to the conclusion that rotationally supported disks align at second order (their spin) with the large-scale structure. This is partially in line with results from hydrodynamical simulations. However, the strength of this blue alignment depends on the implementation of baryonic physics and varies between different hydrodynamical simulations. 
In Millenium TNG \citep{Hernandez2023}, \cite{Delgado2023} found a torquing (i.e. quadratic) alignment for blue galaxies between redshift $z=0$ and $z=1$. However, \cite{Zjupa2022} found a quadratic alignment in Illustris TNG-100 \citep{TNG2018} only for the most massive blue galaxies at redshift $z=1$, while a non-zero linear alignment was detected over a broad redshift range. The presence of both alignment mechanisms suggests that a coherent and unified framework of alignment such as TATT, could be applied in principle to any galaxy, independently of its type. However, it is important to note that the two latter studies investigated the torquing of blue galaxies as a function of the estimated tidal field within the simulation. In \cite{Samuroff21}, where NLA and TATT models are used to model the projected alignment $w_{\rm{g}+}$, they detected a non-zero linear alignment in Illustris TNG-300 for blue galaxies at redshift $z > 0.5$, with a luminosity dependence described by the coefficients $A_{\beta} = 2.5 \pm 0.7$ and $\beta = 0.24 \pm 0.21$. These values were derived from the faint end of the luminosity function, suggesting constant amplitude with luminosity. Regarding the tidal torquing amplitude, \cite{Samuroff21} found a signal consistent with zero in Illustris TNG-300, but detected one in MassiveBlack II \citep[MB-II;][]{MASSIVE2015}. In the Horizon-AGN simulation, on which we performed the high-$z$ calibration, \cite{Chisari15} found that disc galaxies tend to be oriented tangentially around spheroidals (elliptical galaxies) in three dimensions, an effect that is suppressed in projection. Similar alignment signals have also been reported in other hydrodynamical simulations, such as SIMBA \citep{Kraljic2020} and Illustris TNG \citep{Shi2021}.

We present in Fig. \ref{fig:LR_vs_A1_blue} our results for the blue population. We detect over the whole considered luminosity range a non-zero signal of linear intrinsic alignment. The amplitudes and power-law indices are given in Table \ref{tab:tab_LNLA}. As expected, the blue population shows a lower alignment than the red population as the value of $p_5$ for centrals (see Table \ref{tab:params}) is negative such that bluer objects will be more strongly misaligned.
The high-$z$ power-law amplitude is $32\%$ smaller (although nearly consistent at the 1$\sigma$ level) while the index is larger by a factor of 2.4 compared to \cite{Samuroff21}, suggesting a slight increase in luminosity.

The amplitude of blue alignment is also stronger than the one detected in \cite{Hoffmann2022}, due to the different implementation of IA between MICE and Flagship. Indeed in \cite{Hoffmann2022}, elliptical and spiral galaxies were treated independently, with red galaxies aligned and blue galaxies oriented randomly with respect to their host dark matter halos. In Flagship, the model has been unified with additional parameters that control the evolution of IA with colour. The high amplitude of tidal alignment recovered here is the consequence of three effects. First, the Flagship galaxy colour distribution is not perfect. As an alternative colour cut to define blue and red samples, we use the colour kind defined by Flagship catalogue. It splits the catalogue of galaxies based on their colour into three types: blue cloud, green valley, and red sequence. The addition of the green valley should in principle reduce the contamination between samples. However, we repeated the same analysis with Flagship colour cuts and found indistinguishable results. This means that our analysis is robust with respect to the way we define the colour bi-modality. However, we are still biased by the intrinsic colour of Flagship galaxies, which are key ingredients of the IA model. We leave the investigation of how Flagship IA measurements are sensitive to the colour distribution of its galaxies to future work. Second, there is a lack of observational alignment of blue galaxies to perform the calibration in \Kai. Upcoming observations of star-forming galaxies with \Euclid and DESI \citep{desi2016} will facilitate the calibration of the model to properly reproduce the IA signal of such galaxies. 
Third, the calibration against Horizon-AGN at $z = 1$ has a fundamental limitation in the current implementation. As already emphasised, in Horizon-AGN, blue/star-forming galaxies exhibit small tangential alignment amplitudes, not radial ones. However, the methodology described in \Kai makes it challenging to reproduce tangential alignments: achieving a negative $A_1$ would require systematically misaligning galaxies by angles greater than 90 degrees, which might not be possible with the current parametrisation of the misalignment strength in Eq \eqref{eq:kaimodel}. Future implementations could address this by allowing the initial alignment direction (radial vs. tangential) to depend on galaxy properties.

We also compare in Fig. \ref{fig:LR_vs_A1_blue} the luminosity power laws between the red and blue populations. The different evolution as a function of luminosity between these two populations is driven by the colour dependence $u - r$ on IA calibrated on observational data, thus making red galaxies more strongly aligned. This effect is stronger at high redshift as the colour distribution of galaxies spans a broader range as illustrated in Fig. \ref{fig:color_distribution}.

\subsubsection{Tidal torquing and density weighting contributions}
\label{sec:nonlinearIA_luminosity}

\begin{table*}
\centering
\caption{Constraints on the power-law amplitudes and indices of $A_1,A_2$, and $b_{\rm TA}$ with luminosity for the TATT model.}
\label{tab:tab_TATT}
\begin{tabular}{p{2cm}|ccc|ccc|ccc}  
\hline
\hline
\text{Sample} & $A_{1\beta}$ &  \phantom{0} $\beta_1$ & $\chi_{\rm r}^2$ & $A_{2\beta}$ &  \phantom{1}  $\beta_2$ & $\chi_{\rm r}^2$ &  \phantom{0} $b_{\rm TA\beta}$ &  \phantom{0} $\beta_{\rm TA}$ & $\chi_{\rm r}^2$ \\
\hline
Red $z < 0.5$  & $3.11 \pm 0.36$ & $2.37 \pm 0.38$ & 0.23 & $2.21 \pm 1.12$ &\phantom{0} $0.41 \pm 0.85$  & 0.24 & \phantom{0} $0.13 \pm 0.32$ & $5.41 \pm 11.96$& 0.14\\
Red $z > 0.5$  & $4.10 \pm 0.14$ & $1.25 \pm 0.09$ & 0.80 & $0.20 \pm 0.32$ & $-2.37 \pm 1.87$ & 0.57 & \phantom{0} $0.20 \pm 0.16$ & $1.43 \pm 2.14$ & 0.42 \\
Blue $z < 0.5$ & $1.78 \pm 1.09$ & $1.80 \pm 1.15$ & 0.18 & $1.90 \pm 2.76$ & \phantom{0} $0.71 \pm 2.25$ & 0.31 & $-2.86 \pm 7.13$ & $5.05 \pm 10.48$ & 0.07  \\
Blue $z > 0.5$ & $1.27 \pm 0.26$ & $0.57 \pm 0.38$ & 0.49 & $2.55 \pm 1.27$ &\phantom{0} $0.50 \pm 0.74$ & 0.59 & $-0.79 \pm 1.79$ & $0.77 \pm 4.49$ & 0.43 \\
\hline
\end{tabular}
\end{table*}
The evolution of $A_2$ and $b_{\rm TA}$ with respect to redshift and luminosity has not been intensively investigated in the literature. One main reason is that the volume probed by hydrodynamical simulations is too small to precisely constrain these parameters. In an attempt to do so, \cite{Samuroff21} determined a redshift and luminosity evolution of $A_2$, for the three hydrodynamical simulations considered in the analysis: Illustris-TNG, MB-II, and Illustris-1. The authors found $A_2$ consistent with zero, with no significant redshift or luminosity evolution\footnote{Except for MB-II at redshift $z=0$, from which $A_2$ is negative at the 3$\sigma$ level. However, MB-II has known limitations, namely the number of elliptical galaxies \citep{MASSIVE2015} and weak AGN feedback \citep{Huang2019}.}. In \cite{Samuroff22}, the authors determined constraints on $A_2$ and $b_{\rm TA}$ on redMaGiC DES Y3 samples. They detected a non-zero value of $b_{\rm TA}$ at the 2.5$\sigma$ level. However, the observed anti-correlation in the derived constraints between $A_2$ and $b_{\rm TA}$ yields different absolute values for these parameters between the low-$z$ and the high-$z$ sample. 
\referee{It is therefore not strictly ruled out that $b_{\rm TA}$ is negative as} the TATT framework does not include second-order expansion that contributes to the $\delta s_{ij}$ term \citep{Bakx2023,Maion2023}.
Additionally, some other sources of contribution which cannot be captured by a perturbative approach might contribute to the $\delta s_{ij}$ term in the nonlinear regime (satellite alignments, non-local tidal effects). 

We present in Table \ref{tab:tab_TATT} the power-law constraints of $A_1,A_2$, and $b_{\rm TA}$. At low redshift ($0.1 < z < 0.5$) and high redshift ($0.5 < z < 0.9$), we detect at the 2$\sigma$ level positive torquing alignment amplitudes for the red and blue galaxy populations, respectively. This explains why the TATT model provides us with a slightly different evolution of $A_1$ with luminosity for the red population at low redshift, as illustrated in Fig. \ref{fig:LR_vs_A1_obs}. Given our statistical precision, we do not detect any significant evolution of $A_2$ and $b_{\rm TA}$ with luminosity.
Interestingly in  \cite{Hoffmann2022}, the authors found consistent results between NLA and TATT constraints across all redshift ranges considered in their analysis. However, in their analysis, they directly fit the matter-alignment projected correlation function $w_{\rm{m}+}$ while we considered here $w_{\rm{g}+}$. The observed differences at low $z$ can be explained in two ways: the different galaxy bias models used between TATT and NLA, and the contribution of nonlinear IA parameters in TATT. We performed the same analysis with the NLA-2 model (see Sect. \ref{sec:mockvalid}) to see which effect has the stronger contribution. The derived power law has a higher amplitude $A_{\beta}=3.83 \pm 0.10$ and a power-law index $\beta = 1.61 \pm 0.13$ consistent with the NLA-1. This means that the extra TATT parameters are the main drivers of the difference between NLA and TATT at low luminosity and at low redshift.

\subsection{Evolution with redshift}
\label{sec:redshiftsection}
\subsubsection{Linear alignment evolution}
\label{sec:redshiftsectionNLA}

\begin{figure*}
  \begin{center}
  \includegraphics[width=1.8\columnwidth]{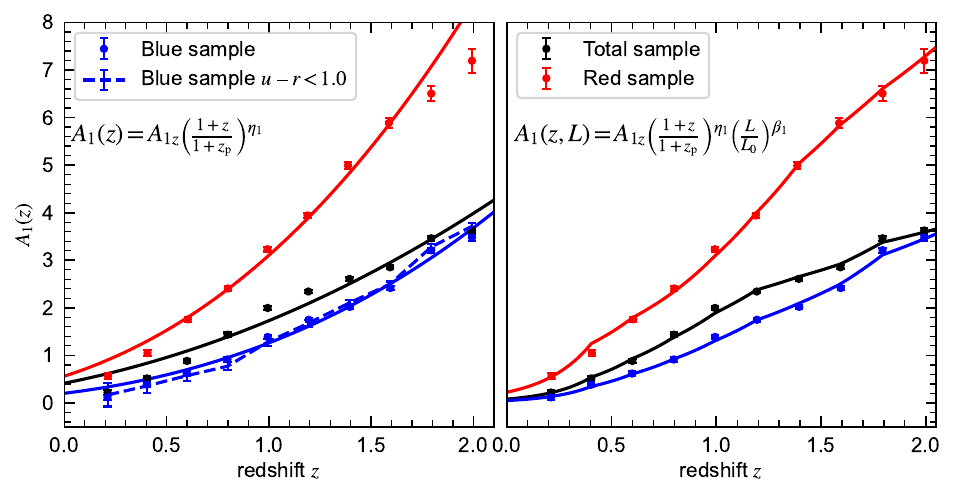}
  \caption{Redshift evolution of the linear IA amplitude. Each point represents the fit performed with the NLA-1 model to either the density samples (black), red samples (red), or blue samples (blue). Here, each sample is subsampled (if possible) to have the same number density, $n = 4 \times 10^{-3} h^{-3}$Mpc$^{3}$. \emph{Left}: the solid lines represent the best-fit power law of Eq. \eqref{eq:znla}, with constraints given in Table \ref{tab:znla}. \emph{Right}: the solid lines now represent the best fit of Eq. \eqref{eq:enla}.}
  \label{fig:IA_vs_red}
  \end{center}
\end{figure*}

The redshift dependence of IA has been quite limitedly studied, with observations of red galaxies \citep{Joachimi11,Singh15,fortuna21b,Samuroff22}. So far, no clear evolution has been observed, with an overall scatter in the IA-redshift relation coming from the diversity of the samples used to derive this relationship. In addition, these samples are not representative of the whole galaxy population and therefore cannot directly probe the redshift evolution of IA. \cite{Yao2020} investigated the redshift evolution of IA amplitude with the Dark Energy
Camera Legacy Survey Data Release 3 (DECaLS DR3) shear
catalogue \citep{Phriskee2020}, with a self-calibration method which was designed to separate the IA signal from cosmic shear. They found that the IA amplitude of red galaxies increases with redshift, ruling out the constant IA amplitude with redshift at the $3.9\sigma$ level. In hydrodynamical simulations \citep{Tenneti2015b,Samuroff21,Zjupa2022,Delgado2023}, elliptical galaxies display a higher alignment amplitude $A_1$ at higher redshifts. However, as highlighted by \cite{Delgado2023}, these studies do not account for the redshift-dependent statistical evolution of galaxy properties, such as galaxy bias and luminosity. The trend is reversed when looking at the projected matter-IA correlation function, and $w_{\mathrm{m}+}$ increases with cosmic time. This agrees with the results of \cite{Bhowmick20}, where the authors showed that the evolution of the ellipticity–direction correlation amplitude of galaxies increases by a factor of four between redshift $z=3$ and $z = 0.6$ on linear scales, implying that IA amplitude increases with cosmic time. Additionally, \cite{Bate2020} investigated the IA evolution of the most massive elliptical progenitors at $z=0$ in Horizon-AGN. The authors found that the alignment increases from $z=3$ until $z=0.5$, below which the alignments of ellipticals stay constant. 

The redshift dependence can also be constrained indirectly in cosmic shear or 3$\times$2pt analyses across multiple photometric bins, by adopting the $z$-NLA model of Eq. \eqref{eq:znla}.
\cite{Secco21} found for the DES survey $\eta_1 = 1.66^{+3.26}_{-1.05}$, showing no strong evidence of evolution with redshift. The recent 3$\times$2pt analysis of the combined KiDS-DES data set \cite[][hereafter \DESKIDS]{2023DESKIDS} yields $\eta_1 = 2.05^{+2.95}_{-0.83}$
with a DES pipeline and $\eta_1 = 2.21^{+2.59}_{-0.96}$ for a KiDS one. These results seem to suggest that the observed IA amplitude increases with redshift. A plausible explanation is galaxy Malmquist bias: in magnitude-limited surveys, higher-redshift samples are biased towards intrinsically brighter galaxies. Because IA amplitude correlates with luminosity, this selection effect boosts the apparent IA signal at higher redshift. However, we have to be cautious as the redshift evolution will be correlated with the faint magnitude cut and will therefore vary as a function of the survey properties. 

The amplitude of the redshift power law has also been constrained in \DESKIDS, but varies by almost $2 \,\sigma$ between KiDS and DES with $A_z = -0.03^{+0.58}_{-0.29}$ for DES versus $A_z = 1.04^{+0.54}_{-0.52}$ for KiDS. As pointed out in \cite{fortuna21a}, the amplitude is mainly sensitive to the signal in the low redshift tomographic bins. Additionally, small differences in the estimation of the model (nonlinear matter power spectrum, baryonic feedback, NLA or TATT) can cause a large shift in the power-law amplitude. As already emphasised, different shape measurement methods will also induce shifts in the IA amplitude. Additionally, \cite{Leonard2024} found that there exists a clear degeneracy between IA parameters and photo-$z$ nuisance parameters in 3$\times$2pt analyses (the mean shift $\Delta \bar{z}$ and scatter $\sigma_z$ in each tomographic bin). Therefore, it is natural that we would observe a distinct redshift evolution of galaxy alignment for different surveys.
\begin{figure}
  \includegraphics[width=0.98\columnwidth]{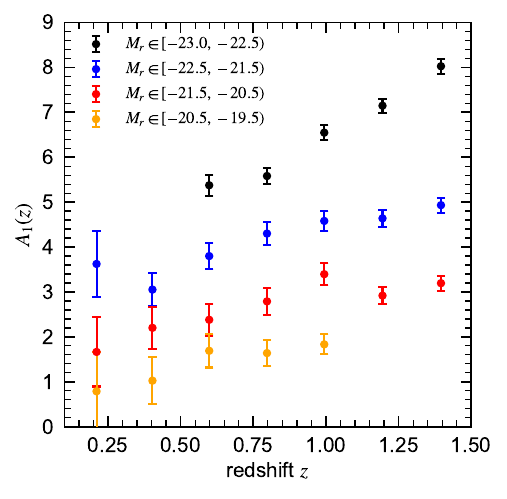}
  \caption{Redshift evolution of the IA amplitude for magnitude-selected samples. Each point corresponds to the median measurement per magnitude bin. We observe a small trend with redshift, as in other simulations (see main text).}
  \label{fig:A1_vs_red_mag}
\end{figure}

We present in Fig. \ref{fig:IA_vs_red} the evolution of IA amplitude as a function of redshift. As a reminder, above redshift $z = 1$, the model in \Kai has not been calibrated against observations, and therefore the observed signal above this redshift threshold corresponds to an extrapolation of the calibrated model in the range $0 < z < 1$. We fit the measurements with either Eq. \eqref{eq:znla} or Eq. \eqref{eq:enla} including or not the luminosity evolution. 
The corresponding constraints are given in Table \ref{tab:znla}. The constraints on both the amplitude and the power-law index of a simple redshift power law are similar compared to the results from \DESKIDS, with an IA amplitude close to KiDS constraints. We cannot however perform a proper comparison, as the different magnitude cut between KiDS/DES and \Euclid will provide distinct redshift evolutions. However, the IA implementation in Flagship seems to provide a realistic evolution with redshift and can be used to forecast the impact of IA on 3$\times$2pt analysis. 

We observe for the blue population a non-zero signal of alignment within the range $0.3 < z < 2.1$. To test the robustness of our results given our colour selection cut, we also selected an even bluer sample by selecting only galaxies with colours $u - r < 1.0$, given the observed colour distribution presented in Fig. \ref{fig:color_distribution}. This cut is enough at high-$z$ to further dissociate between the red and blue populations. Our results show that the redshift evolution of blue galaxy samples defined in this way is fully consistent with a cut at $u - r = 1.32$. However, within the range used for calibration, the amplitude of the blue alignment peak at around a value $A_1 \approx 1.2$, which is fully consistent with upper limits set by observational measurements, as illustrated in Fig. \ref{fig:LR_vs_A1_blue}.

Overall, we observe that the $z$-NLA cannot describe the measured alignment of the simulation. Instead, the e-NLA model provides a better fit to the observed alignment in the full redshift range explored in this analysis, $0.1 < z < 2.1$, as observed on the right panel of Fig. \ref{fig:IA_vs_red}. This is expected, as the magnitude cut in the visible will have an impact on the mean \textit{r}-band magnitude within each bin. Therefore, observed high-redshift galaxies will populate higher mass halos which align more strongly with the large-scale structure. This is why we observed an increase of $A_1$ with redshift in Fig. \ref{fig:IA_vs_red}. Our results also indicate that high-redshift galaxies in the simulation align less strongly (within the e-NLA that accounts for varying magnitude), with a  $\eta_1$  measured to be negative at the $3\sigma$ level for the total population. This agrees with the results of \citep{Bhowmick20,Bate2020}: galaxy alignment increases with cosmic time, i.e. decreases with redshift. However, the e-NLA model still yields large residuals with $\chi^{2}_{\rm r}>2$, due to its inability to model the $u-r$ misalignment dependence modelled in Flagship galaxies. Since Flagship has been calibrated up to redshift of $z=1$, it is also interesting to look at IA redshift evolution in the range $ z \in [0.1,1.1]$. We provide this analysis in Appendix \ref{sec:redshiftprior}. 

\begin{figure*}
    \begin{center}
  \includegraphics[width=1.8\columnwidth]{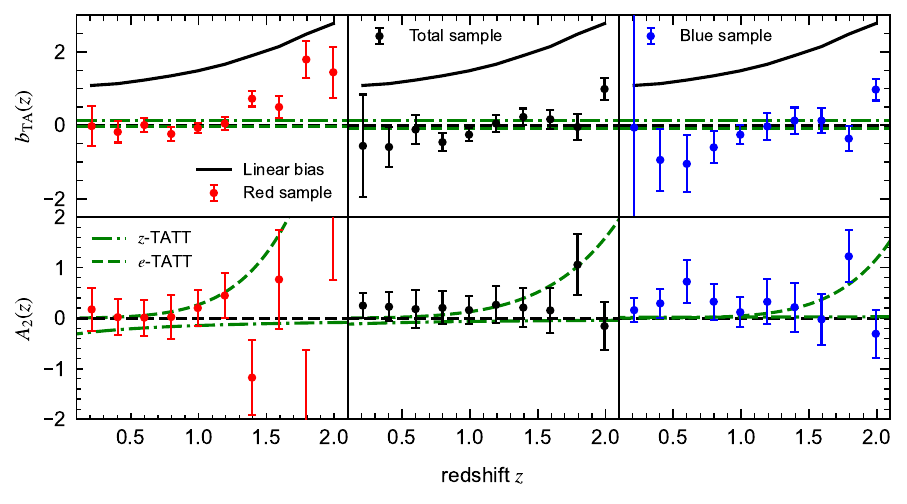}
  \caption{Evolution of $b_{\rm TA}$ (\emph{top}) and  $A_2$ (\emph{bottom})  as a function of redshift. These constraints were determined with the same samples presented in Fig. \ref{fig:IA_vs_red}. We also show the linear bias evolution of the total Flagship sample. If only the density weighting contributes to the $\delta s_{i j}$ term (see Sect. \ref{sec:TATT}.) then $b_{\rm TA}$ should be equal to $b_1$.}
  \label{fig:A2_vs_z}
  \end{center}
\end{figure*}

Alternatively, we can look at the evolution of IA for galaxy samples of similar magnitude. We performed this analysis by considering only the red population to avoid the mixing of galaxies with similar masses but different physical properties. This is illustrated in  Fig. \ref{fig:A1_vs_red_mag}, where each point corresponds to the median $A_1$ measured within thick magnitude bins. Overall, our results suggest a small trend in the redshift range $0.3 < z < 1.0$, with galaxies (with similar properties) having a higher alignment at high redshift. Above a redshift of $z=1$, the alignment of galaxies stays constant while increasing for the brightest galaxies. This is the consequence of two effects. In \cite{Kurita2021}, the authors investigated the evolution of dark matter halo alignments. They found an increase of alignment up to redshift $z=1$. Above this threshold, the alignment of low-mass halos keeps increasing while the alignment of high-mass halos stays constant. However, in Flagship, the negative value of $p_1$ for central indicates that galaxies are misaligned less strongly at higher redshift. As the alignment of massive halos stays constant above $z=1$, this translates into an increase in alignment we observe for the brightest population. 

Below a redshift of $z=1$, our results are consistent with earlier works \citep{Tenneti2015b,Yao2020,Samuroff21,Zjupa2022,Delgado2023}, which have shown that the IA signal of red galaxies increases with redshift. However, it is important to clarify that these findings are tightly linked to the methodology that compares galaxies with similar properties at different redshifts. As already emphasised, galaxies with a given luminosity will be more strongly biased at higher redshift, which makes the comparison not straightforward. When looking at progenitor like in \cite{Bhowmick20} and in \cite{Bate2020}, galaxy IA decreases with redshift, i.e. increases with cosmic time.

\begin{table}
\centering
\caption{Constraints on the redshift evolution of the NLA model with Eqs. \eqref{eq:znla} and \eqref{eq:enla}.}
\setlength{\tabcolsep}{2pt} 
\begin{tabular}{c|cccc}
\hline \hline
Model & Constraint 
& Total & Red & Blue \\
\hline
\noalign{\vskip 1pt}
\noalign{\vskip 1pt}
\multirow{3}{*}{$z$-NLA} & \text{ $A_{1z}$} & $1.12 \pm 0.02 $  & $1.85 \pm  0.02$ & $0.72 \pm  0.02$ \\
 & $\eta_1$ & $2.06 \pm 0.03 $  & $2.46 \pm 0.04$ & $2.64 \pm 0.05 $  \\
 & $\chi^2_{\rm r}$ & 33.8 & 12.4  & 9.1 \\
 \hline
\multirow{4}{*}{e-NLA}  & \text{ $A_{1z}$} & $0.97 \pm 0.03 $ & $1.84 \pm 0.03 $  & $0.64 \pm 0.02 $   \\
 & $\eta_1$ & $-1.09 \pm 0.23 $ & $-0.80 \pm 0.45$ & $-0.87 \pm 0.50 $ \\
 & $\beta_1$ & $1.80 \pm 0.14 $ & $1.69 \pm 0.24$ &  $1.81 \pm 0.30 $   \\
  & $\chi^2_{\rm r}$ & 2.49 & 3.65 & 2.01 \\
 \hline
\end{tabular}
\tablefoot{We observe high $\chi_{\rm r}^2$ values for the $z-$NLA model $(N_{\rm dof}=8)$, as this model fails to fit the low-redshift bins accurately (see Fig. \ref{fig:IA_vs_red}). Moreover, the high-redshift IA measurements are extremely precise, meaning even small discrepancies between model predictions and observations significantly impact the goodness of fit. For the e-NLA model $(N_{\rm dof}=7)$, we assume as pivot luminosity the interpolated luminosity determined at the pivot redshift $z=0.62$ for each sample.}
 \label{tab:znla}
\end{table}
\begin{table}
\centering
\caption{Constraints on the redshift evolution of the TATT model.}
\setlength{\tabcolsep}{2pt} 
\begin{tabular}{c|cccc}
\hline \hline
Model & Constraint 
& Total & Red & Blue \\
\hline
\noalign{\vskip 1pt}
\noalign{\vskip 1pt}
\multirow{6}{*}{$z$-TATT} & \text{ $A_{1z}$} & $1.21 \pm 0.04 $  & $2.00 \pm  0.05$ & $0.75 \pm  0.03$ \\
 & $\eta_1$ & $1.74 \pm 0.05 $  & $2.05 \pm 0.06$ & $2.45 \pm 0.08 $  \\
  &  \text{$A_{2z}$} & $-0.08^{+0.23}_{-0.09}$  & $-0.19^{+0.37}_{-0.14}$ & $0.02^{+0.10}_{-0.13}$  \\
 & $\eta_2$ & $-1.02^{+5.81}_{-1.81}$  & $-1.28^{+6.20}_{-1.41}$ & $0.52^{+2.17}_{-4.40}$  \\
 &  \text{$b_{\rm TA}$} & $0.13^{+0.07}_{-0.22}$  & $0.14^{+0.08}_{-0.16}$ & $0.13^{+0.09}_{-0.16}$   \\

 & $\chi^2_{\rm r}$ & 6.09 & 4.05 & 3.44 \\
 \hline
\multirow{7}{*}{$e$-TATT} & \text{ $A_{1z}$} & $1.04 \pm 0.04 $  & $1.95 \pm  0.05$ & $0.67 \pm  0.03$ \\
 & $\eta_1$ & $-1.60^{+0.38}_{-0.37}$  & $-1.72^{+0.69}_{-0.65}$ & $-1.35^{+0.83}_{-0.83}$  \\
  & $\beta_1$ & $1.88^{+0.22}_{-0.21}$  & $1.90^{+0.33}_{-0.35}$ & $1.92^{+0.32}_{-0.33}$  \\
  &  \text{$A_{2z}$} & $0.06^{+0.06}_{-0.03}$  & $0.10^{+0.06}_{-0.04}$ & $0.05^{+0.04}_{-0.02}$  \\
 & $\eta_2$ & $5.57^{+1.17}_{-1.04}$  & $6.26^{+0.91}_{-0.85}$ & $5.63^{+1.40}_{-1.38}$  \\
 &  \text{ $b_{\rm TA}$} & $-0.08^{+0.07}_{-0.07}$  & $-0.03^{+0.06}_{-0.06}$ & $-0.09^{+0.09}_{-0.10}$   \\

 & $\chi^2_{\rm r}$ & 1.75 & 2.75 & 1.24 \\
 \hline
\end{tabular}
\tablefoot{The $z$-TATT model $(N_{\rm dof}=25)$ corresponds to the most widely used model (see main text) where $A_1$ and $A_2$ are modelled as redshift power laws while $b_{\rm TA}$. The $e$-TATT model $(N_{\rm dof}=24)$ also included a luminosity dependence on $A_1$, with pivot luminosity defined in the same way as in the e-NLA model. The values quoted here correspond to the $16\%$,
    $50\%$, and $84 \%$ percentile of the distribution determined with the \texttt{nautilus} sampler.}
 \label{tab:zTATT}
\end{table}

\subsubsection{TATT redshift evolution}
\label{sec:A2_vs_z_sec}

In 3$\times$2pt analyses where the TATT model has been used to model IA, the redshift dependence of $A_2$ is modelled as a power law across the tomographic bins as in Eq. \eqref{eq:A2_z}, while $b_{\rm TA}$ is assumed constant across redshift: the redshift evolution of $A_{1\delta}$ is modulated by the evolution of $A_1(z)$. We assumed this redshift evolution of TATT, which has five free parameters $(A_1,\eta_1, A_2, \eta_2, b_{\rm TA})$, and to which we refer as $z$-TATT. We present our result in Table \ref{tab:zTATT} and in Fig \ref{fig:A2_vs_z}, determined by minimising the following $\chi^2$ 
\begin{equation}
\chi^2 = \sum_{z_k}\sum_{ij}(A_{i,\text{obs}}-A_{i,\text{mod}})\hat{\Psi}_{z,i j} (A_{j,\text{obs}}-A_{j,\text{mod}}) \, , 
\end{equation}
where the $A_i$ corresponds to $A_1, A_2,b_{\rm TA}$. $\hat{\Psi}_{z,i j}$ is the inverse of the covariance matrix of $(A_1,A_2,b_{\rm TA})$ we derived at redshift $z_k$ from fitting the IA correlation functions. We used this $\chi^2$ definition to incorporate intrinsic covariances between the IA parameters when fitting the redshift evolution. 

The $z-$TATT model does not fit the simulation well ($\chi^2_{\rm r}>3.4$), as $A_1$ cannot be modelled by a simple power law within NLA. Although TATT introduces higher order terms, these do not significantly change the $A_1$ amplitude. Instead, inaccuracies in modelling the redshift evolution of $A_1$ propagate into shifts in the nonlinear IA terms, as shown in the bottom panels of Fig. \ref{fig:A2_vs_z}. \referee{Additionally, mismodelling $A_1$ can lead to further unconstrained TATT parameters with bimodal contours for $A_2$ and $b_{\rm TA}$ (see Fig. \ref{fig:constraintappendix})}. To account for this, we also provide constraints for the $e$-TATT model, which has six parameters $(A_1,\eta_1,\beta_1,A_2,\eta_2,b_{\rm TA})$ and includes an explicit luminosity dependence. This choice is motivated by the findings of Sects. \ref{sec:nonlinearIA_luminosity} and   \ref{sec:redshiftsectionNLA}: the nonlinear IA amplitudes show no luminosity dependence, while the evolution of $A_1$ is better captured by a double power law in redshift and luminosity.

\referee{Even with the $e$-TATT model, $A_2$ remains mismodelled for the total and blue samples. This issue stems from the assumed redshift evolution of $b_{\rm TA}$, with the constant hypothesis rejected beyond $z>1.1$. Because $A_2$ and $b_{\rm TA}$ are strongly anti-correlated, any prior imposed on the redshift evolution of $b_{\rm TA}$ will directly bias the inferred constraints on $A_2$ and its redshift evolution.} It is therefore crucial in $3\times2$pt analyses to accurately model the redshift evolution of the IA parameters, to avoid misinterpreting IA evolution as biases in cosmological parameter constraints. 

\referee{We note that the near-zero constraints on  $b_{\rm TA}$ do not simply reflect its degeneracy with $A_2$. As visible in the joint ($A_2$,  $b_{\rm TA}$) posterior of Fig. \ref{fig:constraintappendix}, the region around  $b_{\rm TA} \approx b_1$ is clearly disfavoured by the data even before marginalisation, suggesting that contributions beyond pure density weighting  effectively shift  $b_{\rm TA}$ away from the $b_1$ expectation.} We provide in Appendix \ref{sec:redshiftprior} an analysis up to redshift $z=1.1$, where a constant evolution $b_{\rm TA}$ with redshift is valid.

\section{Conclusions}\label{sec:Conclusions}

In this paper, we provide numerical constraints on the evolution of IA with galaxy redshift, colour, and luminosity on cosmological scales. This analysis was carried out with the \Euclid $N$-body simulated light-cone Flagship \citep{EuclidSkyFlagship}, followed by the implementation of an IA model according to the process described in \Kai. This model was calibrated against SDSS observations at low redshift and against the Horizon-AGN hydrodynamical simulation at redshift $z=1$. 

We analysed the Flagship simulation with the most widely used IA models in the literature: NLA and TATT. We have shown that both NLA and TATT can correctly describe the IA signal in the simulation over a broad range of redshifts down to small scales, $r_{\mathrm{min}} = 6$--$7 \, \, h^{-1}$Mpc. Additionally, we determined IA constraints on LOWZ-like samples used in the calibration procedure and found good agreement with observational constraints determined in \cite{Singh15}.  
We forecast the evolution of galaxy IA with luminosity over the range $0.1 < z < 0.9$ and we found that galaxy linear alignment, $A_1$, is tightly correlated with its luminosity, as demonstrated in previous studies. This relationship can be modelled within Flagship as a single power law from which we derived amplitudes and indices for the red and blue populations, defined by a redshift-independent cut in the $u-r$ band. We then compared our results against observational measurements not used in our calibration procedure.

For the red galaxy population, our constraints in the low redshift bins $0.1 < z < 0.5$ slightly underestimate the majority of observational data, with an overall amplitude of linear alignment $20\%$ and $38 \%$ lower than measured in \cite{Singh15} and \cite{Joachimi11}. At higher redshifts, however, we found good agreement in the $A_1$-luminosity relationship between observations and Flagship. One major source of uncertainty in our results is the satellite fraction. As satellite galaxies only align with respect to the halo centre at small radii, their random orientations at larger scales modulate the linear amplitude of alignment. This effect is stronger for faint galaxies and it partly explains the scatter in the $A_1$-luminosity relation we observe in the data. Future models that aim to implement realistic IA signals within $N$-body simulations should include the satellite fraction when calibrating against observations and/or hydrodynamical simulations. The observed IA signal also depends on the method used to measure galaxy shapes \citep{Singh16, Georgiou19, Macmahon2024, Leonard2024}, which can shift the constraints on the IA amplitude by up to 1$\sigma$ \citep{FABIBI2024}. Therefore, to properly forecast the impact of IA in future analyses, we need to update our calibration procedure with upcoming IA observations measured with \Euclid shapes. 

For the blue population, we detected a non-zero signal of linear alignment over the redshift range $0.1 < z < 0.9$, consistent with the upper limit set by observational measurements. We checked that our results are not sensitive to the way we defined and split galaxies in colour type. Our results on the $A_1$-luminosity relation for the blue population are similar to the constraints determined in \cite{Samuroff21}. Although the amplitude of galaxy alignment for blue galaxies has always been consistent with zero in the observations \citep{mandelbaum11,Johnston19}, a linear alignment for blue galaxies was observed in hydrodynamical simulations \citep{Chisari15,Samuroff19,Zjupa2022,Delgado2023}. In particular, \cite{Chisari15} showed that blue galaxies align tangentially with elliptical galaxies, a signal being suppressed in projection. Future \Euclid observations will enable us to further understand and constrain the alignment signal of blue galaxies. 

We then investigated how galaxy linear alignment evolves with redshift over the range $0.1 < z < 2.1$. As for the luminosity dependence, the redshift evolution was modelled as well as a power law. Since magnitude-limited surveys observe brighter galaxies at high redshift, this leads to an increase in alignment with redshift. We found good agreement between our derived constraints and the redshift evolution of linear alignment determined for KiDS in \DESKIDS. However, we cannot directly compare these results as different surveys provide a different redshift evolution scenario depending on the magnitude cut of source galaxies. Our results additionally show that a simple power law fails to reproduce the Flagship's IA redshift evolution. The agreement improves when an additional luminosity power-law is included, capturing variations in the observed \textit{r}-band magnitude introduced by the \Euclid selection cut. Provided that the mean absolute magnitude-over-halo mass can be measured accurately in each tomographic bin, 3$\times$2pt analyses should be able to account for the IA redshift evolution using Eq.~\eqref{eq:enla}. An alternative approach would be to implicitly model the redshift dependence through the evolution in luminosity/halo mass within the tomographic bins as was done, for example, in the KiDS legacy analysis with the NLA-M model \citep{NLA-M}.

Overall, using the TATT model instead of the NLA model does not significantly change the redshift evolution of the linear alignment parameter, $A_1$. Using the TATT model instead of the NLA model weakly affects the $A_1$-luminosity relationship in the low-redshift bins $0.1 < z < 0.5$ for the red population. This is the consequence of a positive torquing amplitude $A_2$ measured at low redshift.
Alternatively, we investigated the evolution of red galaxy alignment with redshift for luminosity-selected samples. We found a bimodal evolution of alignment. In the range $0 < z < 1$, IA of galaxies increases with redshift in agreement with previous studies on hydrodynamical simulations \citep{Tenneti2015b,Samuroff21,Delgado2023}. One reason for this behaviour is that galaxies with a similar mass and luminosity are more strongly biased at high redshift and, thus, they tend to populate higher mass halos that are more strongly aligned with the large-scale structure. At higher redshift, $z > 1$, the alignment remains constant, although it rises for the brightest objects. 

We then looked at the evolution of nonlinear IA parameters of the TATT models, the tidal torquing amplitude, $A_2$, and the density weighting term, $b_{\rm TA}$, with luminosity and redshift. Given our statistical precision, we did not find any significant evolution for $b_{\rm TA}$ and $A_2$ with luminosity. 
In \cite{DES21} and \DESKIDS, $b_{\rm TA}$ is assumed to be positive and independent of redshift, while the redshift evolution of $A_2$ is modelled as a power law as for $A_1$. We showed that each assumptions fail to capture the behaviour observed in the Flagship simulation. In particular, $A_1$ cannot be modelled as a single power-law. In addition, $b_{\rm TA}$ exhibits a clear redshift dependence, with the constant hypothesis rejected beyond $z>1.1$. The strong degeneracy with $A_2$ means that mismodelling $b_{\rm TA}$ directly biases the inferred redshift evolution of $A_2$. Finally, we also performed an analysis in the restricted redshift range $0.1 < z < 1.1$, where $b_{\rm TA}$ is constant. We measured in Flagship a negative value for $b_{\rm TA}$, at the 2.8$\sigma$ level. Although, strictly speaking, the density weighting contribution should be positive, other sources of contributions, not modelled in the TATT framework, can contribute to the alignment mechanism of galaxies \citep{Samuroff21,Maion2023}. In addition, there is a non-negligible degeneracy between $A_2$ and $b_{\rm TA}$ that can yield a negative value of $b_{\rm TA}$ in observations \citep{Samuroff22}. Therefore, it cannot be strictly ruled out that $b_{\rm TA}$ is negative. 
Regardless of the torquing alignment, we found $A_2$ to be consistent with zero at the 1$\sigma$ level across the whole redshift range, with a preference for slight positive values and a constant evolution with redshift. 

To summarise our results, we found that analytical IA models can accurately fit the IA signal in the Flagship simulation, which includes a realistic luminosity and redshift evolution of alignment. We conclude that the simulation can be used to both investigate the precision of theoretical IA models and forecast the impact of IA
on cosmological analysis.
In particular, our analysis reveals that TATT and NLA models yield similar goodness of fit and redshift evolution. 
However, we found that a mismodelling of the redshift evolution of IA, as observed with the $z$-NLA and $z$-TATT models, can act as an additional source of systematic bias. Moreover, in this work, we did not conduct a full cosmological 3$\times$2pt analysis, which could possibly highlight further differences between those models. Indeed, such analysis typically reveals a clear degeneracy between cosmological parameters and IA nuisance terms that require a judicious IA model to avoid introducing bias, while at the same time bringing the tightest possible cosmological constraints.  This type of analysis will be the topic of a forthcoming work.


\begin{acknowledgements}
RP's work is partially funded by DIM-ACAV. Author RP acknowledges financial support from the Centre national d’études spatiales (CNES),
France (ROR: https://ror.org/04h1h0y33) within the framework of the Euclid mission. This work has received support under the program “Investissement d’Avenir” launched by the French Government and implemented by ANR, with the reference « ANR‐18‐IdEx‐0001 » as part of its program « Emergence » (projet COMMANDO). BJ acknowledges support by the ERC-selected UKRI Frontier Research Grant EP/Y03015X/1. SC acknowledges support from Fondation Merac and the Agence Nationale de la Recherche (ANR-18-CE31-0009 SPHERES).
\AckEC
\AckCosmoHub
In addition to the packages already quoted in the main text, we also acknowledge the use of \texttt{numpy} \citep{numpy} and \texttt{scipy} \citep{2020SciPy-NMeth}. Plots were made with \texttt{matplotlib}  \citep{matplotlib} while corner plots and density plots were made with \texttt{pygtc}, 
 \citep{pygtc}. 
\end{acknowledgements}

\bibliography{references}

@ARTICLE{2024OJAp....7E..14L,
       author = {{Lamman}, Claire and {Tsaprazi}, Eleni and {Shi}, Jingjing and {{\v{S}}ar{\v{c}}evi{\'c}}, Nikolina Niko and {Pyne}, Susan and {Legnani}, Elisa and {Ferreira}, Tassia},
        title = "{The IA Guide: A Breakdown of Intrinsic Alignment Formalisms}",
      journal = {The Open Journal of Astrophysics},
         year = 2024,
        month = feb,
       volume = {7},
          eid = {14},
        pages = {14},
          doi = {10.21105/astro.2309.08605},
archivePrefix = {arXiv},
       eprint = {2309.08605},
 primaryClass = {astro-ph.CO},
       adsurl = {https://ui.adsabs.harvard.edu/abs/2024OJAp....7E..14L}
}

@ARTICLE{2020MNRAS.491.5330S,
       author = {{Samuroff}, S. and {Mandelbaum}, R. and {Di Matteo}, T.},
        title = "{Testing the impact of satellite anisotropy on large- and small-scale intrinsic alignments using hydrodynamical simulations}",
      journal = {\mnras},
         year = 2020,
        month = feb,
       volume = {491},
       number = {4},
        pages = {5330-5350},
          doi = {10.1093/mnras/stz3114},
archivePrefix = {arXiv},
       eprint = {1901.09925},
 primaryClass = {astro-ph.CO},
       adsurl = {https://ui.adsabs.harvard.edu/abs/2020MNRAS.491.5330S}
}

@ARTICLE{2017arXiv171207818W,
       author = {{Welker}, Charlotte and {Power}, Chris and {Pichon}, Christophe and {Dubois}, Yohan and {Devriendt}, Julien and {Codis}, Sandrine},
        title = "{Caught in the rhythm II: Competitive alignments of satellites with their inner halo and central galaxy}",
      journal = {arXiv},
         year = 2017,
        month = dec,
          eid = {arXiv:1712.07818},
        pages = {arXiv:1712.07818},
          doi = {10.48550/arXiv.1712.07818},
archivePrefix = {arXiv},
       eprint = {1712.07818},
 primaryClass = {astro-ph.GA},
       adsurl = {https://ui.adsabs.harvard.edu/abs/2017arXiv171207818W}
}

@ARTICLE{2009IJMPD..18..173S,
       author = {{Sch{\"a}fer}, Bj{\"o}rn Malte},
        title = "{Galactic Angular Momenta and Angular Momentum Correlations in the Cosmological Large-Scale Structure}",
      journal = {International Journal of Modern Physics D},
         year = 2009,
        month = jan,
       volume = {18},
       number = {2},
        pages = {173-222},
          doi = {10.1142/S0218271809014388},
archivePrefix = {arXiv},
       eprint = {0808.0203},
 primaryClass = {astro-ph},
       adsurl = {https://ui.adsabs.harvard.edu/abs/2009IJMPD..18..173S}
}

@ARTICLE{TreeCorr,
       author = {{Jarvis}, M. and {Bernstein}, G. and {Jain}, B.},
        title = "{The skewness of the aperture mass statistic}",
      journal = {\mnras},
         year = 2004,
        month = jul,
       volume = {352},
       number = {1},
        pages = {338-352},
          doi = {10.1111/j.1365-2966.2004.07926.x},
archivePrefix = {arXiv},
       eprint = {astro-ph/0307393},
 primaryClass = {astro-ph},
       adsurl = {https://ui.adsabs.harvard.edu/abs/2004MNRAS.352..338J}
}

@ARTICLE{DES21,
       author = {{Abbott}, T.~M.~C. and {Aguena}, M. and {Alarcon}, A. and {Allam}, S. and {Alves}, O. and {Amon}, A. and {Andrade-Oliveira}, F. and {Annis}, J. and {Avila}, S. and {Bacon}, D. and {Baxter}, E. and {Bechtol}, K. and {Becker}, M.~R. and {Bernstein}, G.~M. and {Bhargava}, S. and {Birrer}, S. and {Blazek}, J. and {Brandao-Souza}, A. and {Bridle}, S.~L. and {Brooks}, D. and {Buckley-Geer}, E. and {Burke}, D.~L. and {Camacho}, H. and {Campos}, A. and {Carnero Rosell}, A. and {Carrasco Kind}, M. and {Carretero}, J. and {Castander}, F.~J. and {Cawthon}, R. and {Chang}, C. and {Chen}, A. and {Chen}, R. and {Choi}, A. and {Conselice}, C. and {Cordero}, J. and {Costanzi}, M. and {Crocce}, M. and {da Costa}, L.~N. and {da Silva Pereira}, M.~E. and {Davis}, C. and {Davis}, T.~M. and {De Vicente}, J. and {DeRose}, J. and {Desai}, S. and {Di Valentino}, E. and {Diehl}, H.~T. and {Dietrich}, J.~P. and {Dodelson}, S. and {Doel}, P. and {Doux}, C. and {Drlica-Wagner}, A. and {Eckert}, K. and {Eifler}, T.~F. and {Elsner}, F. and {Elvin-Poole}, J. and {Everett}, S. and {Evrard}, A.~E. and {Fang}, X. and {Farahi}, A. and {Fernandez}, E. and {Ferrero}, I. and {Fert{\'e}}, A. and {Fosalba}, P. and {Friedrich}, O. and {Frieman}, J. and {Garc{\'\i}a-Bellido}, J. and {Gatti}, M. and {Gaztanaga}, E. and {Gerdes}, D.~W. and {Giannantonio}, T. and {Giannini}, G. and {Gruen}, D. and {Gruendl}, R.~A. and {Gschwend}, J. and {Gutierrez}, G. and {Harrison}, I. and {Hartley}, W.~G. and {Herner}, K. and {Hinton}, S.~R. and {Hollowood}, D.~L. and {Honscheid}, K. and {Hoyle}, B. and {Huff}, E.~M. and {Huterer}, D. and {Jain}, B. and {James}, D.~J. and {Jarvis}, M. and {Jeffrey}, N. and {Jeltema}, T. and {Kovacs}, A. and {Krause}, E. and {Kron}, R. and {Kuehn}, K. and {Kuropatkin}, N. and {Lahav}, O. and {Leget}, P. -F. and {Lemos}, P. and {Liddle}, A.~R. and {Lidman}, C. and {Lima}, M. and {Lin}, H. and {MacCrann}, N. and {Maia}, M.~A.~G. and {Marshall}, J.~L. and {Martini}, P. and {McCullough}, J. and {Melchior}, P. and {Mena-Fern{\'a}ndez}, J. and {Menanteau}, F. and {Miquel}, R. and {Mohr}, J.~J. and {Morgan}, R. and {Muir}, J. and {Myles}, J. and {Nadathur}, S. and {Navarro-Alsina}, A. and {Nichol}, R.~C. and {Ogando}, R.~L.~C. and {Omori}, Y. and {Palmese}, A. and {Pandey}, S. and {Park}, Y. and {Paz-Chinch{\'o}n}, F. and {Petravick}, D. and {Pieres}, A. and {Plazas Malag{\'o}n}, A.~A. and {Porredon}, A. and {Prat}, J. and {Raveri}, M. and {Rodriguez-Monroy}, M. and {Rollins}, R.~P. and {Romer}, A.~K. and {Roodman}, A. and {Rosenfeld}, R. and {Ross}, A.~J. and {Rykoff}, E.~S. and {Samuroff}, S. and {S{\'a}nchez}, C. and {Sanchez}, E. and {Sanchez}, J. and {Sanchez Cid}, D. and {Scarpine}, V. and {Schubnell}, M. and {Scolnic}, D. and {Secco}, L.~F. and {Serrano}, S. and {Sevilla-Noarbe}, I. and {Sheldon}, E. and {Shin}, T. and {Smith}, M. and {Soares-Santos}, M. and {Suchyta}, E. and {Swanson}, M.~E.~C. and {Tabbutt}, M. and {Tarle}, G. and {Thomas}, D. and {To}, C. and {Troja}, A. and {Troxel}, M.~A. and {Tucker}, D.~L. and {Tutusaus}, I. and {Varga}, T.~N. and {Walker}, A.~R. and {Weaverdyck}, N. and {Wechsler}, R. and {Weller}, J. and {Yanny}, B. and {Yin}, B. and {Zhang}, Y. and {Zuntz}, J. and {DES Collaboration}},
        title = "{Dark Energy Survey Year 3 results: Cosmological constraints from galaxy clustering and weak lensing}",
      journal = {\prd},
         year = 2022,
        month = jan,
       volume = {105},
       number = {2},
          eid = {023520},
        pages = {023520},
          doi = {10.1103/PhysRevD.105.023520},
archivePrefix = {arXiv},
       eprint = {2105.13549},
 primaryClass = {astro-ph.CO},
       adsurl = {https://ui.adsabs.harvard.edu/abs/2022PhRvD.105b3520A}
}

@ARTICLE{Amon21,
       author = {{Amon}, A. and {Gruen}, D. and {Troxel}, M.~A. and {MacCrann}, N. and {Dodelson}, S. and {Choi}, A. and {Doux}, C. and {Secco}, L.~F. and {Samuroff}, S. and {Krause}, E. and {Cordero}, J. and {Myles}, J. and {DeRose}, J. and {Wechsler}, R.~H. and {Gatti}, M. and {Navarro-Alsina}, A. and {Bernstein}, G.~M. and {Jain}, B. and {Blazek}, J. and {Alarcon}, A. and {Fert{\'e}}, A. and {Lemos}, P. and {Raveri}, M. and {Campos}, A. and {Prat}, J. and {S{\'a}nchez}, C. and {Jarvis}, M. and {Alves}, O. and {Andrade-Oliveira}, F. and {Baxter}, E. and {Bechtol}, K. and {Becker}, M.~R. and {Bridle}, S.~L. and {Camacho}, H. and {Carnero Rosell}, A. and {Carrasco Kind}, M. and {Cawthon}, R. and {Chang}, C. and {Chen}, R. and {Chintalapati}, P. and {Crocce}, M. and {Davis}, C. and {Diehl}, H.~T. and {Drlica-Wagner}, A. and {Eckert}, K. and {Eifler}, T.~F. and {Elvin-Poole}, J. and {Everett}, S. and {Fang}, X. and {Fosalba}, P. and {Friedrich}, O. and {Gaztanaga}, E. and {Giannini}, G. and {Gruendl}, R.~A. and {Harrison}, I. and {Hartley}, W.~G. and {Herner}, K. and {Huang}, H. and {Huff}, E.~M. and {Huterer}, D. and {Kuropatkin}, N. and {Leget}, P. and {Liddle}, A.~R. and {McCullough}, J. and {Muir}, J. and {Pandey}, S. and {Park}, Y. and {Porredon}, A. and {Refregier}, A. and {Rollins}, R.~P. and {Roodman}, A. and {Rosenfeld}, R. and {Ross}, A.~J. and {Rykoff}, E.~S. and {Sanchez}, J. and {Sevilla-Noarbe}, I. and {Sheldon}, E. and {Shin}, T. and {Troja}, A. and {Tutusaus}, I. and {Tutusaus}, I. and {Varga}, T.~N. and {Weaverdyck}, N. and {Yanny}, B. and {Yin}, B. and {Zhang}, Y. and {Zuntz}, J. and {Aguena}, M. and {Allam}, S. and {Annis}, J. and {Bacon}, D. and {Bertin}, E. and {Bhargava}, S. and {Brooks}, D. and {Buckley-Geer}, E. and {Burke}, D.~L. and {Carretero}, J. and {Costanzi}, M. and {da Costa}, L.~N. and {Pereira}, M.~E.~S. and {De Vicente}, J. and {Desai}, S. and {Dietrich}, J.~P. and {Doel}, P. and {Ferrero}, I. and {Flaugher}, B. and {Frieman}, J. and {Garc{\'\i}a-Bellido}, J. and {Gaztanaga}, E. and {Gerdes}, D.~W. and {Giannantonio}, T. and {Gschwend}, J. and {Gutierrez}, G. and {Hinton}, S.~R. and {Hollowood}, D.~L. and {Honscheid}, K. and {Hoyle}, B. and {James}, D.~J. and {Kron}, R. and {Kuehn}, K. and {Lahav}, O. and {Lima}, M. and {Lin}, H. and {Maia}, M.~A.~G. and {Marshall}, J.~L. and {Martini}, P. and {Melchior}, P. and {Menanteau}, F. and {Miquel}, R. and {Mohr}, J.~J. and {Morgan}, R. and {Ogando}, R.~L.~C. and {Palmese}, A. and {Paz-Chinch{\'o}n}, F. and {Petravick}, D. and {Pieres}, A. and {Romer}, A.~K. and {Sanchez}, E. and {Scarpine}, V. and {Schubnell}, M. and {Serrano}, S. and {Smith}, M. and {Soares-Santos}, M. and {Tarle}, G. and {Thomas}, D. and {To}, C. and {Weller}, J. and {DES Collaboration}},
        title = "{Dark Energy Survey Year 3 results: Cosmology from cosmic shear and robustness to data calibration}",
      journal = {\prd},
         year = 2022,
        month = jan,
       volume = {105},
       number = {2},
          eid = {023514},
        pages = {023514},
          doi = {10.1103/PhysRevD.105.023514},
archivePrefix = {arXiv},
       eprint = {2105.13543},
 primaryClass = {astro-ph.CO},
       adsurl = {https://ui.adsabs.harvard.edu/abs/2022PhRvD.105b3514A}
}

@ARTICLE{Baldauf12,
       author = {{Baldauf}, Tobias and {Seljak}, Uro{\v{s}} and {Desjacques}, Vincent and {McDonald}, Patrick},
        title = "{Evidence for quadratic tidal tensor bias from the halo bispectrum}",
      journal = {\prd},
         year = 2012,
        month = oct,
       volume = {86},
       number = {8},
          eid = {083540},
        pages = {083540},
          doi = {10.1103/PhysRevD.86.083540},
archivePrefix = {arXiv},
       eprint = {1201.4827},
 primaryClass = {astro-ph.CO},
       adsurl = {https://ui.adsabs.harvard.edu/abs/2012PhRvD..86h3540B}
}

@ARTICLE{Vlah2019,
       author = {{Vlah}, Zvonimir and {Chisari}, Nora Elisa and {Schmidt}, Fabian},
        title = "{An EFT description of galaxy intrinsic alignments}",
      journal = {\jcap},
         year = 2020,
        month = jan,
       volume = {1},
       number = {1},
          eid = {025},
        pages = {025},
          doi = {10.1088/1475-7516/2020/01/025},
archivePrefix = {arXiv},
       eprint = {1910.08085},
 primaryClass = {astro-ph.CO},
       adsurl = {https://ui.adsabs.harvard.edu/abs/2020JCAP...01..025V}
}

@ARTICLE{Vlah2020,
       author = {{Vlah}, Zvonimir and {Chisari}, Nora Elisa and {Schmidt}, Fabian},
        title = "{Galaxy shape statistics in the effective field theory}",
      journal = {\jcap},
         year = 2021,
        month = may,
       volume = {5},
       number = {5},
          eid = {061},
        pages = {061},
          doi = {10.1088/1475-7516/2021/05/061},
archivePrefix = {arXiv},
       eprint = {2012.04114},
 primaryClass = {astro-ph.CO},
       adsurl = {https://ui.adsabs.harvard.edu/abs/2021JCAP...05..061V}
}

@ARTICLE{Bate2020,
       author = {{Bate}, James and {Chisari}, Nora Elisa and {Codis}, Sandrine and {Martin}, Garreth and {Dubois}, Yohan and {Devriendt}, Julien and {Pichon}, Christophe and {Slyz}, Adrianne},
        title = "{When galaxies align: intrinsic alignments of the progenitors of elliptical galaxies in the Horizon-AGN simulation}",
      journal = {\mnras},
         year = 2020,
        month = jan,
       volume = {491},
       number = {3},
        pages = {4057-4068},
          doi = {10.1093/mnras/stz3166},
archivePrefix = {arXiv},
       eprint = {1911.04213},
 primaryClass = {astro-ph.CO},
       adsurl = {https://ui.adsabs.harvard.edu/abs/2020MNRAS.491.4057B}
}

@ARTICLE{Blazek15,
       author = {{Blazek}, Jonathan and {Vlah}, Zvonimir and {Seljak}, Uro{\v{s}}},
        title = "{Tidal alignment of galaxies}",
      journal = {JCAP},
         year = 2015,
        month = aug,
       volume = {8},
       number = {8},
          eid = {015},
        pages = {015},
          doi = {10.1088/1475-7516/2015/08/015},
archivePrefix = {arXiv},
       eprint = {1504.02510},
 primaryClass = {astro-ph.CO},
       adsurl = {https://ui.adsabs.harvard.edu/abs/2015JCAP...08..015B}
}

@ARTICLE{Blazek19,
       author = {{Blazek}, Jonathan A. and {MacCrann}, Niall and {Troxel}, M.~A. and {Fang}, Xiao},
        title = "{Beyond linear galaxy alignments}",
      journal = {\prd},
         year = 2019,
        month = nov,
       volume = {100},
       number = {10},
          eid = {103506},
        pages = {103506},
          doi = {10.1103/PhysRevD.100.103506},
archivePrefix = {arXiv},
       eprint = {1708.09247},
 primaryClass = {astro-ph.CO},
       adsurl = {https://ui.adsabs.harvard.edu/abs/2019PhRvD.100j3506B}
}

@ARTICLE{Bhowmick20,
       author = {{Bhowmick}, Aklant K. and {Chen}, Yingzhang and {Tenneti}, Ananth and {Di Matteo}, Tiziana and {Mandelbaum}, Rachel},
        title = "{The evolution of galaxy intrinsic alignments in the MassiveBlackII universe}",
      journal = {\mnras},
         year = 2020,
        month = jan,
       volume = {491},
       number = {3},
        pages = {4116-4130},
          doi = {10.1093/mnras/stz3240},
archivePrefix = {arXiv},
       eprint = {1905.00906},
 primaryClass = {astro-ph.CO},
       adsurl = {https://ui.adsabs.harvard.edu/abs/2020MNRAS.491.4116B}
}

@ARTICLE{Bridle07,
       author = {{Bridle}, Sarah and {King}, Lindsay},
        title = "{Dark energy constraints from cosmic shear power spectra: impact of intrinsic alignments on photometric redshift requirements}",
      journal = {New Journal of Physics},
         year = 2007,
        month = dec,
       volume = {9},
       number = {12},
        pages = {444},
          doi = {10.1088/1367-2630/9/12/444},
archivePrefix = {arXiv},
       eprint = {0705.0166},
 primaryClass = {astro-ph},
       adsurl = {https://ui.adsabs.harvard.edu/abs/2007NJPh....9..444B}
}

@ARTICLE{Carretero15,
       author = {{Carretero}, J. and {Castander}, F.~J. and {Gazta{\~n}aga}, E. and {Crocce}, M. and {Fosalba}, P.},
        title = "{An algorithm to build mock galaxy catalogues using MICE simulations}",
      journal = {\mnras},
         year = 2015,
        month = feb,
       volume = {447},
       number = {1},
        pages = {646-670},
          doi = {10.1093/mnras/stu2402},
archivePrefix = {arXiv},
       eprint = {1411.3286},
 primaryClass = {astro-ph.GA},
       adsurl = {https://ui.adsabs.harvard.edu/abs/2015MNRAS.447..646C}
}

@ARTICLE{Chisari15,
       author = {{Chisari}, N. and {Codis}, S. and {Laigle}, C. and {Dubois}, Y. and
         {Pichon}, C. and {Devriendt}, J. and {Slyz}, A. and {Miller}, L. and
         {Gavazzi}, R. and {Benabed}, K.},
        title = "{Intrinsic alignments of galaxies in the Horizon-AGN cosmological hydrodynamical simulation}",
      journal = {\mnras},
         year = "2015",
        month = "Dec",
       volume = {454},
       number = {3},
        pages = {2736-2753},
          doi = {10.1093/mnras/stv2154},
archivePrefix = {arXiv},
       eprint = {1507.07843},
 primaryClass = {astro-ph.CO},
       adsurl = {https://ui.adsabs.harvard.edu/abs/2015MNRAS.454.2736C}
}

@ARTICLE{Chisari17,
       author = {{Chisari}, N.~E. and {Koukoufilippas}, N. and {Jindal}, A. and {Peirani}, S. and {Beckmann}, R.~S. and {Codis}, S. and {Devriendt}, J. and {Miller}, L. and {Dubois}, Y. and {Laigle}, C. and {Slyz}, A. and {Pichon}, C.},
        title = "{Galaxy-halo alignments in the Horizon-AGN cosmological hydrodynamical simulation}",
      journal = {\mnras},
         year = 2017,
        month = nov,
       volume = {472},
       number = {1},
        pages = {1163-1181},
          doi = {10.1093/mnras/stx1998},
archivePrefix = {arXiv},
       eprint = {1702.03913},
 primaryClass = {astro-ph.CO},
       adsurl = {https://ui.adsabs.harvard.edu/abs/2017MNRAS.472.1163C}
}

@ARTICLE{1984ApJ...284L...9K,
       author = {{Kaiser}, N.},
        title = "{On the spatial correlations of Abell clusters.}",
      journal = {\apjl},
         year = 1984,
        month = sep,
       volume = {284},
        pages = {L9-L12},
          doi = {10.1086/184341},
       adsurl = {https://ui.adsabs.harvard.edu/abs/1984ApJ...284L...9K}
}

@ARTICLE{Codis18,
       author = {{Codis}, S. and {Jindal}, A. and {Chisari}, N.~E. and {Vibert}, D. and {Dubois}, Y. and {Pichon}, C. and {Devriendt}, J.},
        title = "{Galaxy orientation with the cosmic web across cosmic time}",
      journal = {\mnras},
         year = 2018,
        month = dec,
       volume = {481},
       number = {4},
        pages = {4753-4774},
          doi = {10.1093/mnras/sty2567},
archivePrefix = {arXiv},
       eprint = {1809.06212},
 primaryClass = {astro-ph.CO},
       adsurl = {https://ui.adsabs.harvard.edu/abs/2018MNRAS.481.4753C}
}

@ARTICLE{Dawson13,
       author = {{Dawson}, Kyle S. and {Schlegel}, David J. and {Ahn}, Christopher P. and {Anderson}, Scott F. and {Aubourg}, {\'E}ric and {Bailey}, Stephen and {Barkhouser}, Robert H. and {Bautista}, Julian E. and {Beifiori}, Alessandra and {Berlind}, Andreas A. and {Bhardwaj}, Vaishali and {Bizyaev}, Dmitry and {Blake}, Cullen H. and {Blanton}, Michael R. and {Blomqvist}, Michael and {Bolton}, Adam S. and {Borde}, Arnaud and {Bovy}, Jo and {Brandt}, W.~N. and {Brewington}, Howard and {Brinkmann}, Jon and {Brown}, Peter J. and {Brownstein}, Joel R. and {Bundy}, Kevin and {Busca}, N.~G. and {Carithers}, William and {Carnero}, Aurelio R. and {Carr}, Michael A. and {Chen}, Yanmei and {Comparat}, Johan and {Connolly}, Natalia and {Cope}, Frances and {Croft}, Rupert A.~C. and {Cuesta}, Antonio J. and {da Costa}, Luiz N. and {Davenport}, James R.~A. and {Delubac}, Timoth{\'e}e and {de Putter}, Roland and {Dhital}, Saurav and {Ealet}, Anne and {Ebelke}, Garrett L. and {Eisenstein}, Daniel J. and {Escoffier}, S. and {Fan}, Xiaohui and {Filiz Ak}, N. and {Finley}, Hayley and {Font-Ribera}, Andreu and {G{\'e}nova-Santos}, R. and {Gunn}, James E. and {Guo}, Hong and {Haggard}, Daryl and {Hall}, Patrick B. and {Hamilton}, Jean-Christophe and {Harris}, Ben and {Harris}, David W. and {Ho}, Shirley and {Hogg}, David W. and {Holder}, Diana and {Honscheid}, Klaus and {Huehnerhoff}, Joe and {Jordan}, Beatrice and {Jordan}, Wendell P. and {Kauffmann}, Guinevere and {Kazin}, Eyal A. and {Kirkby}, David and {Klaene}, Mark A. and {Kneib}, Jean-Paul and {Le Goff}, Jean-Marc and {Lee}, Khee-Gan and {Long}, Daniel C. and {Loomis}, Craig P. and {Lundgren}, Britt and {Lupton}, Robert H. and {Maia}, Marcio A.~G. and {Makler}, Martin and {Malanushenko}, Elena and {Malanushenko}, Viktor and {Mandelbaum}, Rachel and {Manera}, Marc and {Maraston}, Claudia and {Margala}, Daniel and {Masters}, Karen L. and {McBride}, Cameron K. and {McDonald}, Patrick and {McGreer}, Ian D. and {McMahon}, Richard G. and {Mena}, Olga and {Miralda-Escud{\'e}}, Jordi and {Montero-Dorta}, Antonio D. and {Montesano}, Francesco and {Muna}, Demitri and {Myers}, Adam D. and {Naugle}, Tracy and {Nichol}, Robert C. and {Noterdaeme}, Pasquier and {Nuza}, Sebasti{\'a}n E. and {Olmstead}, Matthew D. and {Oravetz}, Audrey and {Oravetz}, Daniel J. and {Owen}, Russell and {Padmanabhan}, Nikhil and {Palanque-Delabrouille}, Nathalie and {Pan}, Kaike and {Parejko}, John K. and {P{\^a}ris}, Isabelle and {Percival}, Will J. and {P{\'e}rez-Fournon}, Ismael and {P{\'e}rez-R{\`a}fols}, Ignasi and {Petitjean}, Patrick and {Pfaffenberger}, Robert and {Pforr}, Janine and {Pieri}, Matthew M. and {Prada}, Francisco and {Price-Whelan}, Adrian M. and {Raddick}, M. Jordan and {Rebolo}, Rafael and {Rich}, James and {Richards}, Gordon T. and {Rockosi}, Constance M. and {Roe}, Natalie A. and {Ross}, Ashley J. and {Ross}, Nicholas P. and {Rossi}, Graziano and {Rubi{\~n}o-Martin}, J.~A. and {Samushia}, Lado and {S{\'a}nchez}, Ariel G. and {Sayres}, Conor and {Schmidt}, Sarah J. and {Schneider}, Donald P. and {Sc{\'o}ccola}, C.~G. and {Seo}, Hee-Jong and {Shelden}, Alaina and {Sheldon}, Erin and {Shen}, Yue and {Shu}, Yiping and {Slosar}, An{\v{z}}e and {Smee}, Stephen A. and {Snedden}, Stephanie A. and {Stauffer}, Fritz and {Steele}, Oliver and {Strauss}, Michael A. and {Streblyanska}, Alina and {Suzuki}, Nao and {Swanson}, Molly E.~C. and {Tal}, Tomer and {Tanaka}, Masayuki and {Thomas}, Daniel and {Tinker}, Jeremy L. and {Tojeiro}, Rita and {Tremonti}, Christy A. and {Vargas Maga{\~n}a}, M. and {Verde}, Licia and {Viel}, Matteo and {Wake}, David A. and {Watson}, Mike and {Weaver}, Benjamin A. and {Weinberg}, David H. and {Weiner}, Benjamin J. and {West}, Andrew A. and {White}, Martin and {Wood-Vasey}, W.~M. and {Yeche}, Christophe and {Zehavi}, Idit and {Zhao}, Gong-Bo and {Zheng}, Zheng},
        title = "{The Baryon Oscillation Spectroscopic Survey of SDSS-III}",
      journal = {\aj},
         year = 2013,
        month = jan,
       volume = {145},
       number = {1},
          eid = {10},
        pages = {10},
          doi = {10.1088/0004-6256/145/1/10},
archivePrefix = {arXiv},
       eprint = {1208.0022},
 primaryClass = {astro-ph.CO},
       adsurl = {https://ui.adsabs.harvard.edu/abs/2013AJ....145...10D}
}

@ARTICLE{Dubois14,
       author = {{Dubois}, Y. and {Pichon}, C. and {Welker}, C. and {Le Borgne}, D. and
         {Devriendt}, J. and {Laigle}, C. and {Codis}, S. and {Pogosyan}, D. and
         {Arnouts}, S. and {Benabed}, K. and {Bertin}, E. and {Blaizot}, J. and
         {Bouchet}, F. and {Cardoso}, J. -F. and {Colombi}, S. and
         {de Lapparent}, V. and {Desjacques}, V. and {Gavazzi}, R. and
         {Kassin}, S. and {Kimm}, T. and {McCracken}, H. and {Milliard}, B. and
         {Peirani}, S. and {Prunet}, S. and {Rouberol}, S. and {Silk}, J. and
         {Slyz}, A. and {Sousbie}, T. and {Teyssier}, R. and {Tresse}, L. and
         {Treyer}, M. and {Vibert}, D. and {Volonteri}, M.},
        title = "{Dancing in the dark: galactic properties trace spin swings along the cosmic web}",
      journal = {\mnras},
         year = 2014,
        month = oct,
       volume = {444},
       number = {2},
        pages = {1453-1468},
          doi = {10.1093/mnras/stu1227},
archivePrefix = {arXiv},
       eprint = {1402.1165},
 primaryClass = {astro-ph.CO},
       adsurl = {https://ui.adsabs.harvard.edu/abs/2014MNRAS.444.1453D}
}

@ARTICLE{Fang17,
       author = {{Fang}, Xiao and {Blazek}, Jonathan A. and {McEwen}, Joseph E. and {Hirata}, Christopher M.},
        title = "{FAST-PT II: an algorithm to calculate convolution integrals of general tensor quantities in cosmological perturbation theory}",
      journal = {JCAP},
         year = 2017,
        month = feb,
       volume = {2},
       number = {2},
          eid = {030},
        pages = {030},
          doi = {10.1088/1475-7516/2017/02/030},
archivePrefix = {arXiv},
       eprint = {1609.05978},
 primaryClass = {astro-ph.CO},
       adsurl = {https://ui.adsabs.harvard.edu/abs/2017JCAP...02..030F}
}

@ARTICLE{David2025,
       author = {{Navarro-Giron{\'e}s}, D. and {Crocce}, M. and {Gazta{\~n}aga}, E. and {Wittje}, A. and {Siudek}, M. and {Hoekstra}, H. and {Hildebrandt}, H. and {Joachimi}, B. and {Paviot}, R. and {Baugh}, C.~M. and {Carretero}, J. and {Casas}, R. and {Castander}, F.~J. and {Eriksen}, M. and {Fernandez}, E. and {Fosalba}, P. and {Garc{\'\i}a-Bellido}, J. and {Miquel}, R. and {Padilla}, C. and {Renard}, P. and {S{\'a}nchez}, E. and {Serrano}, S. and {Sevilla-Noarbe}, I. and {Tallada-Cresp{\'\i}}, P.},
        title = "{The PAU Survey: measuring intrinsic galaxy alignments in deep wide fields as a function of colour, luminosity, stellar mass, and redshift}",
      journal = {\mnras},
         year = 2026,
        month = jan,
       volume = {545},
       number = {2},
          eid = {staf1630},
        pages = {staf1630},
          doi = {10.1093/mnras/staf1630},
archivePrefix = {arXiv},
       eprint = {2505.15470},
 primaryClass = {astro-ph.CO},
       adsurl = {https://ui.adsabs.harvard.edu/abs/2026MNRAS.545f1630N}
}

@ARTICLE{Georgiou19,
       author = {{Georgiou}, Christos and {Johnston}, Harry and {Hoekstra}, Henk and
         {Viola}, Massimo and {Kuijken}, Konrad and {Joachimi}, Benjamin and
         {Chisari}, Nora Elisa and {Farrow}, Daniel J. and {Hildebrand
        t}, Hendrik and {Holwerda}, Benne W. and {Kannawadi}, Arun},
        title = "{The dependence of intrinsic alignment of galaxies on wavelength using KiDS and GAMA}",
      journal = {\aap},
         year = 2019,
        month = feb,
       volume = {622},
          eid = {A90},
        pages = {A90},
          doi = {10.1051/0004-6361/201834219},
archivePrefix = {arXiv},
       eprint = {1809.03602},
 primaryClass = {astro-ph.CO},
       adsurl = {https://ui.adsabs.harvard.edu/abs/2019A&A...622A..90G}
}

@ARTICLE{Hartlap07,
       author = {{Hartlap}, J. and {Simon}, P. and {Schneider}, P.},
        title = "{Why your model parameter confidences might be too optimistic. Unbiased estimation of the inverse covariance matrix}",
      journal = {\aap},
         year = 2007,
        month = mar,
       volume = {464},
       number = {1},
        pages = {399-404},
          doi = {10.1051/0004-6361:20066170},
archivePrefix = {arXiv},
       eprint = {astro-ph/0608064},
 primaryClass = {astro-ph},
       adsurl = {https://ui.adsabs.harvard.edu/abs/2007A&A...464..399H}
}

@ARTICLE{Heymans21,
       author = {{Heymans}, Catherine and {Tr{\"o}ster}, Tilman and {Asgari}, Marika and {Blake}, Chris and {Hildebrandt}, Hendrik and {Joachimi}, Benjamin and {Kuijken}, Konrad and {Lin}, Chieh-An and {S{\'a}nchez}, Ariel G. and {van den Busch}, Jan Luca and {Wright}, Angus H. and {Amon}, Alexandra and {Bilicki}, Maciej and {de Jong}, Jelte and {Crocce}, Martin and {Dvornik}, Andrej and {Erben}, Thomas and {Fortuna}, Maria Cristina and {Getman}, Fedor and {Giblin}, Benjamin and {Glazebrook}, Karl and {Hoekstra}, Henk and {Joudaki}, Shahab and {Kannawadi}, Arun and {K{\"o}hlinger}, Fabian and {Lidman}, Chris and {Miller}, Lance and {Napolitano}, Nicola R. and {Parkinson}, David and {Schneider}, Peter and {Shan}, HuanYuan and {Valentijn}, Edwin A. and {Verdoes Kleijn}, Gijs and {Wolf}, Christian},
        title = "{KiDS-1000 Cosmology: Multi-probe weak gravitational lensing and spectroscopic galaxy clustering constraints}",
      journal = {\aap},
         year = 2021,
        month = feb,
       volume = {646},
          eid = {A140},
        pages = {A140},
          doi = {10.1051/0004-6361/202039063},
archivePrefix = {arXiv},
       eprint = {2007.15632},
 primaryClass = {astro-ph.CO},
       adsurl = {https://ui.adsabs.harvard.edu/abs/2021A&A...646A.140H}
}

@ARTICLE{Hilbert17,
       author = {{Hilbert}, Stefan and {Xu}, Dandan and {Schneider}, Peter and {Springel}, Volker and {Vogelsberger}, Mark and {Hernquist}, Lars},
        title = "{Intrinsic alignments of galaxies in the Illustris simulation}",
      journal = {\mnras},
         year = 2017,
        month = jun,
       volume = {468},
       number = {1},
        pages = {790-823},
          doi = {10.1093/mnras/stx482},
archivePrefix = {arXiv},
       eprint = {1606.03216},
 primaryClass = {astro-ph.CO},
       adsurl = {https://ui.adsabs.harvard.edu/abs/2017MNRAS.468..790H}
}

@ARTICLE{Hildebrandt17,
       author = {{Hildebrandt}, H. and {Viola}, M. and {Heymans}, C. and {Joudaki}, S. and {Kuijken}, K. and {Blake}, C. and {Erben}, T. and {Joachimi}, B. and {Klaes}, D. and {Miller}, L. and {Morrison}, C.~B. and {Nakajima}, R. and {Verdoes Kleijn}, G. and {Amon}, A. and {Choi}, A. and {Covone}, G. and {de Jong}, J.~T.~A. and {Dvornik}, A. and {Fenech Conti}, I. and {Grado}, A. and {Harnois-D{\'e}raps}, J. and {Herbonnet}, R. and {Hoekstra}, H. and {K{\"o}hlinger}, F. and {McFarland}, J. and {Mead}, A. and {Merten}, J. and {Napolitano}, N. and {Peacock}, J.~A. and {Radovich}, M. and {Schneider}, P. and {Simon}, P. and {Valentijn}, E.~A. and {van den Busch}, J.~L. and {van Uitert}, E. and {Van Waerbeke}, L.},
        title = "{KiDS-450: cosmological parameter constraints from tomographic weak gravitational lensing}",
      journal = {\mnras},
         year = 2017,
        month = feb,
       volume = {465},
       number = {2},
        pages = {1454-1498},
          doi = {10.1093/mnras/stw2805},
archivePrefix = {arXiv},
       eprint = {1606.05338},
 primaryClass = {astro-ph.CO},
       adsurl = {https://ui.adsabs.harvard.edu/abs/2017MNRAS.465.1454H}
}

@ARTICLE{Hirata04,
       author = {{Hirata}, Christopher M. and {Seljak}, Uro{\v{s}}},
        title = "{Intrinsic alignment-lensing interference as a contaminant of cosmic shear}",
      journal = {\prd},
         year = 2004,
        month = sep,
       volume = {70},
       number = {6},
          eid = {063526},
        pages = {063526},
          doi = {10.1103/PhysRevD.70.063526},
archivePrefix = {arXiv},
       eprint = {astro-ph/0406275},
 primaryClass = {astro-ph},
       adsurl = {https://ui.adsabs.harvard.edu/abs/2004PhRvD..70f3526H}
}

@ARTICLE{Hirata07,
       author = {{Hirata}, Christopher M. and {Mandelbaum}, Rachel and {Ishak}, Mustapha and {Seljak}, Uro{\v{s}} and {Nichol}, Robert and {Pimbblet}, Kevin A. and {Ross}, Nicholas P. and {Wake}, David},
        title = "{Intrinsic galaxy alignments from the 2SLAQ and SDSS surveys: luminosity and redshift scalings and implications for weak lensing surveys}",
      journal = {\mnras},
         year = 2007,
        month = nov,
       volume = {381},
       number = {3},
        pages = {1197-1218},
          doi = {10.1111/j.1365-2966.2007.12312.x},
archivePrefix = {arXiv},
       eprint = {astro-ph/0701671},
 primaryClass = {astro-ph},
       adsurl = {https://ui.adsabs.harvard.edu/abs/2007MNRAS.381.1197H}
}

@ARTICLE{Joachimi11,
       author = {{Joachimi}, B. and {Mandelbaum}, R. and {Abdalla}, F.~B. and {Bridle}, S.~L.},
        title = "{Constraints on intrinsic alignment contamination of weak lensing surveys using the MegaZ-LRG sample}",
      journal = {\aap},
         year = 2011,
        month = mar,
       volume = {527},
          eid = {A26},
        pages = {A26},
          doi = {10.1051/0004-6361/201015621},
archivePrefix = {arXiv},
       eprint = {1008.3491},
 primaryClass = {astro-ph.CO},
       adsurl = {https://ui.adsabs.harvard.edu/abs/2011A&A...527A..26J}
}

@ARTICLE{Joachimi15,
       author = {{Joachimi}, Benjamin and {Cacciato}, Marcello and {Kitching}, Thomas D. and {Leonard}, Adrienne and {Mandelbaum}, Rachel and {Sch{\"a}fer}, Bj{\"o}rn Malte and {Sif{\'o}n}, Crist{\'o}bal and {Hoekstra}, Henk and {Kiessling}, Alina and {Kirk}, Donnacha and {Rassat}, Anais},
        title = "{Galaxy Alignments: An Overview}",
      journal = {\ssr},
         year = 2015,
        month = nov,
       volume = {193},
       number = {1-4},
        pages = {1-65},
          doi = {10.1007/s11214-015-0177-4},
archivePrefix = {arXiv},
       eprint = {1504.05456},
 primaryClass = {astro-ph.GA},
       adsurl = {https://ui.adsabs.harvard.edu/abs/2015SSRv..193....1J}
}

@ARTICLE{Johnston19,
       author = {{Johnston}, Harry and {Georgiou}, Christos and {Joachimi}, Benjamin and {Hoekstra}, Henk and {Chisari}, Nora Elisa and {Farrow}, Daniel and {Fortuna}, Maria Cristina and {Heymans}, Catherine and {Joudaki}, Shahab and {Kuijken}, Konrad and {Wright}, Angus},
        title = "{KiDS+GAMA: Intrinsic alignment model constraints for current and future weak lensing cosmology}",
      journal = {\aap},
         year = 2019,
        month = apr,
       volume = {624},
          eid = {A30},
        pages = {A30},
          doi = {10.1051/0004-6361/201834714},
archivePrefix = {arXiv},
       eprint = {1811.09598},
 primaryClass = {astro-ph.CO},
       adsurl = {https://ui.adsabs.harvard.edu/abs/2019A&A...624A..30J}
}

@ARTICLE{Johnston21,
       author = {{Johnston}, Harry and {Joachimi}, Benjamin and {Norberg}, Peder and {Hoekstra}, Henk and {Eriksen}, Martin and {Fortuna}, Maria Cristina and {Manzoni}, Giorgio and {Serrano}, Santiago and {Siudek}, Malgorzata and {Tortorelli}, Luca and {Asorey}, Jacobo and {Cabayol}, Laura and {Carretero}, Jorge and {Casas}, Ricard and {Castander}, Francisco and {Crocce}, Martin and {Fernandez}, Enrique and {Garc{\'\i}a-Bellido}, Juan and {Gaztanaga}, Enrique and {Hildebrandt}, Hendrik and {Miquel}, Ramon and {Navarro-Girones}, David and {Padilla}, Cristobal and {Sanchez}, Eusebio and {Sevilla-Noarbe}, Ignacio and {Tallada-Cresp{\'\i}}, Pau},
        title = "{The PAU Survey: Intrinsic alignments and clustering of narrow-band photometric galaxies}",
      journal = {\aap},
         year = 2021,
        month = feb,
       volume = {646},
          eid = {A147},
        pages = {A147},
          doi = {10.1051/0004-6361/202039682},
archivePrefix = {arXiv},
       eprint = {2010.09696},
 primaryClass = {astro-ph.GA},
       adsurl = {https://ui.adsabs.harvard.edu/abs/2021A&A...646A.147J}
}

@ARTICLE{Kiessling15,
       author = {{Kiessling}, Alina and {Cacciato}, Marcello and {Joachimi}, Benjamin and {Kirk}, Donnacha and {Kitching}, Thomas D. and {Leonard}, Adrienne and {Mandelbaum}, Rachel and {Sch{\"a}fer}, Bj{\"o}rn Malte and {Sif{\'o}n}, Crist{\'o}bal and {Brown}, Michael L. and {Rassat}, Anais},
        title = "{Galaxy Alignments: Theory, Modelling \& Simulations}",
      journal = {\ssr},
         year = 2015,
        month = nov,
       volume = {193},
       number = {1-4},
        pages = {67-136},
          doi = {10.1007/s11214-015-0203-6},
archivePrefix = {arXiv},
       eprint = {1504.05546},
 primaryClass = {astro-ph.GA},
       adsurl = {https://ui.adsabs.harvard.edu/abs/2015SSRv..193...67K}
}

@ARTICLE{Kirk15,
       author = {{Kirk}, Donnacha and {Brown}, Michael L. and {Hoekstra}, Henk and
         {Joachimi}, Benjamin and {Kitching}, Thomas D. and {Mand
        elbaum}, Rachel and {Sif{\'o}n}, Crist{\'o}bal and {Cacciato}, Marcello and
         {Choi}, Ami and {Kiessling}, Alina and {Leonard}, Adrienne and
         {Rassat}, Anais and {Sch{\"a}fer}, Bj{\"o}rn Malte},
        title = "{Galaxy Alignments: Observations and Impact on Cosmology}",
      journal = {\ssr},
         year = 2015,
        month = nov,
       volume = {193},
       number = {1-4},
        pages = {139-211},
          doi = {10.1007/s11214-015-0213-4},
archivePrefix = {arXiv},
       eprint = {1504.05465},
 primaryClass = {astro-ph.GA},
       adsurl = {https://ui.adsabs.harvard.edu/abs/2015SSRv..193..139K}
}

@ARTICLE{Landy93,
       author = {{Landy}, Stephen D. and {Szalay}, Alexander S.},
        title = "{Bias and Variance of Angular Correlation Functions}",
      journal = {\apj},
         year = "1993",
        month = "Jul",
       volume = {412},
        pages = {64},
          doi = {10.1086/172900},
       adsurl = {https://ui.adsabs.harvard.edu/abs/1993ApJ...412...64L}
}

@ARTICLE{Mandelbaum06,
       author = {{Mandelbaum}, Rachel and {Hirata}, Christopher M. and {Ishak}, Mustapha and {Seljak}, Uro{\v{s}} and {Brinkmann}, Jonathan},
        title = "{Detection of large-scale intrinsic ellipticity-density correlation from the Sloan Digital Sky Survey and implications for weak lensing surveys}",
      journal = {\mnras},
         year = 2006,
        month = apr,
       volume = {367},
       number = {2},
        pages = {611-626},
          doi = {10.1111/j.1365-2966.2005.09946.x},
archivePrefix = {arXiv},
       eprint = {astro-ph/0509026},
 primaryClass = {astro-ph},
       adsurl = {https://ui.adsabs.harvard.edu/abs/2006MNRAS.367..611M}
}

@ARTICLE{Mandelbaum11,
       author = {{Mandelbaum}, Rachel and {Blake}, Chris and {Bridle}, Sarah and {Abdalla}, Filipe B. and {Brough}, Sarah and {Colless}, Matthew and {Couch}, Warrick and {Croom}, Scott and {Davis}, Tamara and {Drinkwater}, Michael J. and {Forster}, Karl and {Glazebrook}, Karl and {Jelliffe}, Ben and {Jurek}, Russell J. and {Li}, I. -Hui and {Madore}, Barry and {Martin}, Chris and {Pimbblet}, Kevin and {Poole}, Gregory B. and {Pracy}, Michael and {Sharp}, Rob and {Wisnioski}, Emily and {Woods}, David and {Wyder}, Ted},
        title = "{The WiggleZ Dark Energy Survey: direct constraints on blue galaxy intrinsic alignments at intermediate redshifts}",
      journal = {\mnras},
         year = 2011,
        month = jan,
       volume = {410},
       number = {2},
        pages = {844-859},
          doi = {10.1111/j.1365-2966.2010.17485.x},
archivePrefix = {arXiv},
       eprint = {0911.5347},
 primaryClass = {astro-ph.CO},
       adsurl = {https://ui.adsabs.harvard.edu/abs/2011MNRAS.410..844M}
}

@ARTICLE{McEwen16,
       author = {{McEwen}, Joseph E. and {Fang}, Xiao and {Hirata}, Christopher M. and {Blazek}, Jonathan A.},
        title = "{FAST-PT: a novel algorithm to calculate convolution integrals in cosmological perturbation theory}",
      journal = {JCAP},
         year = 2016,
        month = sep,
       volume = {9},
       number = {9},
          eid = {015},
        pages = {015},
          doi = {10.1088/1475-7516/2016/09/015},
archivePrefix = {arXiv},
       eprint = {1603.04826},
 primaryClass = {astro-ph.CO},
       adsurl = {https://ui.adsabs.harvard.edu/abs/2016JCAP...09..015M}
}

@ARTICLE{Blanton2005,
       author = {{Blanton}, Michael R. and {Schlegel}, David J. and {Strauss}, Michael A. and {Brinkmann}, J. and {Finkbeiner}, Douglas and {Fukugita}, Masataka and {Gunn}, James E. and {Hogg}, David W. and {Ivezi{\'c}}, {\v{Z}}eljko and {Knapp}, G.~R. and {Lupton}, Robert H. and {Munn}, Jeffrey A. and {Schneider}, Donald P. and {Tegmark}, Max and {Zehavi}, Idit},
        title = "{New York University Value-Added Galaxy Catalog: A Galaxy Catalog Based on New Public Surveys}",
      journal = {\aj},
         year = 2005,
        month = jun,
       volume = {129},
       number = {6},
        pages = {2562-2578},
          doi = {10.1086/429803},
archivePrefix = {arXiv},
       eprint = {astro-ph/0410166},
 primaryClass = {astro-ph},
       adsurl = {https://ui.adsabs.harvard.edu/abs/2005AJ....129.2562B}
}

@ARTICLE{Okumura09b,
       author = {{Okumura}, Teppei and {Jing}, Y.~P. and {Li}, Cheng},
        title = "{Intrinsic Ellipticity Correlation of SDSS Luminous Red Galaxies and Misalignment with Their Host Dark Matter Halos}",
      journal = {\apj},
         year = 2009,
        month = mar,
       volume = {694},
       number = {1},
        pages = {214-221},
          doi = {10.1088/0004-637X/694/1/214},
archivePrefix = {arXiv},
       eprint = {0809.3790},
 primaryClass = {astro-ph},
       adsurl = {https://ui.adsabs.harvard.edu/abs/2009ApJ...694..214O}
}

@ARTICLE{Piras18,
       author = {{Piras}, Davide and {Joachimi}, Benjamin and {Sch{\"a}fer}, Bj{\"o}rn Malte and {Bonamigo}, Mario and {Hilbert}, Stefan and {van Uitert}, Edo},
        title = "{The mass dependence of dark matter halo alignments with large-scale structure}",
      journal = {\mnras},
         year = 2018,
        month = feb,
       volume = {474},
       number = {1},
        pages = {1165-1175},
          doi = {10.1093/mnras/stx2846},
archivePrefix = {arXiv},
       eprint = {1707.06559},
 primaryClass = {astro-ph.CO},
       adsurl = {https://ui.adsabs.harvard.edu/abs/2018MNRAS.474.1165P}
}

@ARTICLE{Reid16,
       author = {{Reid}, Beth and {Ho}, Shirley and {Padmanabhan}, Nikhil and {Percival}, Will J. and {Tinker}, Jeremy and {Tojeiro}, Rita and {White}, Martin and {Eisenstein}, Daniel J. and {Maraston}, Claudia and {Ross}, Ashley J. and {S{\'a}nchez}, Ariel G. and {Schlegel}, David and {Sheldon}, Erin and {Strauss}, Michael A. and {Thomas}, Daniel and {Wake}, David and {Beutler}, Florian and {Bizyaev}, Dmitry and {Bolton}, Adam S. and {Brownstein}, Joel R. and {Chuang}, Chia-Hsun and {Dawson}, Kyle and {Harding}, Paul and {Kitaura}, Francisco-Shu and {Leauthaud}, Alexie and {Masters}, Karen and {McBride}, Cameron K. and {More}, Surhud and {Olmstead}, Matthew D. and {Oravetz}, Daniel and {Nuza}, Sebasti{\'a}n E. and {Pan}, Kaike and {Parejko}, John and {Pforr}, Janine and {Prada}, Francisco and {Rodr{\'\i}guez-Torres}, Sergio and {Salazar-Albornoz}, Salvador and {Samushia}, Lado and {Schneider}, Donald P. and {Sc{\'o}ccola}, Claudia G. and {Simmons}, Audrey and {Vargas-Magana}, Mariana},
        title = "{SDSS-III Baryon Oscillation Spectroscopic Survey Data Release 12: galaxy target selection and large-scale structure catalogues}",
      journal = {\mnras},
         year = 2016,
        month = jan,
       volume = {455},
       number = {2},
        pages = {1553-1573},
          doi = {10.1093/mnras/stv2382},
archivePrefix = {arXiv},
       eprint = {1509.06529},
 primaryClass = {astro-ph.CO},
       adsurl = {https://ui.adsabs.harvard.edu/abs/2016MNRAS.455.1553R}
}

@ARTICLE{Saito14,
       author = {{Saito}, Shun and {Baldauf}, Tobias and {Vlah}, Zvonimir and {Seljak}, Uro{\v{s}} and {Okumura}, Teppei and {McDonald}, Patrick},
        title = "{Understanding higher-order nonlocal halo bias at large scales by combining the power spectrum with the bispectrum}",
      journal = {\prd},
         year = 2014,
        month = dec,
       volume = {90},
       number = {12},
          eid = {123522},
        pages = {123522},
          doi = {10.1103/PhysRevD.90.123522},
archivePrefix = {arXiv},
       eprint = {1405.1447},
 primaryClass = {astro-ph.CO},
       adsurl = {https://ui.adsabs.harvard.edu/abs/2014PhRvD..90l3522S}
}

@ARTICLE{Samuroff19,
       author = {{Samuroff}, S. and {Blazek}, J. and {Troxel}, M.~A. and {MacCrann}, N. and {Krause}, E. and {Leonard}, C.~D. and {Prat}, J. and {Gruen}, D. and {Dodelson}, S. and {Eifler}, T.~F. and {Gatti}, M. and {Hartley}, W.~G. and {Hoyle}, B. and {Larsen}, P. and {Zuntz}, J. and {Abbott}, T.~M.~C. and {Allam}, S. and {Annis}, J. and {Bernstein}, G.~M. and {Bertin}, E. and {Bridle}, S.~L. and {Brooks}, D. and {Carnero Rosell}, A. and {Carrasco Kind}, M. and {Carretero}, J. and {Castander}, F.~J. and {Cunha}, C.~E. and {da Costa}, L.~N. and {Davis}, C. and {De Vicente}, J. and {DePoy}, D.~L. and {Desai}, S. and {Diehl}, H.~T. and {Dietrich}, J.~P. and {Doel}, P. and {Flaugher}, B. and {Fosalba}, P. and {Frieman}, J. and {Garc{\'\i}a-Bellido}, J. and {Gaztanaga}, E. and {Gerdes}, D.~W. and {Gruendl}, R.~A. and {Gschwend}, J. and {Gutierrez}, G. and {Hollowood}, D.~L. and {Honscheid}, K. and {James}, D.~J. and {Kuehn}, K. and {Kuropatkin}, N. and {Lima}, M. and {Maia}, M.~A.~G. and {March}, M. and {Marshall}, J.~L. and {Martini}, P. and {Melchior}, P. and {Menanteau}, F. and {Miller}, C.~J. and {Miquel}, R. and {Ogando}, R.~L.~C. and {Plazas}, A.~A. and {Sanchez}, E. and {Scarpine}, V. and {Schindler}, R. and {Schubnell}, M. and {Serrano}, S. and {Sevilla-Noarbe}, I. and {Sheldon}, E. and {Smith}, M. and {Sobreira}, F. and {Suchyta}, E. and {Tarle}, G. and {Thomas}, D. and {Vikram}, V. and {DES Collaboration}},
        title = "{Dark Energy Survey Year 1 results: constraints on intrinsic alignments and their colour dependence from galaxy clustering and weak lensing}",
      journal = {\mnras},
         year = 2019,
        month = nov,
       volume = {489},
       number = {4},
        pages = {5453-5482},
          doi = {10.1093/mnras/stz2197},
archivePrefix = {arXiv},
       eprint = {1811.06989},
 primaryClass = {astro-ph.CO},
       adsurl = {https://ui.adsabs.harvard.edu/abs/2019MNRAS.489.5453S}
}

@ARTICLE{Samuroff21,
       author = {{Samuroff}, S. and {Mandelbaum}, R. and {Blazek}, J.},
        title = "{Advances in constraining intrinsic alignment models with hydrodynamic simulations}",
      journal = {\mnras},
         year = 2021,
        month = nov,
       volume = {508},
       number = {1},
        pages = {637-664},
          doi = {10.1093/mnras/stab2520},
archivePrefix = {arXiv},
       eprint = {2009.10735},
 primaryClass = {astro-ph.CO},
       adsurl = {https://ui.adsabs.harvard.edu/abs/2021MNRAS.508..637S}
}

@ARTICLE{Singh15,
       author = {{Singh}, Sukhdeep and {Mandelbaum}, Rachel and {More}, Surhud},
        title = "{Intrinsic alignments of SDSS-III BOSS LOWZ sample galaxies}",
      journal = {\mnras},
         year = "2015",
        month = "Jun",
       volume = {450},
       number = {2},
        pages = {2195-2216},
          doi = {10.1093/mnras/stv778},
archivePrefix = {arXiv},
       eprint = {1411.1755},
 primaryClass = {astro-ph.CO},
       adsurl = {https://ui.adsabs.harvard.edu/abs/2015MNRAS.450.2195S}
}

@ARTICLE{Heymans2012,
       author = {{Heymans}, Catherine and {Van Waerbeke}, Ludovic and {Miller}, Lance and {Erben}, Thomas and {Hildebrandt}, Hendrik and {Hoekstra}, Henk and {Kitching}, Thomas D. and {Mellier}, Yannick and {Simon}, Patrick and {Bonnett}, Christopher and {Coupon}, Jean and {Fu}, Liping and {Harnois D{\'e}raps}, Joachim and {Hudson}, Michael J. and {Kilbinger}, Martin and {Kuijken}, Koenraad and {Rowe}, Barnaby and {Schrabback}, Tim and {Semboloni}, Elisabetta and {van Uitert}, Edo and {Vafaei}, Sanaz and {Velander}, Malin},
        title = "{CFHTLenS: the Canada-France-Hawaii Telescope Lensing Survey}",
      journal = {\mnras},
         year = 2012,
        month = nov,
       volume = {427},
       number = {1},
        pages = {146-166},
          doi = {10.1111/j.1365-2966.2012.21952.x},
archivePrefix = {arXiv},
       eprint = {1210.0032},
 primaryClass = {astro-ph.CO},
       adsurl = {https://ui.adsabs.harvard.edu/abs/2012MNRAS.427..146H}
}

@ARTICLE{Singh16,
       author = {{Singh}, Sukhdeep and {Mandelbaum}, Rachel},
        title = "{Intrinsic alignments of BOSS LOWZ galaxies - II. Impact of shape measurement methods}",
      journal = {\mnras},
         year = "2016",
        month = "Apr",
       volume = {457},
       number = {3},
        pages = {2301-2317},
          doi = {10.1093/mnras/stw144},
archivePrefix = {arXiv},
       eprint = {1510.06752},
 primaryClass = {astro-ph.CO},
       adsurl = {https://ui.adsabs.harvard.edu/abs/2016MNRAS.457.2301S}
}

@ARTICLE{Tenneti15,
       author = {{Tenneti}, Ananth and {Mandelbaum}, Rachel and {Di Matteo}, Tiziana and
         {Kiessling}, Alina and {Khandai}, Nishikanta},
        title = "{Galaxy shapes and alignments in the MassiveBlack-II hydrodynamic and dark matter-only simulations}",
      journal = {\mnras},
         year = "2015",
        month = "Oct",
       volume = {453},
       number = {1},
        pages = {469-482},
          doi = {10.1093/mnras/stv1625},
archivePrefix = {arXiv},
       eprint = {1505.03124},
 primaryClass = {astro-ph.CO},
       adsurl = {https://ui.adsabs.harvard.edu/abs/2015MNRAS.453..469T}
}

@ARTICLE{Tenneti2015b,
       author = {{Tenneti}, Ananth and {Singh}, Sukhdeep and {Mandelbaum}, Rachel and {di Matteo}, Tiziana and {Feng}, Yu and {Khandai}, Nishikanta},
        title = "{Intrinsic alignments of galaxies in the MassiveBlack-II simulation: analysis of two-point statistics}",
      journal = {\mnras},
         year = 2015,
        month = apr,
       volume = {448},
       number = {4},
        pages = {3522-3544},
          doi = {10.1093/mnras/stv272},
archivePrefix = {arXiv},
       eprint = {1409.7297},
 primaryClass = {astro-ph.CO},
       adsurl = {https://ui.adsabs.harvard.edu/abs/2015MNRAS.448.3522T}
}

@ARTICLE{Troxel15,
       author = {{Troxel}, M.~A. and {Ishak}, Mustapha},
        title = "{The intrinsic alignment of galaxies and its impact on weak gravitational lensing in an era of precision cosmology}",
      journal = {\physrep},
         year = 2015,
        month = feb,
       volume = {558},
        pages = {1-59},
          doi = {10.1016/j.physrep.2014.11.001},
archivePrefix = {arXiv},
       eprint = {1407.6990},
 primaryClass = {astro-ph.CO},
       adsurl = {https://ui.adsabs.harvard.edu/abs/2015PhR...558....1T}
}

@ARTICLE{Secco21,
       author = {{Secco}, L.~F. and {Samuroff}, S. and {Krause}, E. and {Jain}, B. and {Blazek}, J. and {Raveri}, M. and {Campos}, A. and {Amon}, A. and {Chen}, A. and {Doux}, C. and {Choi}, A. and {Gruen}, D. and {Bernstein}, G.~M. and {Chang}, C. and {DeRose}, J. and {Myles}, J. and {Fert{\'e}}, A. and {Lemos}, P. and {Huterer}, D. and {Prat}, J. and {Troxel}, M.~A. and {MacCrann}, N. and {Liddle}, A.~R. and {Kacprzak}, T. and {Fang}, X. and {S{\'a}nchez}, C. and {Pandey}, S. and {Dodelson}, S. and {Chintalapati}, P. and {Hoffmann}, K. and {Alarcon}, A. and {Alves}, O. and {Andrade-Oliveira}, F. and {Baxter}, E.~J. and {Bechtol}, K. and {Becker}, M.~R. and {Brandao-Souza}, A. and {Camacho}, H. and {Carnero Rosell}, A. and {Carrasco Kind}, M. and {Cawthon}, R. and {Cordero}, J.~P. and {Crocce}, M. and {Davis}, C. and {Di Valentino}, E. and {Drlica-Wagner}, A. and {Eckert}, K. and {Eifler}, T.~F. and {Elidaiana}, M. and {Elsner}, F. and {Elvin-Poole}, J. and {Everett}, S. and {Fosalba}, P. and {Friedrich}, O. and {Gatti}, M. and {Giannini}, G. and {Gruendl}, R.~A. and {Harrison}, I. and {Hartley}, W.~G. and {Herner}, K. and {Huang}, H. and {Huff}, E.~M. and {Jarvis}, M. and {Jeffrey}, N. and {Kuropatkin}, N. and {Leget}, P. -F. and {Muir}, J. and {Mccullough}, J. and {Navarro Alsina}, A. and {Omori}, Y. and {Park}, Y. and {Porredon}, A. and {Rollins}, R. and {Roodman}, A. and {Rosenfeld}, R. and {Ross}, A.~J. and {Rykoff}, E.~S. and {Sanchez}, J. and {Sevilla-Noarbe}, I. and {Sheldon}, E.~S. and {Shin}, T. and {Tutusaus}, I. and {Varga}, T.~N. and {Weaverdyck}, N. and {Wechsler}, R.~H. and {Yanny}, B. and {Yin}, B. and {Zhang}, Y. and {Zuntz}, J. and {Abbott}, T.~M.~C. and {Aguena}, M. and {Allam}, S. and {Annis}, J. and {Bacon}, D. and {Bertin}, E. and {Bhargava}, S. and {Bridle}, S.~L. and {Brooks}, D. and {Buckley-Geer}, E. and {Burke}, D.~L. and {Carretero}, J. and {Costanzi}, M. and {da Costa}, L.~N. and {De Vicente}, J. and {Diehl}, H.~T. and {Dietrich}, J.~P. and {Doel}, P. and {Ferrero}, I. and {Flaugher}, B. and {Frieman}, J. and {Garc{\'\i}a-Bellido}, J. and {Gaztanaga}, E. and {Gerdes}, D.~W. and {Giannantonio}, T. and {Gschwend}, J. and {Gutierrez}, G. and {Hinton}, S.~R. and {Hollowood}, D.~L. and {Honscheid}, K. and {Hoyle}, B. and {James}, D.~J. and {Jeltema}, T. and {Kuehn}, K. and {Lahav}, O. and {Lima}, M. and {Lin}, H. and {Maia}, M.~A.~G. and {Marshall}, J.~L. and {Martini}, P. and {Melchior}, P. and {Menanteau}, F. and {Miquel}, R. and {Mohr}, J.~J. and {Morgan}, R. and {Ogando}, R.~L.~C. and {Palmese}, A. and {Paz-Chinch{\'o}n}, F. and {Petravick}, D. and {Pieres}, A. and {Plazas Malag{\'o}n}, A.~A. and {Rodriguez-Monroy}, M. and {Romer}, A.~K. and {Sanchez}, E. and {Scarpine}, V. and {Schubnell}, M. and {Scolnic}, D. and {Serrano}, S. and {Smith}, M. and {Soares-Santos}, M. and {Suchyta}, E. and {Swanson}, M.~E.~C. and {Tarle}, G. and {Thomas}, D. and {To}, C. and {DES Collaboration}},
        title = "{Dark Energy Survey Year 3 results: Cosmology from cosmic shear and robustness to modeling uncertainty}",
      journal = {\prd},
         year = 2022,
        month = jan,
       volume = {105},
       number = {2},
          eid = {023515},
        pages = {023515},
          doi = {10.1103/PhysRevD.105.023515},
archivePrefix = {arXiv},
       eprint = {2105.13544},
 primaryClass = {astro-ph.CO},
       adsurl = {https://ui.adsabs.harvard.edu/abs/2022PhRvD.105b3515S}
}

@ARTICLE{Catelan01,
       author = {{Catelan}, Paolo and {Kamionkowski}, Marc and {Blandford}, Roger D.},
        title = "{Intrinsic and extrinsic galaxy alignment}",
      journal = {\mnras},
         year = 2001,
        month = jan,
       volume = {320},
       number = {1},
        pages = {L7-L13},
          doi = {10.1046/j.1365-8711.2001.04105.x},
archivePrefix = {arXiv},
       eprint = {astro-ph/0005470},
 primaryClass = {astro-ph},
       adsurl = {https://ui.adsabs.harvard.edu/abs/2001MNRAS.320L...7C}
}

@ARTICLE{Troxel18,
       author = {{Troxel}, M.~A. and {MacCrann}, N. and {Zuntz}, J. and {Eifler}, T.~F. and {Krause}, E. and {Dodelson}, S. and {Gruen}, D. and {Blazek}, J. and {Friedrich}, O. and {Samuroff}, S. and {Prat}, J. and {Secco}, L.~F. and {Davis}, C. and {Fert{\'e}}, A. and {DeRose}, J. and {Alarcon}, A. and {Amara}, A. and {Baxter}, E. and {Becker}, M.~R. and {Bernstein}, G.~M. and {Bridle}, S.~L. and {Cawthon}, R. and {Chang}, C. and {Choi}, A. and {De Vicente}, J. and {Drlica-Wagner}, A. and {Elvin-Poole}, J. and {Frieman}, J. and {Gatti}, M. and {Hartley}, W.~G. and {Honscheid}, K. and {Hoyle}, B. and {Huff}, E.~M. and {Huterer}, D. and {Jain}, B. and {Jarvis}, M. and {Kacprzak}, T. and {Kirk}, D. and {Kokron}, N. and {Krawiec}, C. and {Lahav}, O. and {Liddle}, A.~R. and {Peacock}, J. and {Rau}, M.~M. and {Refregier}, A. and {Rollins}, R.~P. and {Rozo}, E. and {Rykoff}, E.~S. and {S{\'a}nchez}, C. and {Sevilla-Noarbe}, I. and {Sheldon}, E. and {Stebbins}, A. and {Varga}, T.~N. and {Vielzeuf}, P. and {Wang}, M. and {Wechsler}, R.~H. and {Yanny}, B. and {Abbott}, T.~M.~C. and {Abdalla}, F.~B. and {Allam}, S. and {Annis}, J. and {Bechtol}, K. and {Benoit-L{\'e}vy}, A. and {Bertin}, E. and {Brooks}, D. and {Buckley-Geer}, E. and {Burke}, D.~L. and {Carnero Rosell}, A. and {Carrasco Kind}, M. and {Carretero}, J. and {Castander}, F.~J. and {Crocce}, M. and {Cunha}, C.~E. and {D'Andrea}, C.~B. and {da Costa}, L.~N. and {DePoy}, D.~L. and {Desai}, S. and {Diehl}, H.~T. and {Dietrich}, J.~P. and {Doel}, P. and {Fernandez}, E. and {Flaugher}, B. and {Fosalba}, P. and {Garc{\'\i}a-Bellido}, J. and {Gaztanaga}, E. and {Gerdes}, D.~W. and {Giannantonio}, T. and {Goldstein}, D.~A. and {Gruendl}, R.~A. and {Gschwend}, J. and {Gutierrez}, G. and {James}, D.~J. and {Jeltema}, T. and {Johnson}, M.~W.~G. and {Johnson}, M.~D. and {Kent}, S. and {Kuehn}, K. and {Kuhlmann}, S. and {Kuropatkin}, N. and {Li}, T.~S. and {Lima}, M. and {Lin}, H. and {Maia}, M.~A.~G. and {March}, M. and {Marshall}, J.~L. and {Martini}, P. and {Melchior}, P. and {Menanteau}, F. and {Miquel}, R. and {Mohr}, J.~J. and {Neilsen}, E. and {Nichol}, R.~C. and {Nord}, B. and {Petravick}, D. and {Plazas}, A.~A. and {Romer}, A.~K. and {Roodman}, A. and {Sako}, M. and {Sanchez}, E. and {Scarpine}, V. and {Schindler}, R. and {Schubnell}, M. and {Smith}, M. and {Smith}, R.~C. and {Soares-Santos}, M. and {Sobreira}, F. and {Suchyta}, E. and {Swanson}, M.~E.~C. and {Tarle}, G. and {Thomas}, D. and {Tucker}, D.~L. and {Vikram}, V. and {Walker}, A.~R. and {Weller}, J. and {Zhang}, Y. and {DES Collaboration}},
        title = "{Dark Energy Survey Year 1 results: Cosmological constraints from cosmic shear}",
      journal = {\prd},
         year = 2018,
        month = aug,
       volume = {98},
       number = {4},
          eid = {043528},
        pages = {043528},
          doi = {10.1103/PhysRevD.98.043528},
archivePrefix = {arXiv},
       eprint = {1708.01538},
 primaryClass = {astro-ph.CO},
       adsurl = {https://ui.adsabs.harvard.edu/abs/2018PhRvD..98d3528T}
}

@ARTICLE{fortuna21a,
       author = {{Fortuna}, Maria Cristina and {Hoekstra}, Henk and {Joachimi}, Benjamin and {Johnston}, Harry and {Chisari}, Nora Elisa and {Georgiou}, Christos and {Mahony}, Constance},
        title = "{The halo model as a versatile tool to predict intrinsic alignments}",
      journal = {\mnras},
         year = 2021,
        month = feb,
       volume = {501},
       number = {2},
        pages = {2983-3002},
          doi = {10.1093/mnras/staa3802},
archivePrefix = {arXiv},
       eprint = {2003.02700},
 primaryClass = {astro-ph.CO},
       adsurl = {https://ui.adsabs.harvard.edu/abs/2021MNRAS.501.2983F}
}

@ARTICLE{fortuna21b,
       author = {{Fortuna}, Maria Cristina and {Hoekstra}, Henk and {Johnston}, Harry and {Vakili}, Mohammadjavad and {Kannawadi}, Arun and {Georgiou}, Christos and {Joachimi}, Benjamin and {Wright}, Angus H. and {Asgari}, Marika and {Bilicki}, Maciej and {Heymans}, Catherine and {Hildebrandt}, Hendrik and {Kuijken}, Konrad and {Von Wietersheim-Kramsta}, Maximilian},
        title = "{KiDS-1000: Constraints on the intrinsic alignment of luminous red galaxies}",
      journal = {\aap},
         year = 2021,
        month = oct,
       volume = {654},
          eid = {A76},
        pages = {A76},
          doi = {10.1051/0004-6361/202140706},
archivePrefix = {arXiv},
       eprint = {2109.02556},
 primaryClass = {astro-ph.CO},
       adsurl = {https://ui.adsabs.harvard.edu/abs/2021A&A...654A..76F}
}

@ARTICLE{PKDGRAV3-potter2016,
       author = {{Potter}, Douglas and {Stadel}, Joachim and {Teyssier}, Romain},
        title = "{PKDGRAV3: beyond trillion particle cosmological simulations for the next era of galaxy surveys}",
      journal = {Computational Astrophysics and Cosmology},
         year = 2017,
        month = may,
       volume = {4},
       number = {1},
          eid = {2},
        pages = {2},
          doi = {10.1186/s40668-017-0021-1},
archivePrefix = {arXiv},
       eprint = {1609.08621},
 primaryClass = {astro-ph.IM},
       adsurl = {https://ui.adsabs.harvard.edu/abs/2017ComAC...4....2P}
}

@ARTICLE{behroozi2012,
       author = {{Behroozi}, Peter S. and {Wechsler}, Risa H. and {Wu}, Hao-Yi},
        title = "{The ROCKSTAR Phase-space Temporal Halo Finder and the Velocity Offsets of Cluster Cores}",
      journal = {\apj},
         year = 2013,
        month = jan,
       volume = {762},
       number = {2},
          eid = {109},
        pages = {109},
          doi = {10.1088/0004-637X/762/2/109},
archivePrefix = {arXiv},
       eprint = {1110.4372},
 primaryClass = {astro-ph.CO},
       adsurl = {https://ui.adsabs.harvard.edu/abs/2013ApJ...762..109B}
}

@ARTICLE{lensing-fosalba2016,
       author = {{Fosalba}, P. and {Gazta{\~n}aga}, E. and {Castander}, F.~J. and {Crocce}, M.},
        title = "{The MICE Grand Challenge light-cone simulation - III. Galaxy lensing mocks from all-sky lensing maps}",
      journal = {\mnras},
         year = 2015,
        month = feb,
       volume = {447},
       number = {2},
        pages = {1319-1332},
          doi = {10.1093/mnras/stu2464},
archivePrefix = {arXiv},
       eprint = {1312.2947},
 primaryClass = {astro-ph.CO},
       adsurl = {https://ui.adsabs.harvard.edu/abs/2015MNRAS.447.1319F}
}

@ARTICLE{Navarro1997,
       author = {{Navarro}, Julio F. and {Frenk}, Carlos S. and {White}, Simon D.~M.},
        title = "{A Universal Density Profile from Hierarchical Clustering}",
      journal = {\apj},
         year = 1997,
        month = dec,
       volume = {490},
       number = {2},
        pages = {493-508},
          doi = {10.1086/304888},
archivePrefix = {arXiv},
       eprint = {astro-ph/9611107},
 primaryClass = {astro-ph},
       adsurl = {https://ui.adsabs.harvard.edu/abs/1997ApJ...490..493N}
}

@ARTICLE{Jagvaral2022b,
       author = {{Jagvaral}, Yesukhei and {Lanusse}, Fran{\c{c}}ois and {Singh}, Sukhdeep and {Mandelbaum}, Rachel and {Ravanbakhsh}, Siamak and {Campbell}, Duncan},
        title = "{Galaxies and haloes on graph neural networks: Deep generative modelling scalar and vector quantities for intrinsic alignment}",
      journal = {\mnras},
         year = 2022,
        month = oct,
       volume = {516},
       number = {2},
        pages = {2406-2419},
          doi = {10.1093/mnras/stac2083},
archivePrefix = {arXiv},
       eprint = {2204.07077},
 primaryClass = {astro-ph.GA},
       adsurl = {https://ui.adsabs.harvard.edu/abs/2022MNRAS.516.2406J}
}

@ARTICLE{Jagvaral2022,
       author = {{Jagvaral}, Yesukhei and {Mandelbaum}, Rachel and {Lanusse}, Francois},
        title = "{Modeling halo and central galaxy orientations on the SO(3) manifold with score-based generative models}",
         year = 2022,
        month = dec,
          eid = {arXiv:2212.05592},
        pages = {arXiv:2212.05592},
          doi = {10.48550/arXiv.2212.05592},
archivePrefix = {arXiv},
       eprint = {2212.05592},
 primaryClass = {astro-ph.GA},
       adsurl = {https://ui.adsabs.harvard.edu/abs/2022arXiv221205592J}
}

@ARTICLE{VanAlfven2023,
       author = {{Van Alfen}, Nicholas and {Campbell}, Duncan and {Blazek}, Jonathan and {Leonard}, C. Danielle and {Lanusse}, Francois and {Hearin}, Andrew and {Mandelbaum}, Rachel and {LSST Dark Energy Science Collaboration}},
        title = "{An Empirical Model For Intrinsic Alignments: Insights From Cosmological Simulations}",
      journal = {The Open Journal of Astrophysics},
         year = 2024,
        month = jun,
       volume = {7},
          eid = {45},
        pages = {45},
          doi = {10.33232/001c.118783},
archivePrefix = {arXiv},
       eprint = {2311.07374},
 primaryClass = {astro-ph.CO},
       adsurl = {https://ui.adsabs.harvard.edu/abs/2024OJAp....7E..45V}
}

@ARTICLE{Mead2021,
       author = {{Mead}, A.~J. and {Brieden}, S. and {Tr{\"o}ster}, T. and {Heymans}, C.},
        title = "{HMCODE-2020: improved modelling of non-linear cosmological power spectra with baryonic feedback}",
      journal = {\mnras},
         year = 2021,
        month = mar,
       volume = {502},
       number = {1},
        pages = {1401-1422},
          doi = {10.1093/mnras/stab082},
archivePrefix = {arXiv},
       eprint = {2009.01858},
 primaryClass = {astro-ph.CO},
       adsurl = {https://ui.adsabs.harvard.edu/abs/2021MNRAS.502.1401M}
}

@ARTICLE{corrfunc,
       author = {{Sinha}, Manodeep and {Garrison}, Lehman H.},
        title = "{CORRFUNC - a suite of blazing fast correlation functions on the CPU}",
      journal = {\mnras},
         year = 2020,
        month = jan,
       volume = {491},
       number = {2},
        pages = {3022-3041},
          doi = {10.1093/mnras/stz3157},
archivePrefix = {arXiv},
       eprint = {1911.03545},
 primaryClass = {astro-ph.CO},
       adsurl = {https://ui.adsabs.harvard.edu/abs/2020MNRAS.491.3022S}
}

@ARTICLE{Marian2015,
       author = {{Marian}, Laura and {Smith}, Robert E. and {Angulo}, Raul E.},
        title = "{An exploration of galaxy-galaxy lensing and galaxy clustering in the Millennium-XXL simulation}",
      journal = {\mnras},
         year = 2015,
        month = aug,
       volume = {451},
       number = {2},
        pages = {1418-1444},
          doi = {10.1093/mnras/stv984},
archivePrefix = {arXiv},
       eprint = {1410.3468},
 primaryClass = {astro-ph.CO},
       adsurl = {https://ui.adsabs.harvard.edu/abs/2015MNRAS.451.1418M}
}

@ARTICLE{Grieb2016,
       author = {{Grieb}, Jan Niklas and {S{\'a}nchez}, Ariel G. and {Salazar-Albornoz}, Salvador and {Dalla Vecchia}, Claudio},
        title = "{Gaussian covariance matrices for anisotropic galaxy clustering measurements}",
      journal = {\mnras},
         year = 2016,
        month = apr,
       volume = {457},
       number = {2},
        pages = {1577-1592},
          doi = {10.1093/mnras/stw065},
archivePrefix = {arXiv},
       eprint = {1509.04293},
 primaryClass = {astro-ph.CO},
       adsurl = {https://ui.adsabs.harvard.edu/abs/2016MNRAS.457.1577G}
}

@ARTICLE{Tonegawa2018,
       author = {{Tonegawa}, Motonari and {Okumura}, Teppei and {Totani}, Tomonori and {Dalton}, Gavin and {Glazebrook}, Karl and {Yabe}, Kiyoto},
        title = "{The Subaru FMOS galaxy redshift survey (FastSound). V. Intrinsic alignments of emission-line galaxies at z {\ensuremath{\sim}} 1.4}",
      journal = {\pasj},
         year = 2018,
        month = jun,
       volume = {70},
       number = {3},
          eid = {41},
        pages = {41},
          doi = {10.1093/pasj/psy030},
archivePrefix = {arXiv},
       eprint = {1708.02224},
 primaryClass = {astro-ph.CO},
       adsurl = {https://ui.adsabs.harvard.edu/abs/2018PASJ...70...41T}
}

@ARTICLE{Kraljic2020,
       author = {{Kraljic}, Katarina and {Pichon}, Christophe and {Codis}, Sandrine and {Laigle}, Clotilde and {Dav{\'e}}, Romeel and {Dubois}, Yohan and {Hwang}, Ho Seong and {Pogosyan}, Dmitri and {Arnouts}, St{\'e}phane and {Devriendt}, Julien and {Musso}, Marcello and {Peirani}, S{\'e}bastien and {Slyz}, Adrianne and {Treyer}, Marie},
        title = "{The impact of the connectivity of the cosmic web on the physical properties of galaxies at its nodes}",
      journal = {\mnras},
         year = 2020,
        month = jan,
       volume = {491},
       number = {3},
        pages = {4294-4309},
          doi = {10.1093/mnras/stz3319},
archivePrefix = {arXiv},
       eprint = {1910.08066},
 primaryClass = {astro-ph.GA},
       adsurl = {https://ui.adsabs.harvard.edu/abs/2020MNRAS.491.4294K}
}

@ARTICLE{2023DESKIDS,
       author = {{Dark Energy Survey and Kilo-Degree Survey Collaboration}},
        title = "{DES Y3 + KiDS-1000: Consistent cosmology combining cosmic shear surveys}",
      journal = {The Open Journal of Astrophysics},
         year = 2023,
        month = oct,
       volume = {6},
          eid = {36},
        pages = {36},
          doi = {10.21105/astro.2305.17173},
archivePrefix = {arXiv},
       eprint = {2305.17173},
 primaryClass = {astro-ph.CO},
       adsurl = {https://ui.adsabs.harvard.edu/abs/2023OJAp....6E..36D}
}

@ARTICLE{Kurita2021,
       author = {{Kurita}, Toshiki and {Takada}, Masahiro and {Nishimichi}, Takahiro and {Takahashi}, Ryuichi and {Osato}, Ken and {Kobayashi}, Yosuke},
        title = "{Power spectrum of halo intrinsic alignments in simulations}",
      journal = {\mnras},
         year = 2021,
        month = feb,
       volume = {501},
       number = {1},
        pages = {833-852},
          doi = {10.1093/mnras/staa3625},
archivePrefix = {arXiv},
       eprint = {2004.12579},
 primaryClass = {astro-ph.CO},
       adsurl = {https://ui.adsabs.harvard.edu/abs/2021MNRAS.501..833K}
}

@ARTICLE{Pereira2008,
       author = {{Pereira}, Maria J. and {Bryan}, Greg L. and {Gill}, Stuart P.~D.},
        title = "{Radial Alignment in Simulated Clusters}",
      journal = {\apj},
         year = 2008,
        month = jan,
       volume = {672},
       number = {2},
        pages = {825-833},
          doi = {10.1086/523830},
archivePrefix = {arXiv},
       eprint = {0707.1702},
 primaryClass = {astro-ph},
       adsurl = {https://ui.adsabs.harvard.edu/abs/2008ApJ...672..825P}
}

@ARTICLE{Pereira2010,
       author = {{Pereira}, Maria J. and {Bryan}, Greg L.},
        title = "{Tidal Torquing of Elliptical Galaxies in Cluster Environments}",
      journal = {\apj},
         year = 2010,
        month = oct,
       volume = {721},
       number = {2},
        pages = {939-955},
          doi = {10.1088/0004-637X/721/2/939},
archivePrefix = {arXiv},
       eprint = {1009.4191},
 primaryClass = {astro-ph.CO},
       adsurl = {https://ui.adsabs.harvard.edu/abs/2010ApJ...721..939P}
}

@ARTICLE{Chisari2014,
       author = {{Chisari}, Nora Elisa and {Mandelbaum}, Rachel and {Strauss}, Michael A. and {Huff}, Eric M. and {Bahcall}, Neta A.},
        title = "{Intrinsic alignments of group and cluster galaxies in photometric surveys}",
      journal = {\mnras},
         year = 2014,
        month = nov,
       volume = {445},
       number = {1},
        pages = {726-748},
          doi = {10.1093/mnras/stu1786},
archivePrefix = {arXiv},
       eprint = {1407.4813},
 primaryClass = {astro-ph.CO},
       adsurl = {https://ui.adsabs.harvard.edu/abs/2014MNRAS.445..726C}
}

@ARTICLE{Maion2023,
       author = {{Maion}, Francisco and {Angulo}, Raul E. and {Bakx}, Thomas and {Chisari}, Nora Elisa and {Kurita}, Toshiki and {Pellejero-Ib{\'a}{\~n}ez}, Marcos},
        title = "{HYMALAIA: a hybrid lagrangian model for intrinsic alignments}",
      journal = {\mnras},
         year = 2024,
        month = jun,
       volume = {531},
       number = {2},
        pages = {2684-2700},
          doi = {10.1093/mnras/stae1331},
archivePrefix = {arXiv},
       eprint = {2307.13754},
 primaryClass = {astro-ph.CO},
       adsurl = {https://ui.adsabs.harvard.edu/abs/2024MNRAS.531.2684M}
}

@ARTICLE{Bakx2023,
       author = {{Bakx}, Thomas and {Kurita}, Toshiki and {Elisa Chisari}, Nora and {Vlah}, Zvonimir and {Schmidt}, Fabian},
        title = "{Effective field theory of intrinsic alignments at one loop order: a comparison to dark matter simulations}",
      journal = {JCAP},
         year = 2023,
        month = oct,
       volume = {10},
       number = {10},
          eid = {005},
        pages = {005},
          doi = {10.1088/1475-7516/2023/10/005},
archivePrefix = {arXiv},
       eprint = {2303.15565},
 primaryClass = {astro-ph.CO},
       adsurl = {https://ui.adsabs.harvard.edu/abs/2023JCAP...10..005B}
}

@ARTICLE{Hamana2020,
       author = {{Hamana}, Takashi and {Shirasaki}, Masato and {Miyazaki}, Satoshi and {Hikage}, Chiaki and {Oguri}, Masamune and {More}, Surhud and {Armstrong}, Robert and {Leauthaud}, Alexie and {Mandelbaum}, Rachel and {Miyatake}, Hironao and {Nishizawa}, Atsushi J. and {Simet}, Melanie and {Takada}, Masahiro and {Aihara}, Hiroaki and {Bosch}, James and {Komiyama}, Yutaka and {Lupton}, Robert and {Murayama}, Hitoshi and {Strauss}, Michael A. and {Tanaka}, Masayuki},
        title = "{Cosmological constraints from cosmic shear two-point correlation functions with HSC survey first-year data}",
      journal = {\pasj},
         year = 2020,
        month = feb,
       volume = {72},
       number = {1},
          eid = {16},
        pages = {16},
          doi = {10.1093/pasj/psz138},
archivePrefix = {arXiv},
       eprint = {1906.06041},
 primaryClass = {astro-ph.CO},
       adsurl = {https://ui.adsabs.harvard.edu/abs/2020PASJ...72...16H}
}

@ARTICLE{Asgari2021,
       author = {{Asgari}, Marika and {Lin}, Chieh-An and {Joachimi}, Benjamin and {Giblin}, Benjamin and {Heymans}, Catherine and {Hildebrandt}, Hendrik and {Kannawadi}, Arun and {St{\"o}lzner}, Benjamin and {Tr{\"o}ster}, Tilman and {van den Busch}, Jan Luca and {Wright}, Angus H. and {Bilicki}, Maciej and {Blake}, Chris and {de Jong}, Jelte and {Dvornik}, Andrej and {Erben}, Thomas and {Getman}, Fedor and {Hoekstra}, Henk and {K{\"o}hlinger}, Fabian and {Kuijken}, Konrad and {Miller}, Lance and {Radovich}, Mario and {Schneider}, Peter and {Shan}, HuanYuan and {Valentijn}, Edwin},
        title = "{KiDS-1000 cosmology: Cosmic shear constraints and comparison between two point statistics}",
      journal = {\aap},
         year = 2021,
        month = jan,
       volume = {645},
          eid = {A104},
        pages = {A104},
          doi = {10.1051/0004-6361/202039070},
archivePrefix = {arXiv},
       eprint = {2007.15633},
 primaryClass = {astro-ph.CO},
       adsurl = {https://ui.adsabs.harvard.edu/abs/2021A&A...645A.104A}
}

@ARTICLE{Sheldon2004,
       author = {{Sheldon}, Erin S. and {Johnston}, David E. and {Frieman}, Joshua A. and {Scranton}, Ryan and {McKay}, Timothy A. and {Connolly}, A.~J. and {Budav{\'a}ri}, Tam{\'a}s and {Zehavi}, Idit and {Bahcall}, Neta A. and {Brinkmann}, J. and {Fukugita}, Masataka},
        title = "{The Galaxy-Mass Correlation Function Measured from Weak Lensing in the Sloan Digital Sky Survey}",
      journal = {\aj},
         year = 2004,
        month = may,
       volume = {127},
       number = {5},
        pages = {2544-2564},
          doi = {10.1086/383293},
archivePrefix = {arXiv},
       eprint = {astro-ph/0312036},
 primaryClass = {astro-ph},
       adsurl = {https://ui.adsabs.harvard.edu/abs/2004AJ....127.2544S}
}

@ARTICLE{Baldauf2010,
       author = {{Baldauf}, Tobias and {Smith}, Robert E. and {Seljak}, Uro{\v{s}} and {Mandelbaum}, Rachel},
        title = "{Algorithm for the direct reconstruction of the dark matter correlation function from weak lensing and galaxy clustering}",
      journal = {\prd},
         year = 2010,
        month = mar,
       volume = {81},
       number = {6},
          eid = {063531},
        pages = {063531},
          doi = {10.1103/PhysRevD.81.063531},
archivePrefix = {arXiv},
       eprint = {0911.4973},
 primaryClass = {astro-ph.CO},
       adsurl = {https://ui.adsabs.harvard.edu/abs/2010PhRvD..81f3531B}
}

@ARTICLE{Mandelbaum2013,
       author = {{Mandelbaum}, Rachel and {Slosar}, An{\v{z}}e and {Baldauf}, Tobias and {Seljak}, Uro{\v{s}} and {Hirata}, Christopher M. and {Nakajima}, Reiko and {Reyes}, Reinabelle and {Smith}, Robert E.},
        title = "{Cosmological parameter constraints from galaxy-galaxy lensing and galaxy clustering with the SDSS DR7}",
      journal = {\mnras},
         year = 2013,
        month = jun,
       volume = {432},
       number = {2},
        pages = {1544-1575},
          doi = {10.1093/mnras/stt572},
archivePrefix = {arXiv},
       eprint = {1207.1120},
 primaryClass = {astro-ph.CO},
       adsurl = {https://ui.adsabs.harvard.edu/abs/2013MNRAS.432.1544M}
}

@ARTICLE{Prat2018,
       author = {{Prat}, J. and {S{\'a}nchez}, C. and {Fang}, Y. and {Gruen}, D. and {Elvin-Poole}, J. and {Kokron}, N. and {Secco}, L.~F. and {Jain}, B. and {Miquel}, R. and {MacCrann}, N. and {Troxel}, M.~A. and {Alarcon}, A. and {Bacon}, D. and {Bernstein}, G.~M. and {Blazek}, J. and {Cawthon}, R. and {Chang}, C. and {Crocce}, M. and {Davis}, C. and {De Vicente}, J. and {Dietrich}, J.~P. and {Drlica-Wagner}, A. and {Friedrich}, O. and {Gatti}, M. and {Hartley}, W.~G. and {Hoyle}, B. and {Huff}, E.~M. and {Jarvis}, M. and {Rau}, M.~M. and {Rollins}, R.~P. and {Ross}, A.~J. and {Rozo}, E. and {Rykoff}, E.~S. and {Samuroff}, S. and {Sheldon}, E. and {Varga}, T.~N. and {Vielzeuf}, P. and {Zuntz}, J. and {Abbott}, T.~M.~C. and {Abdalla}, F.~B. and {Allam}, S. and {Annis}, J. and {Bechtol}, K. and {Benoit-L{\'e}vy}, A. and {Bertin}, E. and {Brooks}, D. and {Buckley-Geer}, E. and {Burke}, D.~L. and {Carnero Rosell}, A. and {Carrasco Kind}, M. and {Carretero}, J. and {Castander}, F.~J. and {Cunha}, C.~E. and {D'Andrea}, C.~B. and {da Costa}, L.~N. and {Desai}, S. and {Diehl}, H.~T. and {Dodelson}, S. and {Eifler}, T.~F. and {Fernandez}, E. and {Flaugher}, B. and {Fosalba}, P. and {Frieman}, J. and {Garc{\'\i}a-Bellido}, J. and {Gaztanaga}, E. and {Gerdes}, D.~W. and {Giannantonio}, T. and {Goldstein}, D.~A. and {Gruendl}, R.~A. and {Gschwend}, J. and {Gutierrez}, G. and {Honscheid}, K. and {James}, D.~J. and {Jeltema}, T. and {Johnson}, M.~W.~G. and {Johnson}, M.~D. and {Kirk}, D. and {Krause}, E. and {Kuehn}, K. and {Kuhlmann}, S. and {Lahav}, O. and {Li}, T.~S. and {Lima}, M. and {Maia}, M.~A.~G. and {March}, M. and {Marshall}, J.~L. and {Martini}, P. and {Melchior}, P. and {Menanteau}, F. and {Mohr}, J.~J. and {Nichol}, R.~C. and {Nord}, B. and {Plazas}, A.~A. and {Romer}, A.~K. and {Roodman}, A. and {Sako}, M. and {Sanchez}, E. and {Scarpine}, V. and {Schindler}, R. and {Schubnell}, M. and {Sevilla-Noarbe}, I. and {Smith}, M. and {Smith}, R.~C. and {Soares-Santos}, M. and {Sobreira}, F. and {Suchyta}, E. and {Swanson}, M.~E.~C. and {Tarle}, G. and {Thomas}, D. and {Tucker}, D.~L. and {Vikram}, V. and {Walker}, A.~R. and {Wechsler}, R.~H. and {Yanny}, B. and {Zhang}, Y. and {DES Collaboration}},
        title = "{Dark Energy Survey year 1 results: Galaxy-galaxy lensing}",
      journal = {\prd},
         year = 2018,
        month = aug,
       volume = {98},
       number = {4},
          eid = {042005},
        pages = {042005},
          doi = {10.1103/PhysRevD.98.042005},
archivePrefix = {arXiv},
       eprint = {1708.01537},
 primaryClass = {astro-ph.CO},
       adsurl = {https://ui.adsabs.harvard.edu/abs/2018PhRvD..98d2005P}
}

@ARTICLE{Blake2020,
       author = {{Blake}, Chris and {Amon}, Alexandra and {Asgari}, Marika and {Bilicki}, Maciej and {Dvornik}, Andrej and {Erben}, Thomas and {Giblin}, Benjamin and {Glazebrook}, Karl and {Heymans}, Catherine and {Hildebrandt}, Hendrik and {Joachimi}, Benjamin and {Joudaki}, Shahab and {Kannawadi}, Arun and {Kuijken}, Konrad and {Lidman}, Chris and {Parkinson}, David and {Shan}, HuanYuan and {Tr{\"o}ster}, Tilman and {van den Busch}, Jan Luca and {Wolf}, Christian and {Wright}, Angus H.},
        title = "{Testing gravity using galaxy-galaxy lensing and clustering amplitudes in KiDS-1000, BOSS, and 2dFLenS}",
      journal = {\aap},
         year = 2020,
        month = oct,
       volume = {642},
          eid = {A158},
        pages = {A158},
          doi = {10.1051/0004-6361/202038505},
archivePrefix = {arXiv},
       eprint = {2005.14351},
 primaryClass = {astro-ph.CO},
       adsurl = {https://ui.adsabs.harvard.edu/abs/2020A&A...642A.158B}
}

@ARTICLE{Pandey2022,
       author = {{Pandey}, S. and {Krause}, E. and {DeRose}, J. and {MacCrann}, N. and {Jain}, B. and {Crocce}, M. and {Blazek}, J. and {Choi}, A. and {Huang}, H. and {To}, C. and {Fang}, X. and {Elvin-Poole}, J. and {Prat}, J. and {Porredon}, A. and {Secco}, L.~F. and {Rodriguez-Monroy}, M. and {Weaverdyck}, N. and {Park}, Y. and {Raveri}, M. and {Rozo}, E. and {Rykoff}, E.~S. and {Bernstein}, G.~M. and {S{\'a}nchez}, C. and {Jarvis}, M. and {Troxel}, M.~A. and {Zacharegkas}, G. and {Chang}, C. and {Alarcon}, A. and {Alves}, O. and {Amon}, A. and {Andrade-Oliveira}, F. and {Baxter}, E. and {Bechtol}, K. and {Becker}, M.~R. and {Camacho}, H. and {Campos}, A. and {Carnero Rosell}, A. and {Carrasco Kind}, M. and {Cawthon}, R. and {Chen}, R. and {Chintalapati}, P. and {Davis}, C. and {Di Valentino}, E. and {Diehl}, H.~T. and {Dodelson}, S. and {Doux}, C. and {Drlica-Wagner}, A. and {Eckert}, K. and {Eifler}, T.~F. and {Elsner}, F. and {Everett}, S. and {Farahi}, A. and {Fert{\'e}}, A. and {Fosalba}, P. and {Friedrich}, O. and {Gatti}, M. and {Giannini}, G. and {Gruen}, D. and {Gruendl}, R.~A. and {Harrison}, I. and {Hartley}, W.~G. and {Huff}, E.~M. and {Huterer}, D. and {Kovacs}, A. and {Leget}, P.~F. and {McCullough}, J. and {Muir}, J. and {Myles}, J. and {Navarro-Alsina}, A. and {Omori}, Y. and {Rollins}, R.~P. and {Roodman}, A. and {Rosenfeld}, R. and {Sevilla-Noarbe}, I. and {Sheldon}, E. and {Shin}, T. and {Troja}, A. and {Tutusaus}, I. and {Varga}, T.~N. and {Wechsler}, R.~H. and {Yanny}, B. and {Yin}, B. and {Zhang}, Y. and {Zuntz}, J. and {Abbott}, T.~M.~C. and {Aguena}, M. and {Allam}, S. and {Annis}, J. and {Bacon}, D. and {Bertin}, E. and {Brooks}, D. and {Burke}, D.~L. and {Carretero}, J. and {Conselice}, C. and {Costanzi}, M. and {da Costa}, L.~N. and {Pereira}, M.~E.~S. and {De Vicente}, J. and {Dietrich}, J.~P. and {Doel}, P. and {Evrard}, A.~E. and {Ferrero}, I. and {Flaugher}, B. and {Frieman}, J. and {Garc{\'\i}a-Bellido}, J. and {Gaztanaga}, E. and {Gerdes}, D.~W. and {Giannantonio}, T. and {Gschwend}, J. and {Gutierrez}, G. and {Hinton}, S.~R. and {Hollowood}, D.~L. and {Honscheid}, K. and {James}, D.~J. and {Jeltema}, T. and {Kuehn}, K. and {Kuropatkin}, N. and {Lahav}, O. and {Lima}, M. and {Lin}, H. and {Maia}, M.~A.~G. and {Marshall}, J.~L. and {Melchior}, P. and {Menanteau}, F. and {Miller}, C.~J. and {Miquel}, R. and {Mohr}, J.~J. and {Morgan}, R. and {Palmese}, A. and {Paz-Chinch{\'o}n}, F. and {Petravick}, D. and {Pieres}, A. and {Plazas Malag{\'o}n}, A.~A. and {Sanchez}, E. and {Scarpine}, V. and {Serrano}, S. and {Smith}, M. and {Soares-Santos}, M. and {Suchyta}, E. and {Tarle}, G. and {Thomas}, D. and {Weller}, J. and {DES Collaboration}},
        title = "{Dark Energy Survey year 3 results: Constraints on cosmological parameters and galaxy-bias models from galaxy clustering and galaxy-galaxy lensing using the redMaGiC sample}",
      journal = {\prd},
         year = 2022,
        month = aug,
       volume = {106},
       number = {4},
          eid = {043520},
        pages = {043520},
          doi = {10.1103/PhysRevD.106.043520},
archivePrefix = {arXiv},
       eprint = {2105.13545},
 primaryClass = {astro-ph.CO},
       adsurl = {https://ui.adsabs.harvard.edu/abs/2022PhRvD.106d3520P}
}

@ARTICLE{Kannadawi2019,
       author = {{Kannawadi}, Arun and {Hoekstra}, Henk and {Miller}, Lance and {Viola}, Massimo and {Fenech Conti}, Ian and {Herbonnet}, Ricardo and {Erben}, Thomas and {Heymans}, Catherine and {Hildebrandt}, Hendrik and {Kuijken}, Konrad and {Vakili}, Mohammadjavad and {Wright}, Angus H.},
        title = "{Towards emulating cosmic shear data: revisiting the calibration of the shear measurements for the Kilo-Degree Survey}",
      journal = {\aap},
         year = 2019,
        month = apr,
       volume = {624},
          eid = {A92},
        pages = {A92},
          doi = {10.1051/0004-6361/201834819},
archivePrefix = {arXiv},
       eprint = {1812.03983},
 primaryClass = {astro-ph.CO},
       adsurl = {https://ui.adsabs.harvard.edu/abs/2019A&A...624A..92K}
}

@ARTICLE{Mandelbaum2018,
       author = {{Mandelbaum}, Rachel},
        title = "{Weak Lensing for Precision Cosmology}",
      journal = {\araa},
         year = 2018,
        month = sep,
       volume = {56},
        pages = {393-433},
          doi = {10.1146/annurev-astro-081817-051928},
archivePrefix = {arXiv},
       eprint = {1710.03235},
 primaryClass = {astro-ph.CO},
       adsurl = {https://ui.adsabs.harvard.edu/abs/2018ARA&A..56..393M}
}

@ARTICLE{Georgiou2019b,
       author = {{Georgiou}, Christos and {Chisari}, Nora Elisa and {Fortuna}, Maria Cristina and {Hoekstra}, Henk and {Kuijken}, Konrad and {Joachimi}, Benjamin and {Vakili}, Mohammadjavad and {Bilicki}, Maciej and {Dvornik}, Andrej and {Erben}, Thomas and {Giblin}, Benjamin and {Heymans}, Catherine and {Napolitano}, Nicola R. and {Shan}, HuanYuan},
        title = "{GAMA+KiDS: Alignment of galaxies in galaxy groups and its dependence on galaxy scale}",
      journal = {\aap},
         year = 2019,
        month = aug,
       volume = {628},
          eid = {A31},
        pages = {A31},
          doi = {10.1051/0004-6361/201935810},
archivePrefix = {arXiv},
       eprint = {1905.00370},
 primaryClass = {astro-ph.CO},
       adsurl = {https://ui.adsabs.harvard.edu/abs/2019A&A...628A..31G}
}

@ARTICLE{Krause2017,
       author = {{Krause}, Elisabeth and {Eifler}, Tim},
        title = "{cosmolike - cosmological likelihood analyses for photometric galaxy surveys}",
      journal = {\mnras},
         year = 2017,
        month = sep,
       volume = {470},
       number = {2},
        pages = {2100-2112},
          doi = {10.1093/mnras/stx1261},
archivePrefix = {arXiv},
       eprint = {1601.05779},
 primaryClass = {astro-ph.CO},
       adsurl = {https://ui.adsabs.harvard.edu/abs/2017MNRAS.470.2100K}
}

@ARTICLE{Delgado2023,
       author = {{Delgado}, Ana Maria and {Hadzhiyska}, Boryana and {Bose}, Sownak and {Springel}, Volker and {Hernquist}, Lars and {Barrera}, Monica and {Pakmor}, R{\"u}diger and {Ferlito}, Fulvio and {Kannan}, Rahul and {Hern{\'a}ndez-Aguayo}, C{\'e}sar and {White}, Simon D.~M. and {Frenk}, Carlos},
        title = "{The MillenniumTNG project: intrinsic alignments of galaxies and haloes}",
      journal = {\mnras},
         year = 2023,
        month = aug,
       volume = {523},
       number = {4},
        pages = {5899-5914},
          doi = {10.1093/mnras/stad1781},
archivePrefix = {arXiv},
       eprint = {2304.12346},
 primaryClass = {astro-ph.CO},
       adsurl = {https://ui.adsabs.harvard.edu/abs/2023MNRAS.523.5899D}
}

@ARTICLE{Hernandez2023,
       author = {{Hern{\'a}ndez-Aguayo}, C{\'e}sar and {Springel}, Volker and {Pakmor}, R{\"u}diger and {Barrera}, Monica and {Ferlito}, Fulvio and {White}, Simon D.~M. and {Hernquist}, Lars and {Hadzhiyska}, Boryana and {Delgado}, Ana Maria and {Kannan}, Rahul and {Bose}, Sownak and {Frenk}, Carlos},
        title = "{The MillenniumTNG Project: high-precision predictions for matter clustering and halo statistics}",
      journal = {\mnras},
         year = 2023,
        month = sep,
       volume = {524},
       number = {2},
        pages = {2556-2578},
          doi = {10.1093/mnras/stad1657},
archivePrefix = {arXiv},
       eprint = {2210.10059},
 primaryClass = {astro-ph.CO},
       adsurl = {https://ui.adsabs.harvard.edu/abs/2023MNRAS.524.2556H}
}

@ARTICLE{Zjupa2022,
       author = {{Zjupa}, Jolanta and {Sch{\"a}fer}, Bj{\"o}rn Malte and {Hahn}, Oliver},
        title = "{Intrinsic alignments in IllustrisTNG and their implications for weak lensing: Tidal shearing and tidal torquing mechanisms put to the test}",
      journal = {\mnras},
         year = 2022,
        month = aug,
       volume = {514},
       number = {2},
        pages = {2049-2072},
          doi = {10.1093/mnras/stac042},
       adsurl = {https://ui.adsabs.harvard.edu/abs/2022MNRAS.514.2049Z}
}

@ARTICLE{TNG2018,
       author = {{Pillepich}, Annalisa and {Springel}, Volker and {Nelson}, Dylan and {Genel}, Shy and {Naiman}, Jill and {Pakmor}, R{\"u}diger and {Hernquist}, Lars and {Torrey}, Paul and {Vogelsberger}, Mark and {Weinberger}, Rainer and {Marinacci}, Federico},
        title = "{Simulating galaxy formation with the IllustrisTNG model}",
      journal = {\mnras},
         year = 2018,
        month = jan,
       volume = {473},
       number = {3},
        pages = {4077-4106},
          doi = {10.1093/mnras/stx2656},
archivePrefix = {arXiv},
       eprint = {1703.02970},
 primaryClass = {astro-ph.GA},
       adsurl = {https://ui.adsabs.harvard.edu/abs/2018MNRAS.473.4077P}
}

@ARTICLE{MASSIVE2015,
       author = {{Khandai}, Nishikanta and {Di Matteo}, Tiziana and {Croft}, Rupert and {Wilkins}, Stephen and {Feng}, Yu and {Tucker}, Evan and {DeGraf}, Colin and {Liu}, Mao-Sheng},
        title = "{The MassiveBlack-II simulation: the evolution of haloes and galaxies to z {\ensuremath{\sim}} 0}",
      journal = {\mnras},
         year = 2015,
        month = jun,
       volume = {450},
       number = {2},
        pages = {1349-1374},
          doi = {10.1093/mnras/stv627},
archivePrefix = {arXiv},
       eprint = {1402.0888},
 primaryClass = {astro-ph.CO},
       adsurl = {https://ui.adsabs.harvard.edu/abs/2015MNRAS.450.1349K}
}

@ARTICLE{Yao2020,
       author = {{Yao}, Ji and {Shan}, Huanyuan and {Zhang}, Pengjie and {Kneib}, Jean-Paul and {Jullo}, Eric},
        title = "{Unveiling the Intrinsic Alignment of Galaxies with Self-calibration and DECaLS DR3 Data}",
      journal = {\apj},
         year = 2020,
        month = dec,
       volume = {904},
       number = {2},
          eid = {135},
        pages = {135},
          doi = {10.3847/1538-4357/abc175},
archivePrefix = {arXiv},
       eprint = {2002.09826},
 primaryClass = {astro-ph.CO},
       adsurl = {https://ui.adsabs.harvard.edu/abs/2020ApJ...904..135Y}
}

@ARTICLE{Huang2019,
       author = {{Huang}, Hung-Jin and {Eifler}, Tim and {Mandelbaum}, Rachel and {Dodelson}, Scott},
        title = "{Modelling baryonic physics in future weak lensing surveys}",
      journal = {\mnras},
         year = 2019,
        month = sep,
       volume = {488},
       number = {2},
        pages = {1652-1678},
          doi = {10.1093/mnras/stz1714},
archivePrefix = {arXiv},
       eprint = {1809.01146},
 primaryClass = {astro-ph.CO},
       adsurl = {https://ui.adsabs.harvard.edu/abs/2019MNRAS.488.1652H}
}

@ARTICLE{Leonard2024,
       author = {{Leonard}, C. Danielle and {Rau}, Markus Michael and {Mandelbaum}, Rachel},
        title = "{Photometric redshifts and intrinsic alignments: Degeneracies and biases in the 3 {\texttimes}2 pt analysis}",
      journal = {\prd},
         year = 2024,
        month = apr,
       volume = {109},
       number = {8},
          eid = {083528},
        pages = {083528},
          doi = {10.1103/PhysRevD.109.083528},
archivePrefix = {arXiv},
       eprint = {2401.06060},
 primaryClass = {astro-ph.CO},
       adsurl = {https://ui.adsabs.harvard.edu/abs/2024PhRvD.109h3528L}
}

@ARTICLE{Macmahon2024,
       author = {{MacMahon-Gell{\'e}r}, Charlie and {Leonard}, C. Danielle},
        title = "{Intrinsic alignment from multiple shear estimates: a first application to data and forecasts for stage IV}",
      journal = {\mnras},
         year = 2024,
        month = feb,
       volume = {528},
       number = {2},
        pages = {2980-2999},
          doi = {10.1093/mnras/stae054},
archivePrefix = {arXiv},
       eprint = {2306.11428},
 primaryClass = {astro-ph.CO},
       adsurl = {https://ui.adsabs.harvard.edu/abs/2024MNRAS.528.2980M}
}

@ARTICLE{Cosmos2016,
       author = {{Laigle}, C. and {McCracken}, H.~J. and {Ilbert}, O. and {Hsieh}, B.~C. and {Davidzon}, I. and {Capak}, P. and {Hasinger}, G. and {Silverman}, J.~D. and {Pichon}, C. and {Coupon}, J. and {Aussel}, H. and {Le Borgne}, D. and {Caputi}, K. and {Cassata}, P. and {Chang}, Y. -Y. and {Civano}, F. and {Dunlop}, J. and {Fynbo}, J. and {Kartaltepe}, J.~S. and {Koekemoer}, A. and {Le F{\`e}vre}, O. and {Le Floc'h}, E. and {Leauthaud}, A. and {Lilly}, S. and {Lin}, L. and {Marchesi}, S. and {Milvang-Jensen}, B. and {Salvato}, M. and {Sanders}, D.~B. and {Scoville}, N. and {Smolcic}, V. and {Stockmann}, M. and {Taniguchi}, Y. and {Tasca}, L. and {Toft}, S. and {Vaccari}, Mattia and {Zabl}, J.},
        title = "{The COSMOS2015 Catalog: Exploring the 1 < z < 6 Universe with Half a Million Galaxies}",
      journal = {\apjs},
         year = 2016,
        month = jun,
       volume = {224},
       number = {2},
          eid = {24},
        pages = {24},
          doi = {10.3847/0067-0049/224/2/24},
archivePrefix = {arXiv},
       eprint = {1604.02350},
 primaryClass = {astro-ph.GA},
       adsurl = {https://ui.adsabs.harvard.edu/abs/2016ApJS..224...24L}
}

@article{nautilus,
    author = {Lange, Johannes U},
    title = "{nautilus: boosting Bayesian importance nested sampling with deep learning}",
    journal = {MNRAS},
    volume = {525},
    number = {2},
    pages = {3181-3194},
    year = {2023},
    month = {08},
    doi = {10.1093/mnras/stad2441},
    url = {https://doi.org/10.1093/mnras/stad2441},
    eprint = {https://academic.oup.com/mnras/article-pdf/525/2/3181/51331635/stad2441.pdf},
}

@ARTICLE{desi2016,
       author = {{DESI Collaboration: Aghamousa}, A. and al},
        title = "{The DESI Experiment Part I: Science,Targeting, and Survey Design}",
         year = 2016,
        month = oct,
          eid = {arXiv:1611.00036},
        pages = {arXiv:1611.00036},
          doi = {10.48550/arXiv.1611.00036},
       eprint = {1611.00036},
 primaryClass = {astro-ph.IM},
       adsurl = {https://ui.adsabs.harvard.edu/abs/2016arXiv161100036D}
}

@ARTICLE{Gama2011,
       author = {{Driver}, S.~P. and {Hill}, D.~T. and {Kelvin}, L.~S. and {Robotham}, A.~S.~G. and {Liske}, J. and {Norberg}, P. and {Baldry}, I.~K. and {Bamford}, S.~P. and {Hopkins}, A.~M. and {Loveday}, J. and {Peacock}, J.~A. and {Andrae}, E. and {Bland-Hawthorn}, J. and {Brough}, S. and {Brown}, M.~J.~I. and {Cameron}, E. and {Ching}, J.~H.~Y. and {Colless}, M. and {Conselice}, C.~J. and {Croom}, S.~M. and {Cross}, N.~J.~G. and {de Propris}, R. and {Dye}, S. and {Drinkwater}, M.~J. and {Ellis}, S. and {Graham}, Alister W. and {Grootes}, M.~W. and {Gunawardhana}, M. and {Jones}, D.~H. and {van Kampen}, E. and {Maraston}, C. and {Nichol}, R.~C. and {Parkinson}, H.~R. and {Phillipps}, S. and {Pimbblet}, K. and {Popescu}, C.~C. and {Prescott}, M. and {Roseboom}, I.~G. and {Sadler}, E.~M. and {Sansom}, A.~E. and {Sharp}, R.~G. and {Smith}, D.~J.~B. and {Taylor}, E. and {Thomas}, D. and {Tuffs}, R.~J. and {Wijesinghe}, D. and {Dunne}, L. and {Frenk}, C.~S. and {Jarvis}, M.~J. and {Madore}, B.~F. and {Meyer}, M.~J. and {Seibert}, M. and {Staveley-Smith}, L. and {Sutherland}, W.~J. and {Warren}, S.~J.},
        title = "{Galaxy and Mass Assembly (GAMA): survey diagnostics and core data release}",
      journal = {\mnras},
         year = 2011,
        month = may,
       volume = {413},
       number = {2},
        pages = {971-995},
          doi = {10.1111/j.1365-2966.2010.18188.x},
archivePrefix = {arXiv},
       eprint = {1009.0614},
 primaryClass = {astro-ph.CO},
       adsurl = {https://ui.adsabs.harvard.edu/abs/2011MNRAS.413..971D}
}

@ARTICLE{York2000,
       author = {{York}, Donald G. and {Adelman}, J. and {Anderson}, John E., Jr. and {Anderson}, Scott F. and {Annis}, James and {Bahcall}, Neta A. and {Bakken}, J.~A. and {Barkhouser}, Robert and {Bastian}, Steven and {Berman}, Eileen and {Boroski}, William N. and {Bracker}, Steve and {Briegel}, Charlie and {Briggs}, John W. and {Brinkmann}, J. and {Brunner}, Robert and {Burles}, Scott and {Carey}, Larry and {Carr}, Michael A. and {Castander}, Francisco J. and {Chen}, Bing and {Colestock}, Patrick L. and {Connolly}, A.~J. and {Crocker}, J.~H. and {Csabai}, Istv{\'a}n and {Czarapata}, Paul C. and {Davis}, John Eric and {Doi}, Mamoru and {Dombeck}, Tom and {Eisenstein}, Daniel and {Ellman}, Nancy and {Elms}, Brian R. and {Evans}, Michael L. and {Fan}, Xiaohui and {Federwitz}, Glenn R. and {Fiscelli}, Larry and {Friedman}, Scott and {Frieman}, Joshua A. and {Fukugita}, Masataka and {Gillespie}, Bruce and {Gunn}, James E. and {Gurbani}, Vijay K. and {de Haas}, Ernst and {Haldeman}, Merle and {Harris}, Frederick H. and {Hayes}, J. and {Heckman}, Timothy M. and {Hennessy}, G.~S. and {Hindsley}, Robert B. and {Holm}, Scott and {Holmgren}, Donald J. and {Huang}, Chi-hao and {Hull}, Charles and {Husby}, Don and {Ichikawa}, Shin-Ichi and {Ichikawa}, Takashi and {Ivezi{\'c}}, {\v{Z}}eljko and {Kent}, Stephen and {Kim}, Rita S.~J. and {Kinney}, E. and {Klaene}, Mark and {Kleinman}, A.~N. and {Kleinman}, S. and {Knapp}, G.~R. and {Korienek}, John and {Kron}, Richard G. and {Kunszt}, Peter Z. and {Lamb}, D.~Q. and {Lee}, B. and {Leger}, R. French and {Limmongkol}, Siriluk and {Lindenmeyer}, Carl and {Long}, Daniel C. and {Loomis}, Craig and {Loveday}, Jon and {Lucinio}, Rich and {Lupton}, Robert H. and {MacKinnon}, Bryan and {Mannery}, Edward J. and {Mantsch}, P.~M. and {Margon}, Bruce and {McGehee}, Peregrine and {McKay}, Timothy A. and {Meiksin}, Avery and {Merelli}, Aronne and {Monet}, David G. and {Munn}, Jeffrey A. and {Narayanan}, Vijay K. and {Nash}, Thomas and {Neilsen}, Eric and {Neswold}, Rich and {Newberg}, Heidi Jo and {Nichol}, R.~C. and {Nicinski}, Tom and {Nonino}, Mario and {Okada}, Norio and {Okamura}, Sadanori and {Ostriker}, Jeremiah P. and {Owen}, Russell and {Pauls}, A. George and {Peoples}, John and {Peterson}, R.~L. and {Petravick}, Donald and {Pier}, Jeffrey R. and {Pope}, Adrian and {Pordes}, Ruth and {Prosapio}, Angela and {Rechenmacher}, Ron and {Quinn}, Thomas R. and {Richards}, Gordon T. and {Richmond}, Michael W. and {Rivetta}, Claudio H. and {Rockosi}, Constance M. and {Ruthmansdorfer}, Kurt and {Sandford}, Dale and {Schlegel}, David J. and {Schneider}, Donald P. and {Sekiguchi}, Maki and {Sergey}, Gary and {Shimasaku}, Kazuhiro and {Siegmund}, Walter A. and {Smee}, Stephen and {Smith}, J. Allyn and {Snedden}, S. and {Stone}, R. and {Stoughton}, Chris and {Strauss}, Michael A. and {Stubbs}, Christopher and {SubbaRao}, Mark and {Szalay}, Alexander S. and {Szapudi}, Istvan and {Szokoly}, Gyula P. and {Thakar}, Anirudda R. and {Tremonti}, Christy and {Tucker}, Douglas L. and {Uomoto}, Alan and {Vanden Berk}, Dan and {Vogeley}, Michael S. and {Waddell}, Patrick and {Wang}, Shu-i. and {Watanabe}, Masaru and {Weinberg}, David H. and {Yanny}, Brian and {Yasuda}, Naoki and {SDSS Collaboration}},
        title = "{The Sloan Digital Sky Survey: Technical Summary}",
      journal = {\aj},
         year = 2000,
        month = sep,
       volume = {120},
       number = {3},
        pages = {1579-1587},
          doi = {10.1086/301513},
archivePrefix = {arXiv},
       eprint = {astro-ph/0006396},
 primaryClass = {astro-ph},
       adsurl = {https://ui.adsabs.harvard.edu/abs/2000AJ....120.1579Y}
}

@ARTICLE{Dawson16,
       author = {{Dawson}, Kyle S. and {Kneib}, Jean-Paul and {Percival}, Will J. and {Alam}, Shadab and {Albareti}, Franco D. and {Anderson}, Scott F. and {Armengaud}, Eric and {Aubourg}, {\'E}ric and {Bailey}, Stephen and {Bautista}, Julian E. and {Berlind}, Andreas A. and {Bershady}, Matthew A. and {Beutler}, Florian and {Bizyaev}, Dmitry and {Blanton}, Michael R. and {Blomqvist}, Michael and {Bolton}, Adam S. and {Bovy}, Jo and {Brandt}, W.~N. and {Brinkmann}, Jon and {Brownstein}, Joel R. and {Burtin}, Etienne and {Busca}, N.~G. and {Cai}, Zheng and {Chuang}, Chia-Hsun and {Clerc}, Nicolas and {Comparat}, Johan and {Cope}, Frances and {Croft}, Rupert A.~C. and {Cruz-Gonzalez}, Irene and {da Costa}, Luiz N. and {Cousinou}, Marie-Claude and {Darling}, Jeremy and {de la Macorra}, Axel and {de la Torre}, Sylvain and {Delubac}, Timoth{\'e}e and {du Mas des Bourboux}, H{\'e}lion and {Dwelly}, Tom and {Ealet}, Anne and {Eisenstein}, Daniel J. and {Eracleous}, Michael and {Escoffier}, S. and {Fan}, Xiaohui and {Finoguenov}, Alexis and {Font-Ribera}, Andreu and {Frinchaboy}, Peter and {Gaulme}, Patrick and {Georgakakis}, Antonis and {Green}, Paul and {Guo}, Hong and {Guy}, Julien and {Ho}, Shirley and {Holder}, Diana and {Huehnerhoff}, Joe and {Hutchinson}, Timothy and {Jing}, Yipeng and {Jullo}, Eric and {Kamble}, Vikrant and {Kinemuchi}, Karen and {Kirkby}, David and {Kitaura}, Francisco-Shu and {Klaene}, Mark A. and {Laher}, Russ R. and {Lang}, Dustin and {Laurent}, Pierre and {Le Goff}, Jean-Marc and {Li}, Cheng and {Liang}, Yu and {Lima}, Marcos and {Lin}, Qiufan and {Lin}, Weipeng and {Lin}, Yen-Ting and {Long}, Daniel C. and {Lundgren}, Britt and {MacDonald}, Nicholas and {Geimba Maia}, Marcio Antonio and {Malanushenko}, Elena and {Malanushenko}, Viktor and {Mariappan}, Vivek and {McBride}, Cameron K. and {McGreer}, Ian D. and {M{\'e}nard}, Brice and {Merloni}, Andrea and {Meza}, Andres and {Montero-Dorta}, Antonio D. and {Muna}, Demitri and {Myers}, Adam D. and {Nandra}, Kirpal and {Naugle}, Tracy and {Newman}, Jeffrey A. and {Noterdaeme}, Pasquier and {Nugent}, Peter and {Ogando}, Ricardo and {Olmstead}, Matthew D. and {Oravetz}, Audrey and {Oravetz}, Daniel J. and {Padmanabhan}, Nikhil and {Palanque-Delabrouille}, Nathalie and {Pan}, Kaike and {Parejko}, John K. and {P{\^a}ris}, Isabelle and {Peacock}, John A. and {Petitjean}, Patrick and {Pieri}, Matthew M. and {Pisani}, Alice and {Prada}, Francisco and {Prakash}, Abhishek and {Raichoor}, Anand and {Reid}, Beth and {Rich}, James and {Ridl}, Jethro and {Rodriguez-Torres}, Sergio and {Carnero Rosell}, Aurelio and {Ross}, Ashley J. and {Rossi}, Graziano and {Ruan}, John and {Salvato}, Mara and {Sayres}, Conor and {Schneider}, Donald P. and {Schlegel}, David J. and {Seljak}, Uros and {Seo}, Hee-Jong and {Sesar}, Branimir and {Shandera}, Sarah and {Shu}, Yiping and {Slosar}, An{\v{z}}e and {Sobreira}, Flavia and {Streblyanska}, Alina and {Suzuki}, Nao and {Taylor}, Donna and {Tao}, Charling and {Tinker}, Jeremy L. and {Tojeiro}, Rita and {Vargas-Maga{\~n}a}, Mariana and {Wang}, Yuting and {Weaver}, Benjamin A. and {Weinberg}, David H. and {White}, Martin and {Wood-Vasey}, W.~M. and {Yeche}, Christophe and {Zhai}, Zhongxu and {Zhao}, Cheng and {Zhao}, Gong-bo and {Zheng}, Zheng and {Ben Zhu}, Guangtun and {Zou}, Hu},
        title = "{The SDSS-IV Extended Baryon Oscillation Spectroscopic Survey: Overview and Early Data}",
      journal = {\aj},
         year = 2016,
        month = feb,
       volume = {151},
       number = {2},
          eid = {44},
        pages = {44},
          doi = {10.3847/0004-6256/151/2/44},
archivePrefix = {arXiv},
       eprint = {1508.04473},
 primaryClass = {astro-ph.CO},
       adsurl = {https://ui.adsabs.harvard.edu/abs/2016AJ....151...44D}
}

@ARTICLE{FABIBI2024,
       author = {{Hervas Peters}, Fabian and {Kilbinger}, Martin and {Paviot}, Romain and {Baumont}, Lucie and {Russier}, Elisa and {Zhang}, Ziwen and {Murray}, Calum and {Pettorino}, Valeria and {de Boer}, Thomas and {Fabbro}, S{\'e}bastien and {Guerrini}, Sacha and {Hildebrandt}, Hendrik and {Hudson}, Mike and {Van Waerbeke}, Ludovic and {Wittje}, Anna},
        title = "{UNIONS: a direct measurement of intrinsic alignment with BOSS/eBOSS spectroscopy}",
      journal = {arXiv},
         year = 2024,
        month = dec,
          eid = {arXiv:2412.01790},
        pages = {arXiv:2412.01790},
archivePrefix = {arXiv},
       eprint = {2412.01790},
 primaryClass = {astro-ph.CO},
       adsurl = {https://ui.adsabs.harvard.edu/abs/2024arXiv241201790H}
}

@ARTICLE{Dvornik2023,
       author = {{Dvornik}, Andrej and {Heymans}, Catherine and {Asgari}, Marika and {Mahony}, Constance and {Joachimi}, Benjamin and {Bilicki}, Maciej and {Chisari}, Elisa and {Hildebrandt}, Hendrik and {Hoekstra}, Henk and {Johnston}, Harry and {Kuijken}, Konrad and {Mead}, Alexander and {Miyatake}, Hironao and {Nishimichi}, Takahiro and {Reischke}, Robert and {Unruh}, Sandra and {Wright}, Angus H.},
        title = "{KiDS-1000: Combined halo-model cosmology constraints from galaxy abundance, galaxy clustering, and galaxy-galaxy lensing}",
      journal = {\aap},
         year = 2023,
        month = jul,
       volume = {675},
          eid = {A189},
        pages = {A189},
          doi = {10.1051/0004-6361/202245158},
archivePrefix = {arXiv},
       eprint = {2210.03110},
 primaryClass = {astro-ph.CO},
       adsurl = {https://ui.adsabs.harvard.edu/abs/2023A&A...675A.189D}
}

@ARTICLE{Fortuna2024,
       author = {{Fortuna}, Maria Cristina and {Dvornik}, Andrej and {Hoekstra}, Henk and {Chisari}, Nora Elisa and {Asgari}, Marika and {Bilicki}, Maciej and {Heymans}, Catherine and {Hildebrandt}, Hendrik and {Kuijken}, Koen and {Wright}, Angus H. and {Yao}, Ji},
        title = "{KiDS-1000: Weak lensing and intrinsic alignment around luminous red galaxies}",
      journal = {\aap},
         year = 2025,
        month = feb,
       volume = {694},
          eid = {A322},
        pages = {A322},
          doi = {10.1051/0004-6361/202452347},
archivePrefix = {arXiv},
       eprint = {2409.15416},
 primaryClass = {astro-ph.CO},
       adsurl = {https://ui.adsabs.harvard.edu/abs/2025A&A...694A.322F}
}

@ARTICLE{Prat2022,
       author = {{Prat}, J. and {Blazek}, J. and {S{\'a}nchez}, C. and {Tutusaus}, I. and {Pandey}, S. and {Elvin-Poole}, J. and {Krause}, E. and {Troxel}, M.~A. and {Secco}, L.~F. and {Amon}, A. and {DeRose}, J. and {Zacharegkas}, G. and {Chang}, C. and {Jain}, B. and {MacCrann}, N. and {Park}, Y. and {Sheldon}, E. and {Giannini}, G. and {Bocquet}, S. and {To}, C. and {Alarcon}, A. and {Alves}, O. and {Andrade-Oliveira}, F. and {Baxter}, E. and {Bechtol}, K. and {Becker}, M.~R. and {Bernstein}, G.~M. and {Camacho}, H. and {Campos}, A. and {Carnero Rosell}, A. and {Carrasco Kind}, M. and {Cawthon}, R. and {Chen}, R. and {Choi}, A. and {Cordero}, J. and {Crocce}, M. and {Davis}, C. and {De Vicente}, J. and {Diehl}, H.~T. and {Dodelson}, S. and {Doux}, C. and {Drlica-Wagner}, A. and {Eckert}, K. and {Eifler}, T.~F. and {Elsner}, F. and {Everett}, S. and {Fang}, X. and {Farahi}, A. and {Fert{\'e}}, A. and {Fosalba}, P. and {Friedrich}, O. and {Gatti}, M. and {Gruen}, D. and {Gruendl}, R.~A. and {Harrison}, I. and {Hartley}, W.~G. and {Herner}, K. and {Huang}, H. and {Huff}, E.~M. and {Huterer}, D. and {Jarvis}, M. and {Kuropatkin}, N. and {Leget}, P. -F. and {Lemos}, P. and {Liddle}, A.~R. and {McCullough}, J. and {Muir}, J. and {Myles}, J. and {Navarro-Alsina}, A. and {Porredon}, A. and {Raveri}, M. and {Rodriguez-Monroy}, M. and {Rollins}, R.~P. and {Roodman}, A. and {Rosenfeld}, R. and {Ross}, A.~J. and {Rykoff}, E.~S. and {Sanchez}, J. and {Sevilla-Noarbe}, I. and {Shin}, T. and {Troja}, A. and {Varga}, T.~N. and {Weaverdyck}, N. and {Wechsler}, R.~H. and {Yanny}, B. and {Yin}, B. and {Zuntz}, J. and {Abbott}, T.~M.~C. and {Aguena}, M. and {Allam}, S. and {Annis}, J. and {Bacon}, D. and {Brooks}, D. and {Burke}, D.~L. and {Carretero}, J. and {Conselice}, C. and {Costanzi}, M. and {da Costa}, L.~N. and {Pereira}, M.~E.~S. and {Desai}, S. and {Dietrich}, J.~P. and {Doel}, P. and {Evrard}, A.~E. and {Ferrero}, I. and {Flaugher}, B. and {Frieman}, J. and {Garc{\'\i}a-Bellido}, J. and {Gaztanaga}, E. and {Gerdes}, D.~W. and {Giannantonio}, T. and {Gschwend}, J. and {Gutierrez}, G. and {Hinton}, S.~R. and {Hollowood}, D.~L. and {Honscheid}, K. and {James}, D.~J. and {Kuehn}, K. and {Lahav}, O. and {Lin}, H. and {Maia}, M.~A.~G. and {Marshall}, J.~L. and {Martini}, P. and {Melchior}, P. and {Menanteau}, F. and {Miller}, C.~J. and {Miquel}, R. and {Mohr}, J.~J. and {Morgan}, R. and {Ogando}, R.~L.~C. and {Palmese}, A. and {Paz-Chinch{\'o}n}, F. and {Petravick}, D. and {Plazas Malag{\'o}n}, A.~A. and {Sanchez}, E. and {Serrano}, S. and {Smith}, M. and {Soares-Santos}, M. and {Suchyta}, E. and {Tarle}, G. and {Thomas}, D. and {Weller}, J. and {DES Collaboration}},
        title = "{Dark energy survey year 3 results: High-precision measurement and modeling of galaxy-galaxy lensing}",
      journal = {\prd},
         year = 2022,
        month = apr,
       volume = {105},
       number = {8},
          eid = {083528},
        pages = {083528},
          doi = {10.1103/PhysRevD.105.083528},
archivePrefix = {arXiv},
       eprint = {2105.13541},
 primaryClass = {astro-ph.CO},
       adsurl = {https://ui.adsabs.harvard.edu/abs/2022PhRvD.105h3528P}
}

@ARTICLE{Vanuitert2018,
       author = {{van Uitert}, Edo and {Joachimi}, Benjamin and {Joudaki}, Shahab and {Amon}, Alexandra and {Heymans}, Catherine and {K{\"o}hlinger}, Fabian and {Asgari}, Marika and {Blake}, Chris and {Choi}, Ami and {Erben}, Thomas and {Farrow}, Daniel J. and {Harnois-D{\'e}raps}, Joachim and {Hildebrandt}, Hendrik and {Hoekstra}, Henk and {Kitching}, Thomas D. and {Klaes}, Dominik and {Kuijken}, Konrad and {Merten}, Julian and {Miller}, Lance and {Nakajima}, Reiko and {Schneider}, Peter and {Valentijn}, Edwin and {Viola}, Massimo},
        title = "{KiDS+GAMA: cosmology constraints from a joint analysis of cosmic shear, galaxy-galaxy lensing, and angular clustering}",
      journal = {\mnras},
         year = 2018,
        month = jun,
       volume = {476},
       number = {4},
        pages = {4662-4689},
          doi = {10.1093/mnras/sty551},
archivePrefix = {arXiv},
       eprint = {1706.05004},
 primaryClass = {astro-ph.CO},
       adsurl = {https://ui.adsabs.harvard.edu/abs/2018MNRAS.476.4662V}
}

@ARTICLE{Porredon2022,
       author = {{Porredon}, A. and {Crocce}, M. and {Elvin-Poole}, J. and {Cawthon}, R. and {Giannini}, G. and {De Vicente}, J. and {Carnero Rosell}, A. and {Ferrero}, I. and {Krause}, E. and {Fang}, X. and {Prat}, J. and {Rodriguez-Monroy}, M. and {Pandey}, S. and {Pocino}, A. and {Castander}, F.~J. and {Choi}, A. and {Amon}, A. and {Tutusaus}, I. and {Dodelson}, S. and {Sevilla-Noarbe}, I. and {Fosalba}, P. and {Gaztanaga}, E. and {Alarcon}, A. and {Alves}, O. and {Andrade-Oliveira}, F. and {Baxter}, E. and {Bechtol}, K. and {Becker}, M.~R. and {Bernstein}, G.~M. and {Blazek}, J. and {Camacho}, H. and {Campos}, A. and {Carrasco Kind}, M. and {Chintalapati}, P. and {Cordero}, J. and {DeRose}, J. and {Di Valentino}, E. and {Doux}, C. and {Eifler}, T.~F. and {Everett}, S. and {Fert{\'e}}, A. and {Friedrich}, O. and {Gatti}, M. and {Gruen}, D. and {Harrison}, I. and {Hartley}, W.~G. and {Herner}, K. and {Huff}, E.~M. and {Huterer}, D. and {Jain}, B. and {Jarvis}, M. and {Lee}, S. and {Lemos}, P. and {MacCrann}, N. and {Mena-Fern{\'a}ndez}, J. and {Muir}, J. and {Myles}, J. and {Park}, Y. and {Raveri}, M. and {Rosenfeld}, R. and {Ross}, A.~J. and {Rykoff}, E.~S. and {Samuroff}, S. and {S{\'a}nchez}, C. and {Sanchez}, E. and {Sanchez}, J. and {Sanchez Cid}, D. and {Scolnic}, D. and {Secco}, L.~F. and {Sheldon}, E. and {Troja}, A. and {Troxel}, M.~A. and {Weaverdyck}, N. and {Yanny}, B. and {Zuntz}, J. and {Abbott}, T.~M.~C. and {Aguena}, M. and {Allam}, S. and {Annis}, J. and {Avila}, S. and {Bacon}, D. and {Bertin}, E. and {Bhargava}, S. and {Brooks}, D. and {Buckley-Geer}, E. and {Burke}, D.~L. and {Carretero}, J. and {Costanzi}, M. and {da Costa}, L.~N. and {Pereira}, M.~E.~S. and {Davis}, T.~M. and {Desai}, S. and {Diehl}, H.~T. and {Dietrich}, J.~P. and {Doel}, P. and {Drlica-Wagner}, A. and {Eckert}, K. and {Evrard}, A.~E. and {Flaugher}, B. and {Frieman}, J. and {Garc{\'\i}a-Bellido}, J. and {Gerdes}, D.~W. and {Giannantonio}, T. and {Gruendl}, R.~A. and {Gschwend}, J. and {Gutierrez}, G. and {Hinton}, S.~R. and {Hollowood}, D.~L. and {Honscheid}, K. and {Hoyle}, B. and {James}, D.~J. and {Kuehn}, K. and {Kuropatkin}, N. and {Lahav}, O. and {Lidman}, C. and {Lima}, M. and {Lin}, H. and {Maia}, M.~A.~G. and {Marshall}, J.~L. and {Martini}, P. and {Melchior}, P. and {Menanteau}, F. and {Miquel}, R. and {Mohr}, J.~J. and {Morgan}, R. and {Ogando}, R.~L.~C. and {Palmese}, A. and {Paz-Chinch{\'o}n}, F. and {Petravick}, D. and {Pieres}, A. and {Plazas Malag{\'o}n}, A.~A. and {Romer}, A.~K. and {Santiago}, B. and {Scarpine}, V. and {Schubnell}, M. and {Serrano}, S. and {Smith}, M. and {Soares-Santos}, M. and {Suchyta}, E. and {Tarle}, G. and {Thomas}, D. and {To}, C. and {Varga}, T.~N. and {Weller}, J. and {DES Collaboration}},
        title = "{Dark Energy Survey Year 3 results: Cosmological constraints from galaxy clustering and galaxy-galaxy lensing using the MAGLIM lens sample}",
      journal = {\prd},
         year = 2022,
        month = nov,
       volume = {106},
       number = {10},
          eid = {103530},
        pages = {103530},
          doi = {10.1103/PhysRevD.106.103530},
archivePrefix = {arXiv},
       eprint = {2105.13546},
 primaryClass = {astro-ph.CO},
       adsurl = {https://ui.adsabs.harvard.edu/abs/2022PhRvD.106j3530P}
}

@ARTICLE{Sugiyama2023,
       author = {{Sugiyama}, Sunao and {Miyatake}, Hironao and {More}, Surhud and {Li}, Xiangchong and {Shirasaki}, Masato and {Takada}, Masahiro and {Kobayashi}, Yosuke and {Takahashi}, Ryuichi and {Nishimichi}, Takahiro and {Nishizawa}, Atsushi J. and {Rau}, Markus M. and {Zhang}, Tianqing and {Dalal}, Roohi and {Mandelbaum}, Rachel and {Strauss}, Michael A. and {Hamana}, Takashi and {Oguri}, Masamune and {Osato}, Ken and {Kannawadi}, Arun and {Hsieh}, Bau-Ching and {Luo}, Wentao and {Armstrong}, Robert and {Bosch}, James and {Komiyama}, Yutaka and {Lupton}, Robert H. and {Lust}, Nate B. and {Miyazaki}, Satoshi and {Murayama}, Hitoshi and {Okura}, Yuki and {Price}, Paul A. and {Tait}, Philip J. and {Tanaka}, Masayuki and {Wang}, Shiang-Yu},
        title = "{Hyper Suprime-Cam Year 3 results: Cosmology from galaxy clustering and weak lensing with HSC and SDSS using the minimal bias model}",
      journal = {\prd},
         year = 2023,
        month = dec,
       volume = {108},
       number = {12},
          eid = {123521},
        pages = {123521},
          doi = {10.1103/PhysRevD.108.123521},
archivePrefix = {arXiv},
       eprint = {2304.00705},
 primaryClass = {astro-ph.CO},
       adsurl = {https://ui.adsabs.harvard.edu/abs/2023PhRvD.108l3521S}
}

@ARTICLE{Phriskee2020,
       author = {{Phriksee}, Anirut and {Jullo}, Eric and {Limousin}, Marceau and {Shan}, HuanYuan and {Finoguenov}, Alexis and {Komonjinda}, Siramas and {Wannawichian}, Suwicha and {Sawangwit}, Utane},
        title = "{Weak lensing analysis of CODEX clusters using dark energy camera legacy survey: mass-richness relation}",
      journal = {\mnras},
         year = 2020,
        month = jan,
       volume = {491},
       number = {2},
        pages = {1643-1655},
          doi = {10.1093/mnras/stz3049},
archivePrefix = {arXiv},
       eprint = {1910.10983},
 primaryClass = {astro-ph.CO},
       adsurl = {https://ui.adsabs.harvard.edu/abs/2020MNRAS.491.1643P}
}

@ARTICLE{White2011,
       author = {{White}, Martin and {Blanton}, M. and {Bolton}, A. and {Schlegel}, D. and {Tinker}, J. and {Berlind}, A. and {da Costa}, L. and {Kazin}, E. and {Lin}, Y. -T. and {Maia}, M. and {McBride}, C.~K. and {Padmanabhan}, N. and {Parejko}, J. and {Percival}, W. and {Prada}, F. and {Ramos}, B. and {Sheldon}, E. and {de Simoni}, F. and {Skibba}, R. and {Thomas}, D. and {Wake}, D. and {Zehavi}, I. and {Zheng}, Z. and {Nichol}, R. and {Schneider}, Donald P. and {Strauss}, Michael A. and {Weaver}, B.~A. and {Weinberg}, David H.},
        title = "{The Clustering of Massive Galaxies at z \raisebox{-0.5ex}\textasciitilde 0.5 from the First Semester of BOSS Data}",
      journal = {\apj},
         year = 2011,
        month = feb,
       volume = {728},
       number = {2},
          eid = {126},
        pages = {126},
          doi = {10.1088/0004-637X/728/2/126},
archivePrefix = {arXiv},
       eprint = {1010.4915},
 primaryClass = {astro-ph.CO},
       adsurl = {https://ui.adsabs.harvard.edu/abs/2011ApJ...728..126W}
}

@ARTICLE{Zhai2017,
       author = {{Zhai}, Zhongxu and {Tinker}, Jeremy L. and {Hahn}, ChangHoon and {Seo}, Hee-Jong and {Blanton}, Michael R. and {Tojeiro}, Rita and {Camacho}, Hugo O. and {Lima}, Marcos and {Carnero Rosell}, Aurelio and {Sobreira}, Flavia and {da Costa}, Luiz N. and {Bautista}, Julian E. and {Brownstein}, Joel R. and {Comparat}, Johan and {Dawson}, Kyle and {Newman}, Jeffrey A. and {Prakash}, Abhishek and {Roman-Lopes}, Alexandre and {Schneider}, Donald P.},
        title = "{The Clustering of Luminous Red Galaxies at z {\ensuremath{\sim}} 0.7 from EBOSS and BOSS Data}",
      journal = {\apj},
         year = 2017,
        month = oct,
       volume = {848},
       number = {2},
          eid = {76},
        pages = {76},
          doi = {10.3847/1538-4357/aa8eee},
archivePrefix = {arXiv},
       eprint = {1607.05383},
 primaryClass = {astro-ph.CO},
       adsurl = {https://ui.adsabs.harvard.edu/abs/2017ApJ...848...76Z}
}

@Article{numpy,
 title         = {Array programming with {NumPy}},
 author        = {Charles R. Harris and K. Jarrod Millman and St{\'{e}}fan J.
                 van der Walt and Ralf Gommers and Pauli Virtanen and David
                 Cournapeau and Eric Wieser and Julian Taylor and Sebastian
                 Berg and Nathaniel J. Smith and Robert Kern and Matti Picus
                 and Stephan Hoyer and Marten H. van Kerkwijk and Matthew
                 Brett and Allan Haldane and Jaime Fern{\'{a}}ndez del
                 R{\'{i}}o and Mark Wiebe and Pearu Peterson and Pierre
                 G{\'{e}}rard-Marchant and Kevin Sheppard and Tyler Reddy and
                 Warren Weckesser and Hameer Abbasi and Christoph Gohlke and
                 Travis E. Oliphant},
 year          = {2020},
 month         = sep,
 journal       = {Nature},
 volume        = {585},
 number        = {7825},
 pages         = {357--362},
 doi           = {10.1038/s41586-020-2649-2},
 publisher     = {Springer Science and Business Media {LLC}},
 url           = {https://doi.org/10.1038/s41586-020-2649-2}
}

@ARTICLE{2020SciPy-NMeth,
  author  = {Virtanen, Pauli and Gommers, Ralf and Oliphant, Travis E. and
            Haberland, Matt and Reddy, Tyler and Cournapeau, David and
            Burovski, Evgeni and Peterson, Pearu and Weckesser, Warren and
            Bright, Jonathan and {van der Walt}, St{\'e}fan J. and
            Brett, Matthew and Wilson, Joshua and Millman, K. Jarrod and
            Mayorov, Nikolay and Nelson, Andrew R. J. and Jones, Eric and
            Kern, Robert and Larson, Eric and Carey, C J and
            Polat, {\.I}lhan and Feng, Yu and Moore, Eric W. and
            {VanderPlas}, Jake and Laxalde, Denis and Perktold, Josef and
            Cimrman, Robert and Henriksen, Ian and Quintero, E. A. and
            Harris, Charles R. and Archibald, Anne M. and
            Ribeiro, Ant{\^o}nio H. and Pedregosa, Fabian and
            {van Mulbregt}, Paul and {SciPy 1.0 Contributors}},
  title   = {{{SciPy} 1.0: Fundamental Algorithms for Scientific
            Computing in Python}},
  journal = {Nature Methods},
  year    = {2020},
  volume  = {17},
  pages   = {261--272},
  adsurl  = {https://rdcu.be/b08Wh},
  doi     = {10.1038/s41592-019-0686-2},
}

@article{pygtc,
  doi = {10.21105/joss.00046},
  url = {http://dx.doi.org/10.21105/joss.00046},
  year  = {2016},
  month = {oct},
  publisher = {The Open Journal},
  volume = {1},
  number = {6},
  author = {Sebastian Bocquet and Faustin W. Carter},
  title = {pygtc: beautiful parameter covariance plots (aka. Giant Triangle Confusograms)},
  journal = {The Journal of Open Source Software}
}

@Article{matplotlib,
  Author    = {Hunter, J. D.},
  Title     = {Matplotlib: A 2D graphics environment},
  Journal   = {Computing in Science \& Engineering},
  Volume    = {9},
  Number    = {3},
  Pages     = {90--95},
  publisher = {IEEE COMPUTER SOC},
  doi       = {10.1109/MCSE.2007.55},
  year      = 2007
}

@ARTICLE{Georgiou2025,
       author = {{Georgiou}, Christos and {Chisari}, Nora Elisa and {Bilicki}, Maciej and {La Barbera}, Francesco and {Napolitano}, Nicola R. and {Roy}, Nivya and {Tortora}, Crescenzo},
        title = "{Intrinsic galaxy alignments in the KiDS-1000 bright sample: Dependence on colour, luminosity, morphology, and galaxy scale}",
      journal = {\aap},
         year = 2025,
        month = jul,
       volume = {699},
          eid = {A252},
        pages = {A252},
          doi = {10.1051/0004-6361/202554134},
archivePrefix = {arXiv},
       eprint = {2502.09452},
 primaryClass = {astro-ph.CO},
       adsurl = {https://ui.adsabs.harvard.edu/abs/2025A&A...699A.252G}
}

@ARTICLE{Siegel2025,
       author = {{Siegel}, J. and {McCullough}, J. and {Amon}, A. and {Lamman}, C. and {Jeffrey}, N. and {Joachimi}, B. and {Hoekstra}, H. and {Heydenreich}, S. and {Ross}, A.~J. and {Aguilar}, J. and {Ahlen}, S. and {Bianchi}, D. and {Blake}, C. and {Brooks}, D. and {Castander}, F.~J. and {Claybaugh}, T. and {de la Macorra}, A. and {DeRose}, J. and {Doel}, P. and {Emas}, N. and {Ferraro}, S. and {Font-Ribera}, A. and {Forero-Romero}, J.~E. and {Gazta{\~n}aga}, E. and {Gontcho}, S. Gontcho A and {Gutierrez}, G. and {Honscheid}, K. and {Ishak}, M. and {Joudaki}, S. and {Kehoe}, R. and {Kirkby}, D. and {Kisner}, T. and {Krolewski}, A. and {Lahav}, O. and {Lambert}, A. and {Landriau}, M. and {Le Guillou}, L. and {Levi}, M.~E. and {Manera}, M. and {Meisner}, A. and {Miquel}, R. and {Moustakas}, J. and {Nadathur}, S. and {Newman}, J.~A. and {Niz}, G. and {Palanque-Delabrouille}, N. and {Percival}, W.~J. and {Porredon}, A. and {Prada}, F. and {P{\'e}rez-R{\`a}fols}, I. and {Rossi}, G. and {Sanchez}, E. and {Saulder}, C. and {Schlegel}, D. and {Schubnell}, M. and {Semenaite}, A. and {Silber}, J. and {Sprayberry}, D. and {Sun}, Z. and {Tarl{\'e}}, G. and {Weaver}, B.~A. and {Zhou}, R. and {Zou}, H.},
        title = "{Intrinsic alignment demographics for next-generation lensing: Revealing galaxy property trends with DESI Y1 direct measurements}",
      journal = {arXiv},
         year = 2025,
        month = jul,
          eid = {arXiv:2507.11530},
        pages = {arXiv:2507.11530},
          doi = {10.48550/arXiv.2507.11530},
}

@ARTICLE{NLA-M,
       author = {{Wright}, Angus H. and {St{\"o}lzner}, Benjamin and {Asgari}, Marika and {Bilicki}, Maciej and {Giblin}, Benjamin and {Heymans}, Catherine and {Hildebrandt}, Hendrik and {Hoekstra}, Henk and {Joachimi}, Benjamin and {Kuijken}, Konrad and {Li}, Shun-Sheng and {Reischke}, Robert and {von Wietersheim-Kramsta}, Maximilian and {Yoon}, Mijin and {Burger}, Pierre and {Chisari}, Nora Elisa and {de Jong}, Jelte and {Dvornik}, Andrej and {Georgiou}, Christos and {Harnois-D{\'e}raps}, Joachim and {Jalan}, Priyanka and {William}, Anjitha John and {Joudaki}, Shahab and {Lesci}, Giorgio Francesco and {Linke}, Laila and {Loureiro}, Arthur and {Mahony}, Constance and {Maturi}, Matteo and {Miller}, Lance and {Moscardini}, Lauro and {Napolitano}, Nicola R. and {Porth}, Lucas and {Radovich}, Mario and {Schneider}, Peter and {Tr{\"o}ster}, Tilman and {Valentijn}, Edwin and {Wittje}, Anna and {Yan}, Ziang and {Zhang}, Yun-Hao},
        title = "{KiDS-Legacy: Cosmological constraints from cosmic shear with the complete Kilo-Degree Survey}",
      journal = {\aap},
         year = 2025,
        month = nov,
       volume = {703},
          eid = {A158},
        pages = {A158},
          doi = {10.1051/0004-6361/202554908},
archivePrefix = {arXiv},
       eprint = {2503.19441},
 primaryClass = {astro-ph.CO},
       adsurl = {https://ui.adsabs.harvard.edu/abs/2025A&A...703A.158W}
}

@ARTICLE{Shi2021,
       author = {{Shi}, Jingjing and {Kurita}, Toshiki and {Takada}, Masahiro and {Osato}, Ken and {Kobayashi}, Yosuke and {Nishimichi}, Takahiro},
        title = "{Power spectrum of intrinsic alignments of galaxies in IllustrisTNG}",
      journal = {JCAP},
         year = 2021,
        month = mar,
       volume = {03},
       number = {03},
          eid = {030},
        pages = {030},
          doi = {10.1088/1475-7516/2021/03/030},
archivePrefix = {arXiv},
       eprint = {2009.00276},
 primaryClass = {astro-ph.GA},
       adsurl = {https://ui.adsabs.harvard.edu/abs/2021JCAP...03..030S}
}

@ARTICLE{Chen2023,
       author = {{Chen}, Shi-Fan and {Kokron}, Nickolas},
        title = "{A Lagrangian theory for galaxy shape statistics}",
      journal = {JCAP},
         year = 2024,
        month = jan,
       volume = {01},
       number = {01},
          eid = {027},
        pages = {027},
          doi = {10.1088/1475-7516/2024/01/027},
archivePrefix = {arXiv},
       eprint = {2309.16761},
 primaryClass = {astro-ph.CO},
       adsurl = {https://ui.adsabs.harvard.edu/abs/2024JCAP...01..027C}
}

@ARTICLE{codis_intrinsic_2015,
       author = {{Codis}, S. and {Gavazzi}, R. and {Dubois}, Y. and {Pichon}, C. and {Benabed}, K. and {Desjacques}, V. and {Pogosyan}, D. and {Devriendt}, J. and {Slyz}, A.},
        title = "{Intrinsic alignment of simulated galaxies in the cosmic web: implications for weak lensing surveys}",
      journal = {\mnras},
         year = 2015,
        month = apr,
       volume = {448},
       number = {4},
        pages = {3391-3404},
          doi = {10.1093/mnras/stv231},
archivePrefix = {arXiv},
       eprint = {1406.4668},
 primaryClass = {astro-ph.CO},
       adsurl = {https://ui.adsabs.harvard.edu/abs/2015MNRAS.448.3391C}
}

@ARTICLE{McDonald2009,
       author = {{McDonald}, Patrick and {Roy}, Arabindo},
        title = "{Clustering of dark matter tracers: generalizing bias for the coming era of precision LSS}",
      journal = {JCAP},
         year = 2009,
        month = aug,
       volume = {8},
       number = {8},
          eid = {020},
        pages = {020},
          doi = {10.1088/1475-7516/2009/08/020},
archivePrefix = {arXiv},
       eprint = {0902.0991},
 primaryClass = {astro-ph.CO},
       adsurl = {https://ui.adsabs.harvard.edu/abs/2009JCAP...08..020M}
}

@ARTICLE{Krause2021,
       author = {{Krause}, E.},
        title = "{Dark Energy Survey Year 3 Results: Multi-Probe Modeling Strategy and Validation}",
      journal = {arXiv e-prints},
         year = 2021,
        month = may,
          eid = {arXiv:2105.13548},
        pages = {arXiv:2105.13548},
          doi = {10.48550/arXiv.2105.13548},
archivePrefix = {arXiv},
       eprint = {2105.13548},
 primaryClass = {astro-ph.CO},
       adsurl = {https://ui.adsabs.harvard.edu/abs/2021arXiv210513548K}
}

@ARTICLE{Takahashi12,
       author = {{Takahashi}, Ryuichi},
        title = "{Revising the Halofit Model for the Nonlinear Matter Power Spectrum}",
      journal = {\apj},
         year = 2012,
        month = dec,
       volume = {761},
       number = {2},
          eid = {152},
        pages = {152},
          doi = {10.1088/0004-637X/761/2/152},
archivePrefix = {arXiv},
       eprint = {1208.2701},
 primaryClass = {astro-ph.CO},
       adsurl = {https://ui.adsabs.harvard.edu/abs/2012ApJ...761..152T}
}

@ARTICLE{Samuroff22,
       author = {{Samuroff}, S. and {Mandelbaum}, R. and {Blazek}, J. and {Campos}, A. and {MacCrann}, N. and {Zacharegkas}, G. and {Amon}, A. and {Prat}, J. and {Singh}, S. and {Elvin-Poole}, J. and {Ross}, A.~J. and {Alarcon}, A. and {Baxter}, E. and {Bechtol}, K. and {Becker}, M.~R. and {Bernstein}, G.~M. and {Rosell}, A. Carnero and {Kind}, M. Carrasco and {Cawthon}, R. and {Chang}, C. and {Chen}, R. and {Choi}, A. and {Crocce}, M. and {Davis}, C. and {DeRose}, J. and {Dodelson}, S. and {Doux}, C. and {Drlica-Wagner}, A. and {Eckert}, K. and {Everett}, S. and {Fert{\'e}}, A. and {Gatti}, M. and {Giannini}, G. and {Gruen}, D. and {Gruendl}, R.~A. and {Harrison}, I. and {Herner}, K. and {Huff}, E.~M. and {Jarvis}, M. and {Kuropatkin}, N. and {Leget}, P. -F. and {Lemos}, P. and {McCullough}, J. and {Myles}, J. and {Navarro-Alsina}, A. and {Pandey}, S. and {Porredon}, A. and {Raveri}, M. and {Rodriguez-Monroy}, M. and {Rollins}, R.~P. and {Roodman}, A. and {Rossi}, G. and {Rykoff}, E.~S. and {S{\'a}nchez}, C. and {Secco}, L.~F. and {Sevilla-Noarbe}, I. and {Sheldon}, E. and {Shin}, T. and {Troxel}, M.~A. and {Tutusaus}, I. and {Weaverdyck}, N. and {Yanny}, B. and {Yin}, B. and {Zhang}, Y. and {Zuntz}, J. and {Aguena}, M. and {Alves}, O. and {Annis}, J. and {Bacon}, D. and {Bertin}, E. and {Bocquet}, S. and {Brooks}, D. and {Burke}, D.~L. and {Carretero}, J. and {Costanzi}, M. and {da Costa}, L.~N. and {Pereira}, M.~E.~S. and {De Vicente}, J. and {Desai}, S. and {Diehl}, H.~T. and {Dietrich}, J.~P. and {Doel}, P. and {Ferrero}, I. and {Flaugher}, B. and {Frieman}, J. and {Garc{\'\i}a-Bellido}, J. and {Hinton}, S.~R. and {Hollowood}, D.~L. and {Honscheid}, K. and {James}, D.~J. and {Kuehn}, K. and {Lahav}, O. and {Marshall}, J.~L. and {Melchior}, P. and {Mena-Fern{\'a}ndez}, J. and {Menanteau}, F. and {Miquel}, R. and {Newman}, J. and {Palmese}, A. and {Pieres}, A. and {Malag{\'o}n}, A.~A. Plazas and {Sanchez}, E. and {Scarpine}, V. and {Smith}, M. and {Suchyta}, E. and {Swanson}, M.~E.~C. and {Tarle}, G. and {To}, C. and {DES Collaboration}},
        title = "{The Dark Energy Survey Year 3 and eBOSS: constraining galaxy intrinsic alignments across luminosity and colour space}",
      journal = {\mnras},
         year = 2023,
        month = sep,
       volume = {524},
       number = {2},
        pages = {2195-2223},
          doi = {10.1093/mnras/stad2013},
archivePrefix = {arXiv},
       eprint = {2212.11319},
 primaryClass = {astro-ph.CO},
       adsurl = {https://ui.adsabs.harvard.edu/abs/2023MNRAS.524.2195S}
}

@ARTICLE{Kaiser87,
       author = {{Kaiser}, Nick},
        title = "{Clustering in real space and in redshift space}",
      journal = {\mnras},
         year = 1987,
        month = jul,
       volume = {227},
        pages = {1-21},
          doi = {10.1093/mnras/227.1.1},
       adsurl = {https://ui.adsabs.harvard.edu/abs/1987MNRAS.227....1K}
}

@ARTICLE{James1975,
       author = {{James}, F. and {Roos}, M.},
        title = "{Minuit - a system for function minimization and analysis of the parameter errors and correlations}",
      journal = {Computer Physics Communications},
         year = 1975,
        month = dec,
       volume = {10},
       number = {6},
        pages = {343-367},
          doi = {10.1016/0010-4655(75)90039-9},
       adsurl = {https://ui.adsabs.harvard.edu/abs/1975CoPhC..10..343J}
}

@ARTICLE{Lewis2000,
       author = {{Lewis}, Antony and {Challinor}, Anthony and {Lasenby}, Anthony},
        title = "{Efficient Computation of Cosmic Microwave Background Anisotropies in Closed Friedmann-Robertson-Walker Models}",
      journal = {\apj},
         year = 2000,
        month = aug,
       volume = {538},
       number = {2},
        pages = {473-476},
          doi = {10.1086/309179},
archivePrefix = {arXiv},
       eprint = {astro-ph/9911177},
 primaryClass = {astro-ph},
       adsurl = {https://ui.adsabs.harvard.edu/abs/2000ApJ...538..473L}
}

@ARTICLE{Sifon2015,
       author = {{Sif{\'o}n}, Crist{\'o}bal and {Hoekstra}, Henk and {Cacciato}, Marcello and {Viola}, Massimo and {K{\"o}hlinger}, Fabian and {van der Burg}, Remco F.~J. and {Sand}, David J. and {Graham}, Melissa L.},
        title = "{Constraints on the alignment of galaxies in galaxy clusters from \raisebox{-0.5ex}\textasciitilde14 000 spectroscopic members}",
      journal = {\aap},
         year = 2015,
        month = mar,
       volume = {575},
          eid = {A48},
        pages = {A48},
          doi = {10.1051/0004-6361/201424435},
archivePrefix = {arXiv},
       eprint = {1406.5196},
 primaryClass = {astro-ph.CO},
       adsurl = {https://ui.adsabs.harvard.edu/abs/2015A&A...575A..48S}
}

@ARTICLE{Hoffmann2022,
       author = {{Hoffmann}, Kai and {Secco}, Lucas F. and {Blazek}, Jonathan and {Crocce}, Martin and {Tallada-Cresp{\'\i}}, Pau and {Samuroff}, Simon and {Prat}, Judit and {Carretero}, Jorge and {Fosalba}, Pablo and {Gazta{\~n}aga}, Enrique and {Castander}, Francisco J. and {DES Collaboration}},
        title = "{Modeling intrinsic galaxy alignment in the MICE simulation}",
      journal = {\prd},
         year = 2022,
        month = dec,
       volume = {106},
       number = {12},
          eid = {123510},
        pages = {123510},
          doi = {10.1103/PhysRevD.106.123510},
archivePrefix = {arXiv},
       eprint = {2206.14219},
 primaryClass = {astro-ph.CO},
       adsurl = {https://ui.adsabs.harvard.edu/abs/2022PhRvD.106l3510H}
}

@ARTICLE{Q1-SP028,
       author = {{Euclid Collaboration: Laigle}, C. and {Gouin}, C. and {Sarron}, F. and others},
       title = "{Euclid Quick Data Release (Q1). Galaxy shapes and alignments in the cosmic web}",
      journal = {A\&A, in press (Euclid Q1 SI), \url{https://doi.org/10.1051/0004-6361/202554651}},
         year = 2025,
        month = mar,
          eid = {arXiv:2503.15333},
        pages = {arXiv:2503.15333},
archivePrefix = {arXiv},
       eprint = {2503.15333},
 primaryClass = {astro-ph.GA},
       adsurl = {https://ui.adsabs.harvard.edu/abs/2025arXiv250315333E}
}

@ARTICLE{EuclidSkyOverview,
author = {{Euclid Collaboration: Mellier}, Y. and {Abdurro'uf} and {Acevedo~Barroso}, J.A. and others},
	title = {Euclid - I. Overview of the Euclid mission},
	DOI= "10.1051/0004-6361/202450810",
	url= "https://doi.org/10.1051/0004-6361/202450810",
	journal = {A\&A},
	year = 2025,
	volume = 697,
	pages = "A1",
}

@ARTICLE{EuclidSkyFlagship,
author = {{Euclid Collaboration: Castander}, F. and {Fosalba}, P. and {Stadel}, J. and others},
	title = {Euclid - V. The Flagship galaxy mock catalogue: A comprehensive simulation for the Euclid},
	DOI= "10.1051/0004-6361/202450853",
	url= "https://doi.org/10.1051/0004-6361/202450853",
	journal = {A\&A},
	year = 2025,
	volume = 697,
	pages = "A5",
}

@ARTICLE{EP-NavarroGirones,
       author = {{Euclid Collaboration: Navarro-Giron{\'e}s}, D. and {Tutusaus}, I. and {Crocce}, M. and others},
       title = "{Euclid preparation. Impact of galaxy intrinsic alignment modelling choices on Euclid 3x2pt cosmology}",
      journal = {A\&A, submitted},
         year = 2026,
        month = feb,
          eid = {arXiv:2602.16448},
        pages = {arXiv:2602.16448},
          doi = {10.48550/arXiv.2602.16448},
archivePrefix = {arXiv},
       eprint = {2602.16448},
 primaryClass = {astro-ph.CO},
       adsurl = {https://ui.adsabs.harvard.edu/abs/2026arXiv260216448E}
}

@ARTICLE{EP-Hoffmann,
       author = {{Euclid Collaboration: Hoffmann}, K. and {Paviot}, R. and {Joachimi}, B. and others},
       title = "{Euclid preparation. Calibrated intrinsic galaxy alignments in the Euclid Flagship simulation}",
      journal = {A\&A, accepted},
         year = 2026,
        month = jan,
          eid = {arXiv:2601.07785},
        pages = {arXiv:2601.07785},
archivePrefix = {arXiv},
       eprint = {2601.07785},
 primaryClass = {astro-ph.CO},
       adsurl = {https://ui.adsabs.harvard.edu/abs/2026arXiv260107785E}
}

\begin{appendix}
\begin{appendix} 

\section{Model implementation}
\label{sec:appendix_theory}
Given Eq. \eqref{eq:nl_bias_pgg}, we can express the galaxy-galaxy power spectrum as \citep{Saito14,Krause2021}
\begin{align}
P_{\rm{gg}}(k)=& \ b_1^2 \, P_{\delta\delta}(k)+b_1 b_2 \, P_{\delta  b_2}(k)+b_1 b_{s^2} \,P_{\delta s^2}(k) \nonumber \\
& +b_1 b_{3 \mathrm{nl}} \, P_{\delta b_{3 \mathrm{nl}}}(k)+\frac{1}{4} b_2^2 \, P_{b_2 b_2}(k) \\
& +\frac{1}{2} b_2 b_{s^2} \, P_{b_2 s^2}(k)+\frac{1}{4} b_s^2 \, P_{s^2 s^2}(k) \, . \nonumber
\end{align}
Our model for the galaxy power spectrum includes all contributions listed above, as implemented in the \texttt{FASTPT} \citep{McEwen16,Fang17} package. The linear power spectrum is calculated with \texttt{camb} \citep{Lewis2000}. The nonlinear matter power spectrum $P_{\delta\delta}(k)$ is estimated with the latest version of \texttt{HMcode} \citep{Mead2021}. We have tested that this choice gives results comparable to other prescriptions, such as the \texttt{Halofit} model \citep{Takahashi12}. The second-order and third-order non-local biases, $b_{s^2}$ and $b_{3 \mathrm{nl}}$, are fixed assuming a local Lagrangian bias prescription \citep{Baldauf12,Saito14,Krause2021,Pandey2022} such that\footnote{In the \texttt{FASTPT} implementation, the 32/315 factor that usually scales $b_{3 \mathrm{nl}}$ is included in the one loop integral contributions.}
\begin{equation}
\label{eq:lagrangian_bias}
 b_{s 2}=-\frac{4}{7} \, \left(b_1-1\right) \, , \\
 b_{3 \mathrm{nl}}=b_1-1 \, . 
\end{equation}
The intrinsic alignment power spectra are given by \citep{Blazek19,Krause2021,Secco21},
\begin{equation}\label{eq:TATT1}
\begin{aligned}
P_{\mathrm{\delta I}}(k) = & \, C_1 P_{\delta\delta}(k)+C_{1 \delta} \, P_{0 \mid 0 E}(k)+C_2 P_{0 \mid E 2}(k) \, , \\
P_{\mathrm{II}, EE}(k)= & \, C_1^2P_{\delta\delta}+2 C_1 C_{1 \delta}  \, P_{0 \mid 0 E}(k)+C_{1 \delta}^2 \, P_{0 E \mid 0 E}(k) \\
& +C_2^2 \, P_{E 2 \mid E 2}(k)+2 C_1 C_2 \, P_{0 \mid E 2}(k) \\
& +2 C_{1 \delta} C_2 \, P_{0 E \mid E 2}(k) \, , \\
P_{\mathrm{II},BB}(k)= & \, C_{1 \delta}^2 \, P_{0 B \mid 0 B}(k)+C_2^2 \, P_{B 2 \mid B 2}(k)+2 C_{1 \delta} A_2 \, P_{0 B \mid B 2}(k) \, . 
\end{aligned}
\end{equation}
For the TATT model, $B$-modes are included in the modelling, as density weighting produces an additional $B$-mode contribution. 
These perturbative contributions can be computed from the \texttt{FASTPT} package, we refer the reader to \cite{Blazek19} for technical details. 

We can determine projected correlation functions given a weighted set of Hankel transforms along the l.o.s.,
\begin{align}
\label{eq:w_gp_estimation}
w_{\rm{g g}}\left(r_{\mathrm{p}}\right)= &\int_{z_{\rm min}}^{z_{\rm max}} \mathrm{d} z \, \mathcal{W}^{ii}(z) \int_{0}^{\infty} \frac{\mathrm{d} k_{\perp} k_{\perp}}{2 \pi} J_0\left(k_{\perp} r_{\mathrm{p}}\right) P_{\rm gg}\left(k_{\perp}, z\right) \, , \\
w_{\rm{g+}}\left(r_{\mathrm{p}}\right)=-&\int_{z_{\rm min}}^{z_{\rm max}}\mathrm{d} z \, \mathcal{W}^{ij}(z) \int_{0}^{\infty} \frac{\mathrm{d} k_{\perp} k_{\perp}}{2 \pi} J_2\left(k_{\perp} r_{\mathrm{p}}\right) P_{\mathrm{gI}}\left(k_{\perp}, z\right) \,,  \\
w_{++,(\times\times)}\left(r_{\mathrm{p}}\right) = &\int_{z_{\rm min}}^{z_{\rm max}} \mathrm{d} z \, \mathcal{W}^{j j}(z) \nonumber \\
\times & \,\Bigg[ \Bigg. \int_{0}^{\infty} \frac{\mathrm{d} k_{\perp} k_{\perp}}{4 \pi}\left[J_0\left(k_{\perp} r_{\mathrm{p}}\right) \pm J_4\left(k_{\perp} r_{\mathrm{p}}\right)\right] P_{\mathrm{II},\mathrm{EE}}\left(k_{\perp}, z\right) \\
+ & \,\left[J_0\left(k_{\perp} r_{\mathrm{p}}\right) \mp J_4\left(k_{\perp} r_{\mathrm{p}}\right) \right]P_{\mathrm{II},\mathrm{BB}}(k_{\perp},z) \nonumber \, \Bigg. \Bigg] \, ,\\
\label{eq:w_gp_estimation2}
\end{align} 
where $z_{\rm min}$ and $z_{\rm max}$ denote the survey boundaries.$\mathcal{W}$ is the projection kernel \citep{mandelbaum11} given by
\begin{equation}
\mathcal{W}^{ij}(z)=\frac{n^i(z) n^j(z)}{\chi^2(z) \mathrm{d} \chi / \mathrm{d} z} \;\left[\int_{z_{\rm min}}^{z_{\rm max}} \mathrm{d} z \frac{n^i(z) n^j(z)}{\chi^2(z) \mathrm{d} \chi / \mathrm{d} z}\right]^{-1}. 
\end{equation}
In the above equation, $n^i(z)$ and $n^j(z)$ represent the redshift distribution of the density and source sample respectively, and $\chi(z)$ is the comoving l.o.s. distance. To model projected quantities for more realistic cases (i.e. for a galaxy photometric redshift survey), we refer the readers to \citet{Joachimi11} and \citet{Samuroff22}.

In addition to clustering and shape alignment correlations, magnification and lensing can have direct effects on IA measurements \citep{Samuroff22}. In Flagship, we do have access directly to both un-magnified galaxy positions, and intrinsic galaxy shapes, such that we do not need to model these effects in projected IA statistics.

To model RSD, we followed a similar approach as in \cite{Samuroff22}. As we are considering projected correlation functions, which are weakly affected by galaxy anisotropies, we used the simple Kaiser model \citep{Kaiser87}
\begin{equation}
P_{\rm gg}^{s}(\textbf{k},z)= b_1^2\left(1+\beta_z \,\mu\right)^2 P_{\rm gg}(k,z) \, , 
\end{equation}
where $\beta_z  \equiv  f(z)/b_1(z)$ is the ratio of the logarithmic growth rate over the bias of the clustering sample. $\mu$ corresponds to the angle between the mode $\textbf{k}$ and the axis of line of sight $\mu=\hat{\mathbf{k}} \cdot \hat{\mathbf{z}}$. The two-dimensional power spectrum can then be decomposed as a sum of Legendre polynomials $L_{\ell}$ as
\begin{equation}
\label{eq:pk_decom}
P_{\rm gg}^{s}(\textbf{k},z)=\sum_{0}^{\ell=4} L_{\ell}(\mu) P_{\ell}^{s}(k) \, .
\end{equation}
In the linear regime, only the three first even multipoles, namely the monopole, quadrupole, and hexadecapole are non-zero and are given by \citep{Kaiser87} and expressed as
\begin{equation}
\begin{aligned}
& P_0^s(k)=\left(1+\frac{2}{3} \beta+\frac{1}{5} \beta^2\right) P(k) \, , \\
& P_2^s(k)=\left(\frac{4}{3} \beta+\frac{4}{7} \beta^2\right) P(k) \, ,  \\
& P_4^s(k)=\left(\frac{8}{35} \beta^2\right) P(k) \,.
\end{aligned}
\end{equation}
The configuration space multipoles can then be estimated by Hankel transforms into
\begin{equation}
\xi_{\ell}^{s}(r)=\frac{\rm{i}^{\ell}}{2 \pi^2} \int_0^{\infty} k^2 j_{\ell}(k r) \,P_{\ell}^{s}(k) \, \mathrm{d} k \, ,
\end{equation}
where $j_{\ell}$ corresponds to the spherical Bessel function of the order of $\ell$.
We can then reconstruct the 2D anisotropic 2-point correlation function as 
\begin{equation}
\xi_{\rm gg}^{s}\left(r_{\mathrm{p}}, \Pi\right)=\sum_{\ell=0,2,4} L_{\ell}(\nu)  \,\xi_{\ell}(s) \, ,
\end{equation}
with $s = \sqrt{r_{\mathrm{p}}^2 +\Pi^2}$ and $\nu = \Pi /s$. The integration of $\xi_{\rm g g}^{s} \left(r_{\mathrm{p}}, \Pi\right)$ along the l.o.s. provides us with an estimate of $w_{\rm gg}(r_{\mathrm{p}})$ in redshift space. Although RSD have a small impact on $w_{\rm{g}+}$, a similar effect exists due to the projection of 3D shapes in 2D shapes \citep{Singh15,Samuroff22}, which suppresses the alignment signal on scales $|\Pi| \gg r_{\mathrm{p}}$. This effect can be modelled in a similar fashion \citep{Singh16}. However, \cite{Samuroff22} demonstrated that this effect is very small (below $1\%$ on scales $r_{\mathrm{p}}< 70 \, h^{-1}$Mpc) and we therefore do not model IA anisotropies in this paper.

\clearpage \onecolumn\section{Full flagship IA measurements} 
\begin{figure}[!htbp] \centering \includegraphics[width=0.85\textwidth]{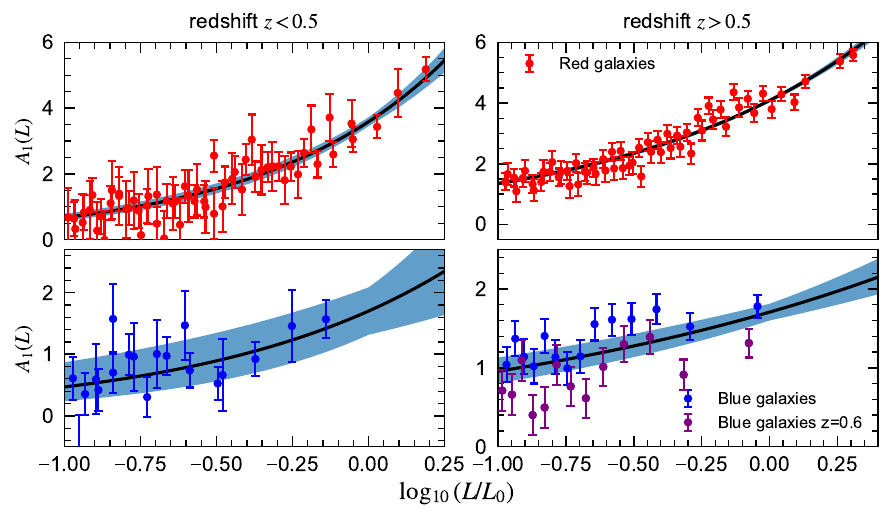} \caption{IA linear amplitude versus luminosity for the NLA model. The low-$z$ sample corresponds to the redshift slices below z = 0.5, while the high-z sample corresponds to the redshift slice 0.5 < z < 0.9. The luminosity is computed as $\logten(L/L_0) = -(M_r - M_0)/2.5$. The solid lines represent the best-fit models, while the shaded areas correspond to the 68$\%$ confidence regions, determined by propagating the errors on the power-law amplitudes and indices. At high redshift, the non-negligible variation in linear amplitude $A_1$ between $z=0.6$ and $z=0.8$ results in a high $\chi^2_{\rm r}$ value.} \label{fig:Lr_vs_A1_red_appendix} \end{figure}

We present in Fig. \ref{fig:Lr_vs_A1_red_appendix} for completeness the full set of the samples used to investigate the luminosity dependence of intrinsic alignments. \label{sec:Appendixb}  \FloatBarrier \begin{figure}[!htbp] \begin{center} \includegraphics[width=0.85\textwidth]{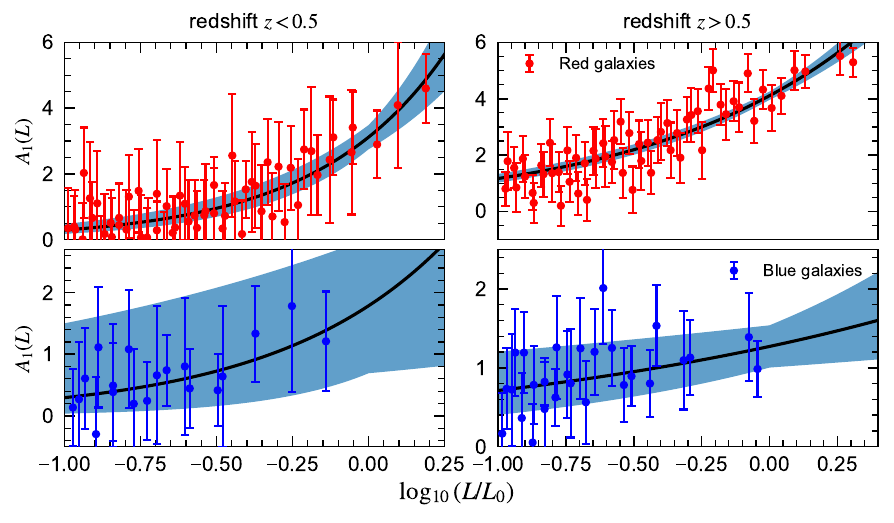} \caption{Same as Fig. \ref{fig:Lr_vs_A1_red_appendix}, but this time presenting all the samples used with the TATT model.} \label{fig:Lr_vs_A1_red_appendix_TATT} \end{center} \end{figure} \FloatBarrier \clearpage
\twocolumn
\section{Dependence of IA amplitude on satellite fraction} \label{sec:fsat}

\begin{figure}[!ht]
  \includegraphics[width=\columnwidth]{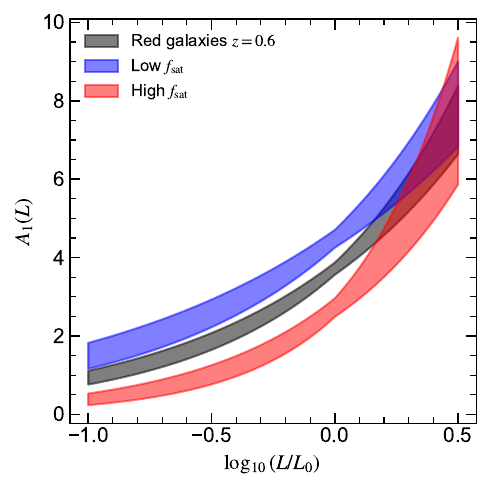}
  \caption{Evolution of the IA-luminosity relation as a function of satellite fraction. The satellite fraction causes a shift in the linear IA amplitude.}
  \label{fig:A1_vs_fsat}
\end{figure}

Studies in simulations \citep{Pereira2008,Pereira2010,2017arXiv171207818W} have shown that satellite galaxies tend to align radially towards their host halo's centre. This is due to a continuous torquing mechanism that
aligns the satellite's major axes towards the direction of the gravitational potential gradient. However, neither \cite{Sifon2015} nor \cite{chisari2014} found evidence of satellite alignment in clusters. Observationally, satellite alignment was detected in the overlapping region between KiDS and GAMA in \cite{Johnston19} and \cite{Georgiou2019b}. In particular, \cite{Georgiou2019b} showed that satellites are radially aligned towards their host dark matter halo at small radii (with a different amplitude for the red and blue populations), with a vanishing signal at larger radii. Since satellites are randomly oriented at large scales, the presence of these galaxies will suppress the large-scale IA signal as observed in \cite{Johnston19}. This means that the satellite fraction affects not only the one-halo term signal but also the two-halo term in the linear regime. Thus, galaxy samples with similar observational properties such as $r$-band magnitude or stellar mass, will yield different alignment amplitudes as a function of satellite fraction when using the NLA and TATT perturbative frameworks (developed for central galaxies) to model the data. This is why an IA halo model has been proposed in \cite{fortuna21a} to provide a unified model to describe the IA signal down to the one-halo regime. Here, we do not aim to use this model but instead investigate how satellite fractions impact the linear IA amplitude as done in \cite{fortuna21a}.

 \begin{figure}[h!]
 \begin{center}
\includegraphics[width=\columnwidth]{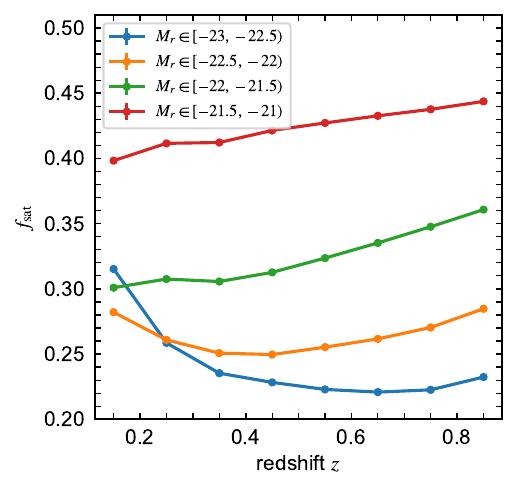}
\caption{Evolution of the Flagship satellite fraction $f_{\rm sat}$ over the redshift range $0.1 < z < 0.9$. We present $f_{\rm sat}$ for the red galaxy population, in bins of absolute rest-frame magnitude in the $r$ band.}
\label{fig:fsat_fs2}
\end{center}
\end{figure}

We present our results in Fig. \ref{fig:A1_vs_fsat} for the red population. We perform our analysis in the redshift bin $[0.5,0.7]$ that has an effective redshift $z=0.6$. In this figure, each coloured area corresponds to a best-fit power law, determined in the same way as in Fig. \ref{fig:Lr_vs_A1_red_appendix}. The low and high $f_{\text{sat}}$ are determined by sub-sampling galaxy samples (with initial number density $n = 3 \times 10^{-4} h^{-3}$ Mpc$^3$) to obtain new samples with satellite fraction $40 \%$ above (high $f_{\text{sat}}$) or below (low $f_{\text{sat}}$) the initial satellite fraction of each sample. Since the galaxy samples are first binned according to magnitude/luminosity, the linear bias remains the same such that any differences in the IA amplitude are a result of a change of the satellite fraction. We observe a correlation with satellite fraction, with faint galaxy samples having fewer satellites exhibiting a higher IA amplitude, in agreement with previous results \citep{Johnston19,fortuna21a}. 
At higher luminosity ($\logten L/L_0 > 0.2$), the satellite fraction on the sample does not significantly impact the recovered IA linear amplitude. The main reason is that the brightest tail of the luminosity function is dominated by central galaxies. The few satellite galaxies with such high luminosity will reside in massive clusters where the alignment mechanism due to tidal fields will be stronger compared to fainter satellites that populate less massive halos \citep{Georgiou2019b}. Therefore, deriving a luminosity-IA relation from samples with different satellite fractions can lead to a regime where no concrete dependence on luminosity is observed for faint galaxies. This explains the measured value of $\beta_1$ consistent with zero in \cite{fortuna21b}, see Sect. \ref{sec:luminosity}. In addition, this feature might explain in part the relatively low IA signal we observed (at least for $L < L_0$) compared to the observations that we find in Fig. \ref{fig:LR_vs_A1_obs}. Therefore, future methods that aim to model realistic galaxy alignments within $N$-body simulations should incorporate satellite fraction as a key parameter to perform calibration on observational and hydrodynamical simulations. 

We present in Fig. \ref{fig:fsat_fs2} the evolution of the satellite fraction for the red galaxy population, in the range $0.1 < z < 0.9$. For reference, the satellite fractions of BOSS LOWZ, BOSS CMASS, and eBOSS LRG samples are approximately 10 $\%$ \citep{White2011,Singh15,Zhai2017}. As we can see, Flagship seems to overestimate the number of satellites for bright galaxies samples ($M_{r} > - 22$), particularly at low redshift. The KiDS samples presented in \cite{fortuna21b,Fortuna2024} have mean absolute $r$-band magnitudes ranging from approximatively $-21.5$ to $-22$ for satellite fractions ranging from 10 $\%$ to 30 $\%$ (from bright to dimmer samples). Only the fainter dense sample has a satellite fraction close to Flagship's predictions. These non-negligible differences between Flagship and observations can explain, in part, why our measurements of $A_1$ are slightly below the observations.

\section{Redshift prior}
\label{sec:redshiftprior}

 \begin{figure}[h!]
 \begin{center}
\includegraphics[width=\columnwidth]{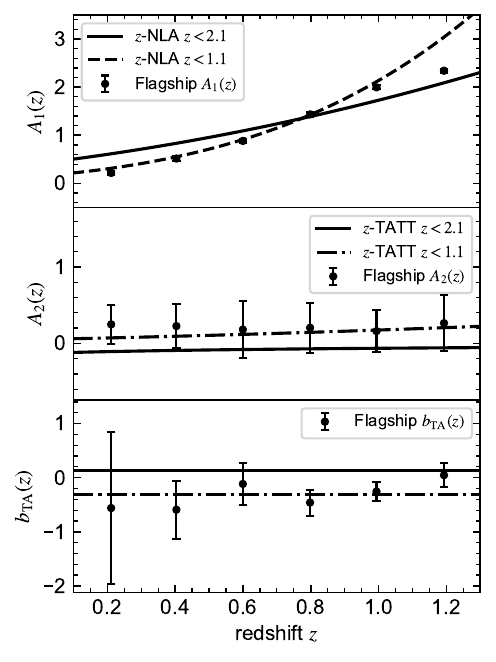}
\caption{Redshift evolution of the linear IA amplitude in Flagship. The points with errors represent the measurement of $A_1$ for the total Flagship sample, while the dashed and continuous lines represent respectively the fits of the data with Eq. \eqref{eq:znla} in the range $0.1 < z < 1.1$ and $0.1 < z < 2.1$.  }
\label{fig:z_evol_cutted_range}
\end{center}
\end{figure}

\begin{table}[h!]
\centering
\caption{Constraints on the redshift evolution of the $z$-NLA and $z$-TATT models in the redshift range $0.1 < z < 1.1$.}
\setlength{\tabcolsep}{2pt} 
\begin{tabular}{c|cccc}
\hline \hline
Model & Constraint 
& Total & Red & Blue \\
\hline
\noalign{\vskip 1pt}
\noalign{\vskip 1pt}
\multirow{3}{*}{$z$-NLA} & \text{ $A_{1z}$} & $0.91 \pm 0.03$  & $1.71 \pm  0.03$ & $0.62 \pm  0.03$ \\
 & $\eta_1$ & $3.85 \pm 0.15 $  & $3.12 \pm 0.12$ & $3.93 \pm 0.24 $  \\
 & $\chi^2_{\rm r}$ & 2.08 & 2.84 & 1.26 \\
 \hline
\multirow{6}{*}{$z$-TATT} & \text{ $A_{1z}$} & $0.98 \pm 0.05 $  & $1.81 \pm  0.06$ & $0.59 \pm  0.05$ \\
 & $\eta_1$ & $3.67 \pm 0.26 $  & $4.38 \pm 0.40$ & $2.45 \pm 0.08 $  \\
  & \text{$A_{2z}$} &  \phantom{0}$0.12^{+0.11}_{-0.10}$  & \phantom{0} $0.03^{+0.13}_{-0.13}$ &  \phantom{0}$0.10^{+0.10}_{-0.06}$  \\
 & $\eta_2$ &  \phantom{0} $1.77^{+2.60}_{-2.39}$  &  \phantom{0} $1.80^{+3.00}_{-2.88}$ &  \phantom{0} $2.47^{+2.54}_{-2.74}$  \\
 &  \text{ $b_{\rm TA}$} & $-0.31^{+0.11}_{-0.12}$  & $-0.10^{+0.08}_{-0.08}$ & $-0.41^{+0.17}_{-0.16}$   \\

 & $\chi^2_{\rm r}$ & 0.57 & 0.86 & 0.37 \\
 \hline
\end{tabular}
\tablefoot{The values quoted here correspond to the $16\%$,
    $50\%$, and $84 \%$ percentile of the distribution determined with the \texttt{nautilus} sampler.}
\label{tab:znla_appendix}
\end{table}

\begin{figure}
 \begin{center}
\includegraphics[width=\columnwidth]{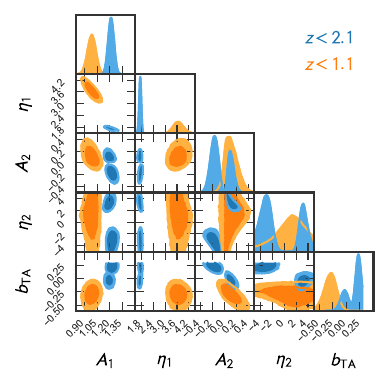}
\caption{Posterior distribution obtained by fitting the observed Flagship's IA redshift evolution with the $z$-TATT model in the redshift range $0.1 < z < 1.1$ and $0.1 < z < 2.1$.}
\label{fig:constraintappendix}
\end{center}
\end{figure}

We present in Fig. \ref{fig:z_evol_cutted_range} the evolution of the Flagship IA in the redshift range $0.1 < z < 1.1$. A single power-law fit (Eq. \ref{eq:znla}) on this restricted redshift range can reproduce more accurately the lower redshift bins but will overestimate the IA amplitude at higher redshift. The constraints for the redshift range $0.1 < z < 1.1$ are shown in Table \ref{tab:znla_appendix}. Figure \ref{fig:constraintappendix} shows how the parameter constraints shift for the total sample between the two analyses. This noticeable shift shows that a simple redshift power-law does not accurately fit the Flagship IA  over the full redshift range $0.1 < z < 2.1$.

\referee{The bimodality visible in Fig. \ref{fig:z_evol_cutted_range} when fitting the full redshift range $0.1 < z < 2.1$ is a direct consequence of model misspecification at high redshift. When restricted to $z < 1.1$, $b_{\mathrm{TA}}$ is approximately constant, and $A_1$ is reasonably well described by a single power law, resulting in a well-behaved unimodal posterior. At higher redshift, however, the $z$-TATT model struggles to simultaneously account for the breakdown of the power-law description of $A_1$ and the clear redshift evolution of $b_{\mathrm{TA}}$ seen in Fig.  \ref{fig:A2_vs_z}. The strong degeneracy between $A_2$ and $b_{\mathrm{TA}}$ amplifies this tension, as the sampler finds two distinct regions of parameter space providing comparably acceptable fits to the data. The bimodality is therefore due to  model misspecification beyond $z \sim 1.1$, further motivating the restricted redshift analysis presented in this appendix.}

\end{appendix}

\end{appendix}

\label{LastPage}

\end{document}